\documentclass[]{aa}
\usepackage{afterpage}
\usepackage{amsmath}
\usepackage{booktabs}
\usepackage{dcolumn}
\newcolumntype{d}[1]{D{.}{.}{#1}}
\newcolumntype{p}[1]{D{-}{-}{#1}}
\usepackage{graphicx}
\usepackage{longtable}
\usepackage{lscape}
\usepackage{rotating}
\usepackage[]{SIunits}
\usepackage[tight]{subfigure}
\usepackage{supertabular}

\begin{document}

\title{The Circumstellar Environments of High-Mass Protostellar Objects I: Submillimetre Continuum Emission}

\author{S.~J.~Williams\inst{1}
  \and G.~A.~Fuller\inst{1}
  \and T~.K.~Sridharan\inst{2}}

\institute{Department of Physics, UMIST, P.O. Box 88, Manchester, M60 1QD, United Kingdom
  \and Harvard-Smithsonian Center for Astrophysics, 60 Garden Street, MS 78, Cambridge, MA 02138}

\offprints{G.~A.~Fuller, 
  \email{g.fuller@umist.ac.uk}}

\date{Received date / Accepted date}

\abstract{We present maps of the 850 $\mu$m and 450 $\mu$m continuum emission
  seen towards a sample of 68 high-mass protostellar candidates with
  luminosities ranging from 10$^{2.5}$ $L_{\odot}$ to $\sim 10^{5}$
  $L_{\odot}$. Most of these candidate high-mass stars are in the earliest
  stages of evolution, and have not yet developed an ultra-compact HII region.
  We observe a variety of continuum emission morphologies, from compact
  symmetric sources through to multiple cores embedded in long filaments of
  emission.  We find on average there is a 65\% probability of an IRAS
  point-source having a companion detection at submillimetre wavelengths. The
  ratio of integrated flux to peak flux for our detections shows no strong
  dependence on distance, suggesting the emission we have observed is
  primarily from scale-free envelopes with power-law density structures.
  Assuming a near kinematic distance projection, the clumps we detect vary in
  mass from $\sim$1 M$_{\odot}$ to over 1000 M$_{\odot}$, with a mean clump
  mass of 330 M$_{\odot}$, column density of $9\times10^{23}$ cm$^{-2}$ and
  diameter of $\sim0.6$ pc. The high luminosity and low mass of the smallest
  clumps suggests they are accompanied by a minimal number of stellar
  companions, while the most massive clumps may be examples of young
  protogroups and protoclusters. We measure the spectral index of the dust
  emission ($\alpha$) and the spectral index of the dust grain opacity
  ($\beta$) towards each object, finding clumps with morphologies suggestive
  of strong temperature gradients, and of grain growth in their dense inner
  regions.  We find a mean value for $\beta$ of 0.9, significantly smaller
  than observed towards UCHII regions.

}

\titlerunning{The Circumstellar Environments of HMPOs I: Submillimetre Continuum Emission} 

\maketitle

\section{Introduction}
Our understanding of the processes involved in low-mass star formation has
matured steadily over the last forty years, and the concepts of gravity-driven
collapse and accretion-driven evolution appear to consistently explain how a
low-mass pre-main-sequence star can form from a cloud core. However, our
knowledge of how high-mass stars form has remained limited, primarily due to a
lack of candidate high-mass protostars to study. At the upper reaches of the
initial mass function, high-mass stars are statistically rare, and coupled
with a characteristically short evolutionary timescale, there have been few
chances to observe massive stars at the instance of formation.

Until recently, most young high-mass stars were first identified through the
detection of a radio-bright ultra-compact HII (UCHII) region, considered a
beacon pointing to the presence of a young high-mass star. As high-mass
protostars increase in mass and luminosity, they emit an ever larger number of
high energy UV photons which ionize the protostar's immediate surroundings,
hence the small, compact nature of a UCHII region is usually considered
evidence of the youthful status of the driving source (although debate
continues about the exact timescale of the UCHII stage; for a review, see
Kurtz et al. \cite{kurtz:ppiv}).

Unfortunately, a powerful protostar and UCHII region soon act to disrupt and
confuse their surroundings, so the initial conditions of the natal cloud and
the mechanisms that led to the formation of the massive protostar cannot be
unambiguously reconstructed. As a result, many questions about high-mass
protostars remain - in particular, do they form via processes similar to their
low-mass counterparts? To address the mechanisms that create and shape
high-mass stars, we must observe \emph{before} they have formed a UCHII
region, during the initial collapse of the star-forming core.

\subsection{The search for precursors of UCHII regions}
Clumps bearing the youngest high-mass protostars have proved particularly
difficult to find, even though their identifying characteristics have been
known for almost 25 years. For example, Habing \& Israel (\cite{habing1979})
predicted that candidate high-mass protostars should be founded embedded in
dense environments, and although highly luminous, they should not at this
stage be associated with HII regions. However, despite knowledge of these
distinguishing features, it was not until the last decade that samples of
candidate high-mass protostellar objects (HMPOs) were finally compiled.

Observations of these preliminary samples of HMPOs have allowed the first
glimpses of high-mass protostars in their earliest evolutionary states: a
typical core not yet associated with an ionised region is found to be larger,
more massive, and more turbulent than a UCHII-class protostar, with a typical
diameter of around 0.5-1.0 pc and a mass that may range from a few tens to a
few thousand solar masses (Brand et al.  \cite{brand2001}; Beuther et al.
\cite{beuther2002a}).  They are cooler, with typical dust temperature
averaging around 30-40 K (Sridharan et al.  \cite{sridharan}; Molinari et al.
\cite{molinari2000}), while the dust opacity usually has a spectral index of
around 2, suggestive of silicate dust grains (Molinari et al.
\cite{molinari2000}). Self-absorption profiles towards a number of candidate
HMPOs suggest infall may be an important part of the formation mechanism (eg.
Brand et al.  \cite{brand2001}; Fuller et al.  \cite{fuller2003}), while
outflow observations suggest that accretion is a significant process (Zhang et
al.  \cite{zhang2001}; Beuther et al.  \cite{beuther2002b}; Molinari et al.
\cite{molinari}). Water maser emission has also been detected towards
candidate protostars, a feature thought to be missing from more evolved
sources (eg. Palla et al. \cite{palla1993}; Sridharan et al.
\cite{sridharan}). Despite these advances, there is still much to be learned
about the pre-UCHII stage of high-mass star formation, and there remains a
need for additional candidates and further observations.

\subsubsection{A new sample of high-mass protostars}
Recently, Sridharan et al. (\cite{sridharan}; SBSMW hereinafter) identified a
new sample of HMPOs. The SBSMW sample is a flux-limited sample, constructed
through an analysis of the IRAS point-source catalogue: as young high-mass
stars are usually associated with UCHII regions, they began by initially
selecting bright IRAS detections ($S_{60}>90$ Jy and $S_{100}>500$ Jy) with
colour characteristics similar to known UCHII regions (they conform to the
Wood \& Churchwell (\cite{wood}) FIR colour criteria that selects UCHII
regions, and they also satisfy the additional Ramesh \& Sridharan
(\cite{ramesh}) criteria). Candidate sources detected in Galaxy-wide 5GHz
continuum surveys were removed, thereby rejecting sources already sufficiently
evolved to have ionized their surroundings. As a final requirement, successful
candidates must also be associated with CS(2-1) emission, an indicator of
dense molecular gas (Bronfman et al. \cite{bronfman}).

In total, sixty-nine IRAS point sources satisfied these cumulative criteria,
identifying these sources as potentially among the most massive and deeply
embedded pre-UCHII protostars in our Galaxy. The SBSMW sample has been studied
in detail over the last few years, and their status as high-mass candidate
protostars has been supported through observations of 1.2mm and 3.6cm
continuum emission (Beuther et al.  \cite{beuther2002a}; SBSMW), molecular
line emission (CS, CO and NH$_{3}$), and H$_{2}$O and CH$_{3}$OH maser
transitions (SBSMW) towards the sources.

This paper presents the results of a new set of submillimetre (submm)
observations of the SBSMW sample of candidate high-mass protostars. All but
one (IRAS 18517+0437) of the SBSMW candidate HMPOs were observed. An
additional source, IRAS 18449-0158, was observed but this source does not
satisfy the SBSMW criteria and is not included in any analysis. Our
observations are detailed in \S\ref{sec:observations}, with maps of the
reduced data found in \S\ref{sec:results}. We measure the multiplicity of the
detections in \S\ref{sec:ccf}, commenting on the position and morphologies of
the sample in \S\ref{sec:position and morphology}. We analyse the dust optical
depth in \S\ref{sec:dust optical depth}, and use the spectral index of the
emission to investigate the nature of the dust in \S\ref{sec:alpha}. We
calculate the mass characteristics of our sample in \S\ref{sec:mass}, and
consider the implications of the cumulative mass spectrum in \S\ref{sec:mass
  spectrum}. After a brief discussion and comparison of our results with the
IRAM 1.2mm continuum observations of Beuther et al.  (\cite{beuther2002a}) in
\S\ref{sec:discussion}, we conclude in \S\ref{sec:conclusion} with a summary
of our results.

This paper presents the first half of our study and analysis of the dust
emission; the companion to this paper presents the results of radiative
transfer modelling of the clumps (Williams et al. \cite{williams2004}).

\section{Observations and data reduction}
\label{sec:observations}
The sample of HMPOs was observed at 850 $\mu$m and 450 $\mu$m between March
2000 and June 2000 using the Submillimetre Common-User Bolometer Array (SCUBA)
on the James Clerk Maxwell Telescope (Holland et al. \cite{holland}). The
SBSMW sample target co-ordinates and date(s) of observation are listed in
Table \ref{tab:source list}. The SCUBA array covers a hexagonal 2.5$'$ field
of view with 97 and 37 pixels at 450 $\mu$m and 850 $\mu$m respectively. Maps
were formed simultaneously at both frequencies using the ``jiggle'' mode, in
which the telescope beam is moved around a 64-point pattern by the secondary
mirror in order to fully sample the sky.

\begin{table}
\begin{scriptsize}
  \begin{center}
    \begin{tabular}{lc@{                  }cd{-1}@{}d{-1}l}
      \toprule
      \toprule
      IRAS Source &\multicolumn{2}{c}{Position (J2000)} &\multicolumn{2}{c}{Distance (kpc)}   &Date(s) of\\
      &\multicolumn{1}{c}{$\alpha$} &\multicolumn{1}{c}{$\delta$} &\multicolumn{1}{c}{far} &\multicolumn{1}{c}{near} & observation \\
      \midrule
05358$+$3543      &05 39 10.4     &$+$35 45 19    &1.8  &       &03/18          \\
05490$+$2658      &05 52 12.9     &$+$26 59 33    &2.1  &       &03/18          \\
05553$+$1631      &05 58 13.9     &$+$16 32 00    &2.5  &       &03/18          \\
18089$-$1732      &18 11 51.3     &$-$17 31 29    &13.0 &3.6    &05/11          \\
18090$-$1832      &18 12 01.9     &$-$18 31 56    &10.0 &6.6    &05/11          \\
18102$-$1800      &18 13 12.2     &$-$17 59 35    &14.0 &2.6    &05/11          \\
18151$-$1208      &18 17 57.1     &$-$12 07 22    &3.0  &       &05/11          \\
18159$-$1550      &18 18 47.6     &$-$15 48 54    &11.7 &4.7    &05/11          \\
18182$-$1433      &18 21 07.9     &$-$14 31 53    &11.8 &4.5    &05/11          \\
18223$-$1243      &18 25 11.1     &$-$12 42 15    &12.4 &3.7    &05/11          \\
18247$-$1147      &18 27 31.1     &$-$11 45 56    &9.3  &6.7    &05/11          \\
18264$-$1152      &18 29 14.3     &$-$11 50 26    &12.5 &3.5    &05/11          \\
18272$-$1217      &18 30 02.7     &$-$12 15 27    &2.9  &       &05/11          \\
18290$-$0924      &18 31 44.8     &$-$09 22 09    &10.5 &5.3    &05/11          \\
18306$-$0835      &18 33 21.8     &$-$08 33 39    &10.7 &4.9    &05/11          \\
18308$-$0841      &18 33 31.9     &$-$08 39 17    &10.7 &4.9    &05/11          \\
18310$-$0825      &18 33 47.2     &$-$08 23 35    &10.4 &5.2    &05/11          \\
18337$-$0743      &18 36 29.0     &$-$07 40 33    &11.5 &4      &05/11, 05/30   \\
18345$-$0641      &18 37 16.8     &$-$06 38 32    &9.5  &       &05/11          \\
18348$-$0616      &18 37 29.0     &$-$06 14 15    &9.0  &6.3    &05/11          \\
18372$-$0541      &18 39 56.0     &$-$05 38 49    &13.4 &1.8    &05/11          \\
18385$-$0512      &18 41 12.0     &$-$05 09 07    &13.1 &2      &05/23          \\
18426$-$0204      &18 45 12.8     &$-$02 01 12    &13.5 &1.1    &05/23          \\
18431$-$0312      &18 45 46.9     &$-$03 09 24    &8.2  &6.7    &05/23          \\
18437$-$0216      &18 46 22.7     &$-$02 13 24    &7.3  &       &05/23          \\
18440$-$0148      &18 46 36.3     &$-$01 45 23    &8.3  &       &05/23          \\
18445$-$0222      &18 47 10.8     &$-$02 19 06    &9.4  &5.3    &06/13          \\
18447$-$0229      &18 47 23.7     &$-$02 25 55    &8.2  &6.6    &05/30          \\
18449$-$0158      &18 47 35.6     &$-$01 55 26    &8.7  &5.9    &06/13          \\
18454$-$0136      &18 48 03.7     &$-$01 33 23    &11.9 &2.7    &05/30          \\
18454$-$0158      &18 48 01.3     &$-$01 54 49    &5.6  &       &06/13          \\
18460$-$0307      &18 48 39.2     &$-$03 03 53    &9.5  &5.2    &05/30          \\
18470$-$0044      &18 49 36.7     &$-$00 41 05    &8.2  &       &06/13          \\
18472$-$0022      &18 49 50.7     &$-$00 19 09    &11.1 &3.2    &05/30          \\
18488$+$0000      &18 51 24.8     &$+$00 04 18    &8.9  &5.4    &06/13          \\
18521$+$0134      &18 54 40.8     &$+$01 38 01    &9.0  &5      &05/30          \\
18530$+$0215      &18 55 34.2     &$+$02 19 08    &8.7  &5.1    &06/13          \\
18540$+$0220      &18 56 35.6     &$+$02 24 54    &10.6 &3.3    &05/30          \\
18553$+$0414      &18 57 53.0     &$+$04 18 06    &12.9 &0.6    &06/19          \\
18566$+$0408      &18 59 09.9     &$+$04 12 14    &6.7  &       &05/30          \\
19012$+$0536      &19 03 45.1     &$+$05 40 40    &8.6  &4.6    &05/23          \\
19035$+$0641      &19 06 01.1     &$+$06 46 35    &2.2  &       &05/23          \\
19074$+$0752      &19 09 53.3     &$+$07 57 22    &8.9  &3.7    &05/23          \\
19175$+$1357      &19 19 49.1     &$+$14 02 46    &10.6 &       &05/23          \\
19217$+$1651      &19 23 58.8     &$+$16 57 36    &10.5 &       &05/23          \\
19220$+$1432      &19 24 19.7     &$+$14 38 03    &5.5  &       &06/13          \\
19266$+$1745      &19 28 54.0     &$+$17 51 56    &10.0 &0.3    &05/30          \\
19282$+$1814      &19 30 28.1     &$+$18 20 53    &8.2  &1.9    &05/30          \\
19403$+$2258      &19 42 27.2     &$+$23 05 12    &6.3  &2.4    &06/13          \\
19410$+$2336      &19 43 11.6     &$+$23 44 06    &6.4  &2.1    &05/30          \\
19411$+$2306      &19 43 18.1     &$+$23 13 59    &5.8  &2.9    &06/13          \\
19413$+$2332      &19 43 29.0     &$+$23 40 04    &6.8  &1.8    &05/30          \\
19471$+$2641      &19 49 09.9     &$+$26 48 51    &2.4  &       &06/13          \\
20051$+$3435      &20 07 03.8     &$+$34 44 35    &3.7  &1.6    &05/11          \\
20081$+$2720      &20 10 11.5     &$+$27 29 06    &0.7  &       &05/11, 05/23   \\
20126$+$4104      &20 14 26.0     &$+$41 13 31    &1.7  &       &05/11          \\
20205$+$3948      &20 22 22.0     &$+$39 58 05    &4.5  &       &05/11, 05/30   \\
20216$+$4107      &20 23 23.8     &$+$41 17 40    &1.7  &       &05/11          \\
20293$+$3952      &20 31 10.7     &$+$40 03 10    &2.0  &1.3    &05/11          \\
20319$+$3958      &20 33 49.4     &$+$40 08 45    &1.6  &       &05/11          \\
20332$+$4124      &20 35 00.5     &$+$41 34 48    &3.9  &       &05/23          \\
20343$+$4129      &20 36 07.1     &$+$41 40 01    &1.4  &       &05/23          \\
22134$+$5834      &22 15 09.1     &$+$58 49 09    &2.6  &       &05/06          \\
22551$+$6221      &22 57 05.2     &$+$62 37 44    &0.7  &       &05/06, 05/23   \\
22570$+$5912      &22 59 06.5     &$+$59 28 28    &5.1  &       &05/06, 05/23   \\
23033$+$5951      &23 05 25.2     &$+$60 08 11    &3.5  &       &05/06          \\
23139$+$5939      &23 16 09.3     &$+$59 55 23    &4.8  &       &05/06          \\
23151$+$5912      &23 17 21.1     &$+$59 28 49    &5.7  &       &05/06          \\
23545$+$6508      &23 57 05.2     &$+$65 25 11    &0.8  &       &05/06, 05/23   \\
\bottomrule
    \end{tabular}
    \end{center}
\end{scriptsize}

\caption{
  Positions of the IRAS point sources satisifying the Sridharan et al. (\cite{sridharan}) criteria, precessed to J2000 co-ordinates, alongside the kinematic distance of the IRAS source and date(s) of observation. All sources were observed during the spring and summer of 2000. All distances are taken from Sridharan et al. \cite{sridharan} with the exception of IRAS 18449-0158, for which we derive the kinematic distance using $V_{LSR}$ from the CS(2-1) observations of Bronfman et al.(\cite{bronfman}). Candidates with only the far kinematic distance listed have had their distance uncertainty resolved.}
\label{tab:source list}
\end{table}

The data were reduced using the SCUBA User Reduction Facility (SURF; Jenness
\& Lightfoot \cite{jenness}). Correlated sky noise was removed using the
REMSKY routine, based on the signal from a hand-picked sample of bolometers
considered free from source emission. Maps were extinction calibrated from
skydips and flux calibrated in terms of Jy beam$^{-1}$ from maps of Uranus,
IRAS 16293-2422, CRL 618, and CRL 2688, following the procedures defined by
Sandell et al. (\cite{sandell}).

Zenith opacities at 225 GHz ranged from 0.05-0.12 during the observations, but
usually averaged around 0.10. Telescope pointing was calibrated many times
during each observing run, and telescope drift was minimal, requiring very
small ($\sigma=1.7''$) corrections overall. We measured the JCMT beam size
from observations of Uranus, finding a full-width half-maximum of
$\theta_{beam}=8.0''$ at 450 $\mu$m and $\theta_{beam}=14.4''$ at 850 $\mu$m.
An average 1-$\sigma$ RMS noise level of 0.03 Jy beam$^{-1}$ and 0.69 Jy
beam$^{-1}$ was found at 850 $\mu$m and 450 $\mu$m respectively. The RMS noise
level measured in each jigglemap is listed in Table \ref{tab:flux} as the
uncertainty in the peak flux.

Clumps were identified using object detection routines in the software package
GAIA (Chipperfield \& Draper \cite{sun226}). We define a positive detection as
a group of pixels subtending at least the area of the JCMT beam with emission
above a 3-$\sigma$ level, where $\sigma$ is the RMS noise level of the
jigglemap. The validity of each detection was also confirmed manually. Clumps
not quite bright enough to be automatically detected were examined, and if
deemed worthy of inclusion, added to the list of detections.  These lower
sigma detections are labelled by a note in Table \ref{tab:flux}.

We list the peak flux per beam and the integrated flux for each detection.
The peak flux per beam gives the peak flux level averaged in a 14.4$''$ beam
for 850 $\mu$m maps and in an 8.0$''$ beam for 450 $\mu$m maps, while the
integrated flux of a detection measures the total flux inside an isophote
tracing the 3-$\sigma$ RMS noise level around the detection. We quote the
position of each detection as the location of peak emission, not as the
centroid of the 3-$\sigma$ isophote.

\begin{figure}
\resizebox{\hsize}{!}{
  \includegraphics[width=0.49\hsize]{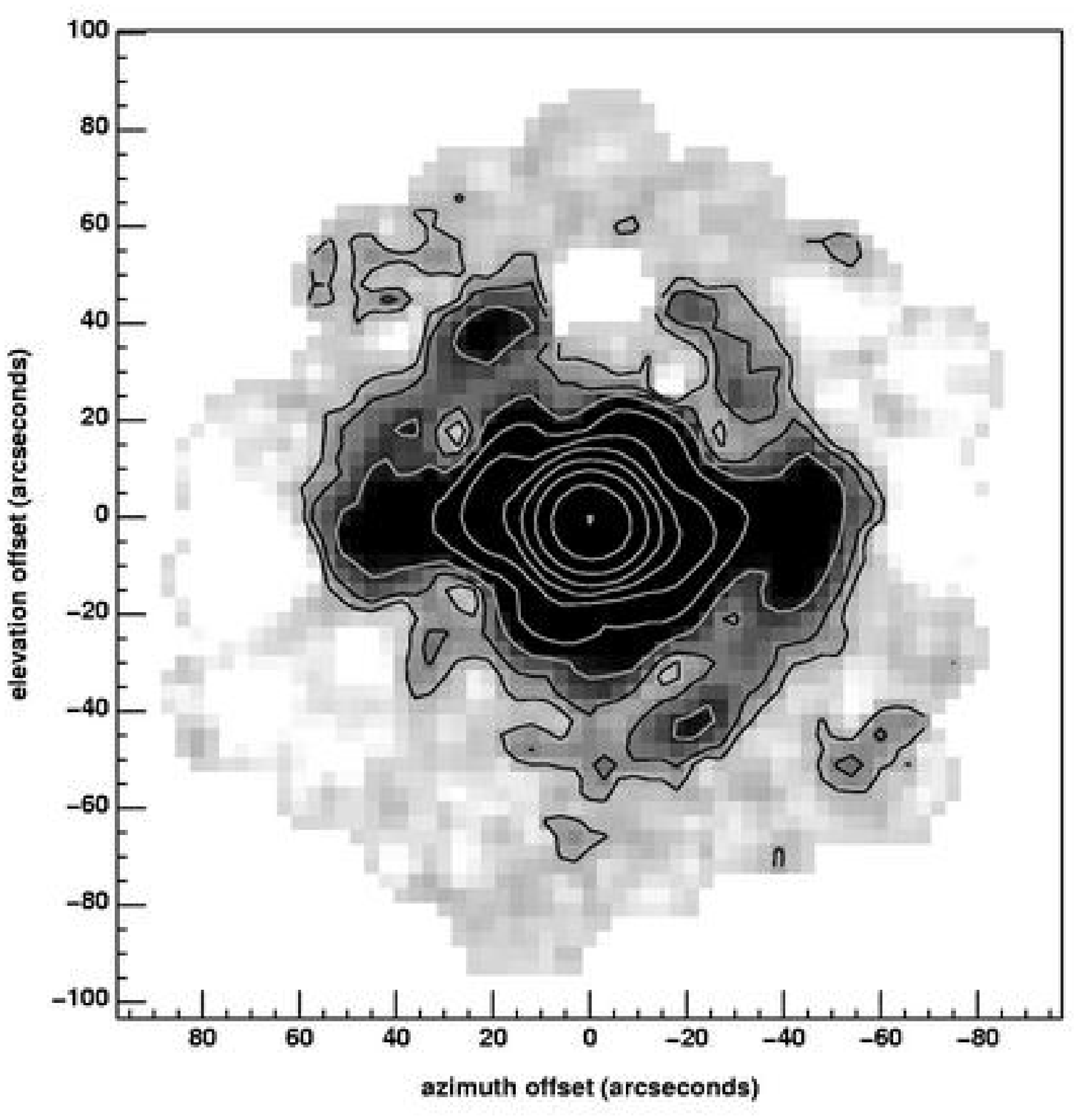}
  \includegraphics[width=0.49\hsize]{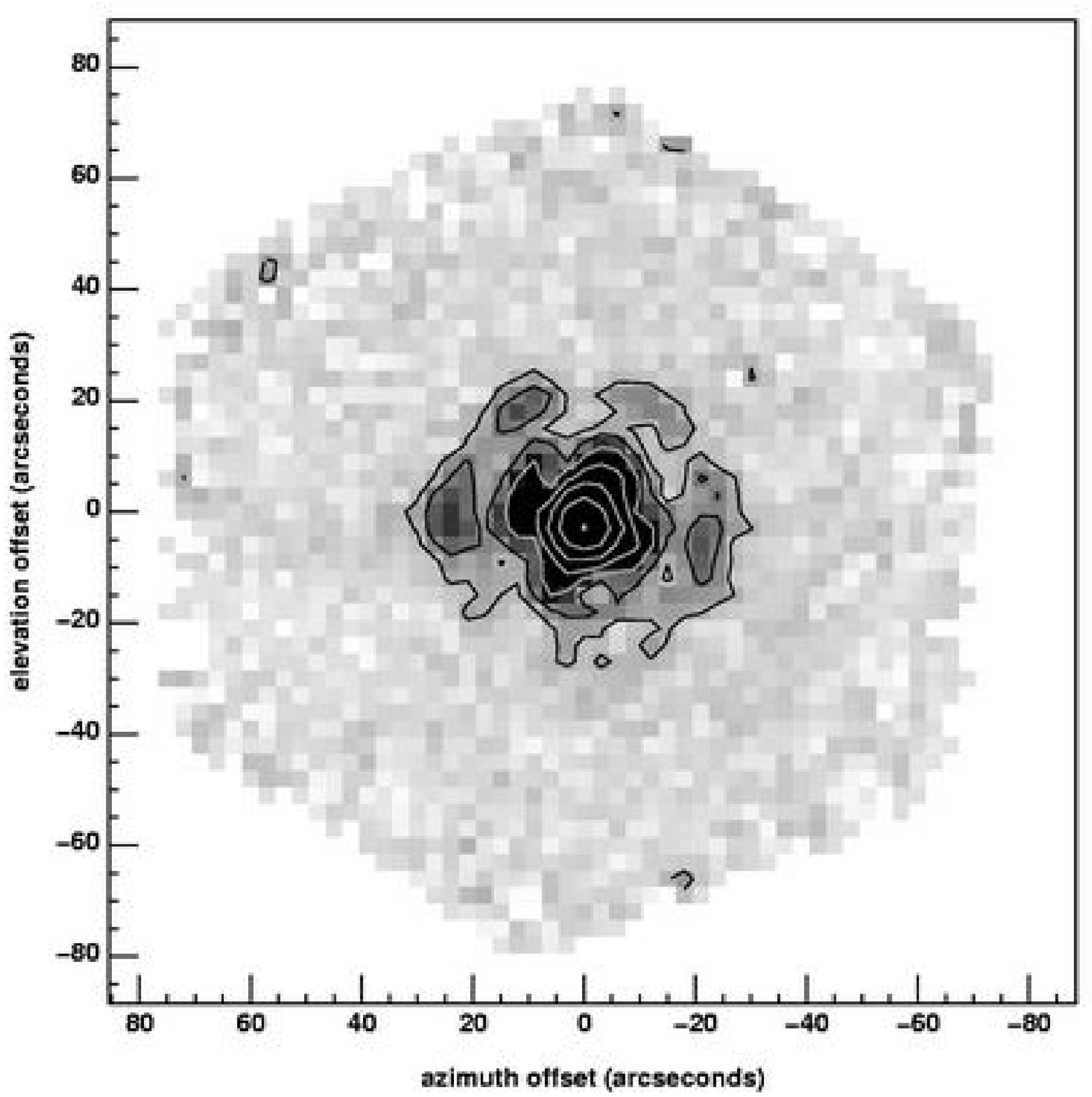}
}
\caption{Jiggle-maps of Uranus measured at 850 $\mu$m (left-hand
  image) and at 450 $\mu$m (right-hand image). Contours were chosen to
  highlight the JCMT beam structure, and are drawn every 0.8 magnitudes below
  65 Jy for the 850 $\mu$m image and every 0.8 magnitudes below 150 Jy for the
  450 $\mu$m map. The primary error beam can clearly be seen as a ring
  encircling the main beam, containing $\sim9\%$ and $\sim25\%$ of the total
  flux at 850 $\mu$m and 450 $\mu$m respectively.}
\label{fig:psf}
\end{figure}

Calibrating the integrated flux of a detection required additional
consideration, as the JCMT beam structure is complex (Figure \ref{fig:psf}),
so the number of detector counts recovered within an aperture is also a
function of aperture size. We quantified the extent of this relationship using
maps of Uranus (which we consider a point source), calibrating detector counts
recovered inside circular apertures of increasing radius. We did not include
the small number of non-planetary flux calibrators in the calibration of
integrated flux. The resulting function measures increasing counts with
aperture size, asymptotically reaching maximum counts once the aperture has
expanded to encompass the JCMT beam and its primary error beam. For each
detection, we then converted $n$ counts recovered inside an isophote of area
$A$ to Janskys by multiplying $n$ by the counts-to-Jy conversion factor
derived from a circular aperture of equivalent area.

Our observations were performed on seven nights over a period of three months.
Despite the protracted nature of our observations, a comparison of the
counts-to-Jy conversion factor calculated for each night showed it usually
remained consistent with the published JCMT
response.\footnote{\scriptsize
  http://www.jach.hawaii.edu/JACpublic/JCMT/\ldots\\
  \ldots Continuum\_observing/SCUBA/astronomy/calibration/gains.html}
Where the conversion factor appeared inconsistent and no other recent flux
calibrator maps were available, we assumed a conversion factor equal to the
mean value for our run. A comparison with the independent 1.2 mm continuum
observations of Beuther et al. (\cite{beuther2002a}) shows the data to be
consistently calibrated (\S\ref{sec:consistent calibration}), and we estimate
the absolute flux uncertainty to be $\pm$ 10\% at 850 $\mu$m and $\pm$ 30\% at
450 $\mu$m.

\subsection{Telescope and reduction artefacts}
We masked a small number of consistently noisy 850 $\mu$m bolometers during
data reduction; these bad bolometers usually fell in a region of background
sky or faint extended emission. Unfortunately, the secondary source seen
towards IRAS 05358+3543 fell on a noisy 850 $\mu$m bolometer, but we consider
the companion source an important feature of the jigglemap and so leave the
bolometer unmasked.  In addition, the map of IRAS 18553+0414 forms an isolated
case that suffers from an unusually large number of bad 850 $\mu$m bolometers;
we still include this data as the 450 $\mu$m map reveals the majority of flux
has been recovered by good bolometers.

Our observations used a 120$''$ chop to sky to measure and remove the
background emission. However, in crowded regions, the 120$''$ chop-throw
sometimes points the telescope towards an occupied region of sky rather than
an empty field. When this occurs, emission from objects in the sky reference
beam is subtracted from the target field emission, resulting in negative
images of clumps seen towards the reference position superimposed onto the
final map.  Some of our maps contain these artefacts, which are usually seen
away from regions of interest (eg. IRAS 18151-1208, IRAS 18431-0312), but
chopping onto emission altered the map of IRAS 18454-0158 to such an extent
that no reliable measurement was possible, and this source was removed from
our analysis.

We occasionally observed additional jigglemaps offset from the target position
to map fields with emission continuing outside the $\sim 120''$ SCUBA field of
view. These additional maps were calibrated as individual jigglemaps before
they were combined into a mosaic, weighting the contribution of each map to
intersecting areas by 1/$\sigma^{2}$, where $\sigma$ is the RMS noise level in
the map. Detections within the mosaic are still defined as clumps with
emission above a 3-$\sigma$ limit over an area the size of the JCMT beam, but
using the RMS local to the section of mosaic being measured.

\section{Results}
\label{sec:results}
Submillimetre emission at 850 $\mu$m was detected towards all the IRAS sources
in our sample, although not all sources were bright enough to be detected
above the increased background emission at 450 $\mu$m. When sources were
bright enough to be detected at 450 $\mu$m, the increased resolution of the
450 $\mu$m observations sometimes resolved additional peaks within the area of
a single 850 $\mu$m detection. A presentation of the reduced jigglemaps,
calibrated in Jy beam$^{-1}$, alongside maps of $\alpha$, the spectral index
of the dust emission (detailed in \S\ref{sec:alpha}), can be found in Figure
\ref{fig:jigglemaps}\footnote{To meet size constraints, Figure
  \ref{fig:jigglemaps} has been truncated to a single page; the complete
  figure displaying all 68 maps can be found in the preprint available at
  http://saturn.phy.umist.ac.uk:8000/$\sim$tjm/hmpoI.ps.gz}. The position and flux
measured for each detection is listed in Table \ref{tab:flux}.

\afterpage{\onecolumn \begin{figure*}
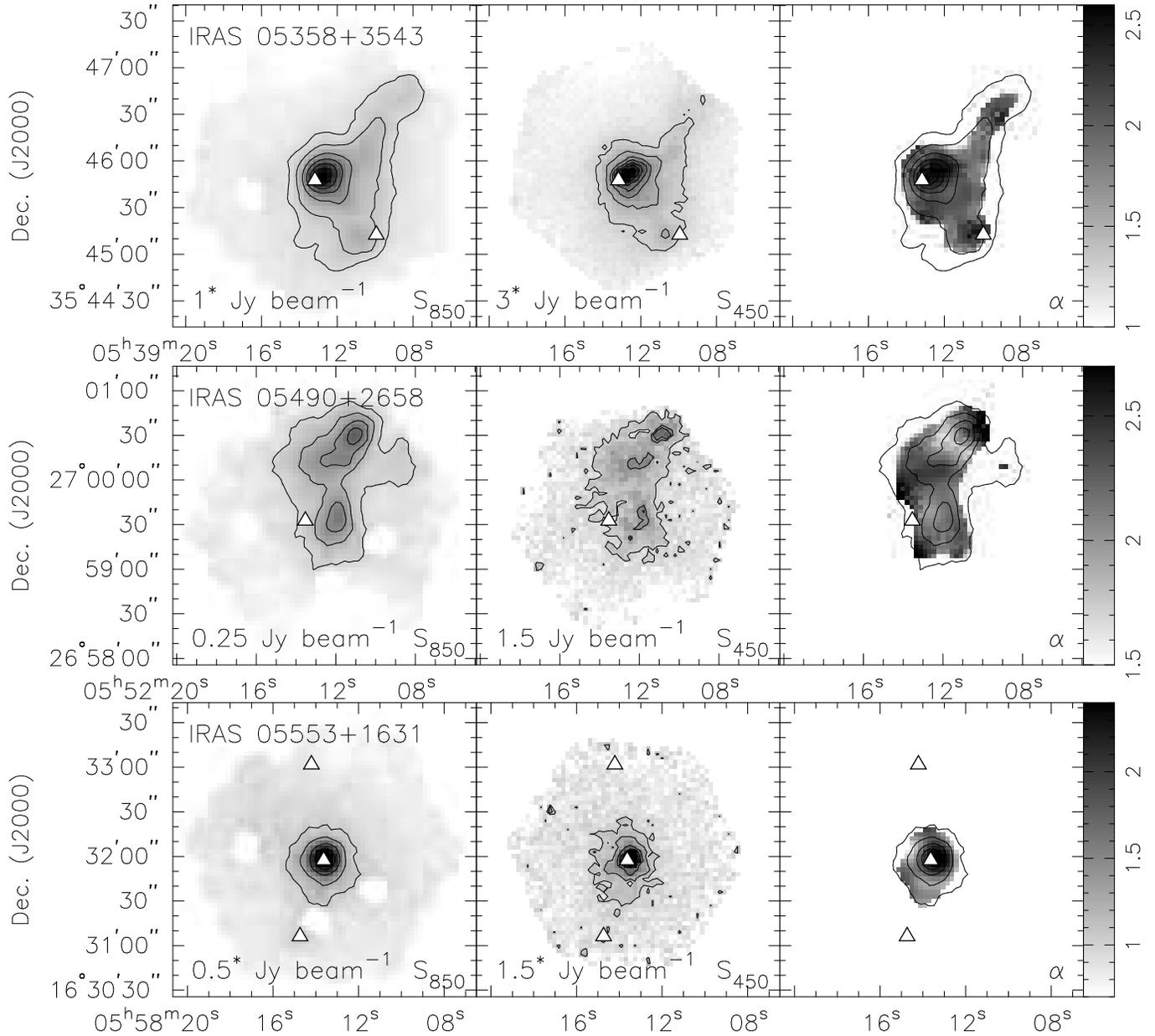

\includegraphics[angle=270,width=\hsize]{figures/postscripts/jigglemaps/05358+3543-0318-050}
\includegraphics[angle=270,width=\hsize]{figures/postscripts/jigglemaps/05490+2658-0318-051}
\includegraphics[angle=270,width=\hsize]{figures/postscripts/jigglemaps/05553+1631-0318-052}
\caption{Maps of $S_{850}$, $S_{450}$, and the $\alpha$ distribution in 
  the left-hand, centre, and right-hand panels respectively. Greyscale limits
  are chosen to emphasise flux levels between -2 $\sigma$ and +7 $\sigma$,
  where $\sigma$ is the RMS noise level in the map.  Contours trace the
  intensity in units of Jy beam$^{-1}$, using the step-size listed in the
  bottom left-hand corner of each map. The first contour is drawn at the first
  step above zero Jy beam$^{-1}$ unless the index is marked with an asterix:
  this signifies that an additional contour is plotted at half a step above
  zero Jy beam$^{-1}$. Triangular symbols plot the location of MSX point
  sources in each field of view. The spectral index of the dust emission
  ($\alpha$) is plotted in the right-hand greyscale maps: $\alpha$ is masked
  outside the boundary of the first 850 $\mu$m and 450 $\mu$m contours.
  Contours on the $\alpha$ map directly mirror those drawn on the 850 $\mu$m
  submillimetre emission map.}
\label{fig:jigglemaps}
\label{fig:jigstart}
\end{figure*}
 \twocolumn}
\afterpage{\onecolumn \begin{landscape}
\begin{scriptsize}
  \begin{center}
    \tablefirsthead{
      \toprule%
      \toprule%
      &
      &
      \multicolumn{2}{c}{Peak Position} &
      \multicolumn{4}{c}{\unit{850}{\micro\metre} Flux} &
      \multicolumn{4}{c}{\unit{450}{\micro\metre} Flux} &
      \multicolumn{1}{c}{$Y=S_{int}/S_{peak}$} &
      \\
      \cmidrule(r){3-4}\cmidrule(lr){5-8}\cmidrule(l){9-12}
      \multicolumn{1}{c}{WFS} &
      \multicolumn{1}{c}{IRAS field} &
      \multicolumn{2}{c}{J2000} &
      \multicolumn{2}{c}{peak} &
      \multicolumn{2}{c}{int.} &
      \multicolumn{2}{c}{peak} &
      \multicolumn{2}{c}{int.} & 
      &
      \multicolumn{1}{c}{Notes} \\
      &
      &
      \multicolumn{1}{c}{h:m:s} &
      \multicolumn{1}{c}{$^{\circ}$ : $'$ : $''$} &
      \multicolumn{2}{c}{Jy/\unit{14.4}{\arcsecond} beam} &
      \multicolumn{2}{c}{Jy} &
      \multicolumn{2}{c}{Jy/\unit{8.0}{\arcsecond} beam} &
      \multicolumn{2}{c}{Jy} &
      \multicolumn{1}{c}{\unit{14.4}{\arcsecond} beams} &
      \\
      \midrule
    }

    \tablehead{%
      \toprule%
      \multicolumn{14}{l}{\small\sl continued from previous page}\\
      \toprule%
      &
      &
      \multicolumn{2}{c}{Peak Position} &
      \multicolumn{4}{c}{\unit{850}{\micro\metre} Flux} &
      \multicolumn{4}{c}{\unit{450}{\micro\metre} Flux} &
      \multicolumn{1}{c}{$Y=S_{int}/S_{peak}$} &
      \\
      \cmidrule(r){3-4}\cmidrule(lr){5-8}\cmidrule(l){9-12}
      \multicolumn{1}{c}{WFS} &
      \multicolumn{1}{c}{IRAS field} &
      \multicolumn{2}{c}{J2000} &
      \multicolumn{2}{c}{peak} &
      \multicolumn{2}{c}{int.} &
      \multicolumn{2}{c}{peak} &
      \multicolumn{2}{c}{int.} & 
      &
      \multicolumn{1}{c}{Notes} \\
      &
      &
      \multicolumn{1}{c}{h:m:s} &
      \multicolumn{1}{c}{$^{\circ}$ : $'$ : $''$} &
      \multicolumn{2}{c}{Jy/\unit{14.4}{\arcsecond} beam} &
      \multicolumn{2}{c}{Jy} &
      \multicolumn{2}{c}{Jy/\unit{8.0}{\arcsecond} beam} &
      \multicolumn{2}{c}{Jy} &
      \multicolumn{1}{c}{\unit{14.4}{\arcsecond} beams} &
      \\
      \midrule
    }

    \tabletail{%
      \midrule
      \multicolumn{14}{r}{\small\sl continued on next page}\\
      \midrule}
    \tablelasttail{\bottomrule}
    
    \tablecaption[Position and flux of the submm detections]{ The
      position and measured flux of the detections resolved by our
      JCMT observations. Positions are measured from the
      \unit{450}{\micro\metre} jigglemaps wherever possible. Unless
      otherwise stated, \unit{450}{\micro\metre} detections without
      corresponding \unit{850}{\micro\metre} detections arise from the
      increased resolution of the \unit{450}{\micro\metre}
      observations.\newline
      $^{1}$\unit{850}{\micro\metre} emission falls on a noisy bolometer.\newline
      $^{2}$2-$\sigma$ detection at \unit{450}{\micro\metre}, where $\sigma$ is the RMS noise level in the map.
      }

    \begin{supertabular}{%
      @{}c@{}
      c
      r
      r
      r@{}l@{}r@{}l
      r@{}l@{}r@{}l@{}
      d{-1}@{}
      c@{}}

1       &05358+3543     &05:39:10.8     &+35:45:16      &1.69    &$\pm$0.06   &3.5    &$\pm$0.2    &6.9    &$\pm$0.3    &18.4   &$\pm$0.7  &2.1    &1 \\
2       &"              &05:39:12.7     &+35:45:51      &5.97    &$\pm$0.06   &28.7   &$\pm$0.7    &26.9   &$\pm$0.3    &190.4  &$\pm$2.8  &4.8    & \\
3       &05490+2658     &05:52:11.0     &+27:00:34      &1.02    &$\pm$0.02   &6.4    &$\pm$0.2    &3.2    &$\pm$0.2    &9.4    &$\pm$0.6  &6.3    & \\
4       &"              &05:52:12.1     &+27:00:11      &        &            &       &            &2.5    &$\pm$0.2    &13.8   &$\pm$0.7  &       & \\
5       &"              &05:52:12.1     &+26:59:38      &0.84    &$\pm$0.02   &3.3    &$\pm$0.1    &2.6    &$\pm$0.2    &15.1   &$\pm$0.8  &3.9    & \\
6       &05553+1631     &05:58:13.4     &+16:32:00      &2.15    &$\pm$0.02   &6.1    &$\pm$0.2    &10.4   &$\pm$0.2    &26.0   &$\pm$1.0  &2.8    & \\
7       &18089-1732     &18:11:45.2     &-17:30:43      &0.80    &$\pm$0.05   &3.9    &$\pm$0.3    &       &            &       &          &4.8    & \\
8       &"              &18:11:51.5     &-17:31:34      &11.08   &$\pm$0.05   &30.0   &$\pm$0.5    &71.7   &$\pm$0.7    &253.9  &$\pm$5.3  &2.7    & \\
9       &"              &18:11:53.9     &-17:30:02      &1.56    &$\pm$0.05   &3.0    &$\pm$0.2    &9.2    &$\pm$0.7    &21.8   &$\pm$1.6  &1.9    & \\
10      &"              &18:11:56.4     &-17:30:07      &0.35    &$\pm$0.05   &0.6    &$\pm$0.1    &       &            &       &          &1.7    & \\
11      &"              &18:11:57.0     &-17:29:34      &0.38    &$\pm$0.05   &1.1    &$\pm$0.2    &       &            &       &          &2.8    & \\
12      &18090-1832     &18:12:02.1     &-18:31:58      &1.31    &$\pm$0.04   &4.5    &$\pm$0.3    &9.0    &$\pm$0.9    &8.6    &$\pm$0.9  &3.5    & \\
13      &18102-1800     &18:13:11.7     &-18:00:04      &3.06    &$\pm$0.05   &13.7   &$\pm$0.5    &6.4    &$\pm$1.0    &13.9   &$\pm$1.8  &4.5    & \\
14      &18151-1208     &18:17:58.2     &-12:07:28      &3.89    &$\pm$0.04   &11.2   &$\pm$0.3    &15.5   &$\pm$0.6    &36.6   &$\pm$1.8  &2.9    & \\
15      &18159-1550     &18:18:48.4     &-15:49:00      &0.86    &$\pm$0.03   &4.5    &$\pm$0.2    &5.3    &$\pm$0.8    &13.6   &$\pm$1.5  &5.3    & \\
16      &18182-1433     &18:21:08.9     &-14:31:46      &5.41    &$\pm$0.04   &13.0   &$\pm$0.4    &45.4   &$\pm$0.9    &116.0  &$\pm$4.0  &2.4    & \\
17      &18223-1243     &18:25:10.6     &-12:42:27      &2.44    &$\pm$0.04   &10.4   &$\pm$0.3    &14.0   &$\pm$0.7    &54.8   &$\pm$2.6  &4.2    & \\
18      &18247-1147     &18:27:31.4     &-11:45:55      &2.02    &$\pm$0.04   &6.9    &$\pm$0.4    &14.5   &$\pm$0.7    &32.1   &$\pm$1.8  &3.4    & \\
19      &18264-1152     &18:29:14.3     &-11:50:22      &7.98    &$\pm$0.03   &20.0   &$\pm$0.3    &40.3   &$\pm$0.6    &106.6  &$\pm$2.8  &2.5    & \\
20      &18272-1217     &18:30:02.2     &-12:15:40      &0.54    &$\pm$0.03   &1.7    &$\pm$0.1    &4.7    &$\pm$0.6    &9.4    &$\pm$0.8  &3.2    &2 \\
21      &"              &18:30:03.2     &-12:15:11      &0.62    &$\pm$0.03   &2.2    &$\pm$0.2    &3.5    &$\pm$0.6    &8.8    &$\pm$1.2  &3.6    & \\
22      &18290-0924     &18:31:43.4     &-09:22:26      &1.82    &$\pm$0.02   &11.0   &$\pm$0.2    &6.7    &$\pm$0.7    &27.1   &$\pm$2.1  &6.1    & \\
23      &"              &18:31:44.0     &-09:22:17      &        &            &       &            &7.3    &$\pm$0.7    &18.2   &$\pm$1.6  &       & \\
24      &18306-0835     &18:33:17.3     &-08:33:28      &0.77    &$\pm$0.02   &1.3    &$\pm$0.1    &       &            &       &          &1.7    & \\
25      &"              &18:33:23.9     &-08:33:33      &2.32    &$\pm$0.02   &6.9    &$\pm$0.2    &15.1   &$\pm$0.6    &55.4   &$\pm$2.4  &3.0    & \\
26      &18308-0841     &18:33:29.8     &-08:38:33      &0.54    &$\pm$0.03   &1.6    &$\pm$0.1    &       &            &       &          &3.0    & \\
27      &"              &18:33:32.9     &-08:39:09      &2.43    &$\pm$0.03   &8.7    &$\pm$0.3    &13.2   &$\pm$0.7    &46.5   &$\pm$2.6  &3.6    & \\
28      &18310-0825     &18:33:47.9     &-08:23:52      &1.48    &$\pm$0.03   &5.2    &$\pm$0.2    &8.6    &$\pm$1.2    &27.3   &$\pm$2.7  &3.5    & \\
29      &18337-0743     &18:36:27.9     &-07:40:25      &0.96    &$\pm$0.02   &3.2    &$\pm$0.2    &4.9    &$\pm$0.2    &31.9   &$\pm$1.1  &3.3    & \\
30      &18345-0641     &18:37:16.8     &-06:38:35      &1.46    &$\pm$0.03   &3.6    &$\pm$0.2    &8.2    &$\pm$1.5    &14.0   &$\pm$2.1  &2.4    & \\
31      &18348-0616     &18:37:26.4     &-06:13:40      &0.64    &$\pm$0.05   &3.4    &$\pm$0.3    &       &            &       &          &5.3    & \\
32      &"              &18:37:27.5     &-06:14:05      &0.79    &$\pm$0.05   &2.0    &$\pm$0.2    &       &            &       &          &2.5    &2 \\
33      &"              &18:37:30.5     &-06:14:13      &1.66    &$\pm$0.05   &8.6    &$\pm$0.4    &9.4    &$\pm$2.2    &20.6   &$\pm$2.6  &5.2    & \\
34      &18372-0541     &18:39:56.0     &-05:38:52      &1.42    &$\pm$0.02   &4.3    &$\pm$0.2    &9.5    &$\pm$1.6    &12.9   &$\pm$2.2  &3.0    &2 \\
35      &18385-0512     &18:41:12.8     &-05:08:58      &3.50    &$\pm$0.02   &7.5    &$\pm$0.2    &23.8   &$\pm$0.2    &47.1   &$\pm$1.1  &2.1    & \\
36      &18426-0204     &18:45:12.1     &-02:01:10      &0.77    &$\pm$0.03   &4.0    &$\pm$0.2    &3.7    &$\pm$0.3    &12.8   &$\pm$0.9  &5.2    & \\
37      &18431-0312     &18:45:45.5     &-03:09:21      &0.97    &$\pm$0.02   &3.1    &$\pm$0.1    &3.7    &$\pm$0.3    &17.1   &$\pm$1.1  &3.2    & \\
38      &18437-0216     &18:46:21.8     &-02:12:20      &0.58    &$\pm$0.02   &1.9    &$\pm$0.1    &2.1    &$\pm$0.2    &8.9    &$\pm$0.7  &3.2    & \\
39      &"              &18:46:22.4     &-02:14:16      &0.86    &$\pm$0.02   &8.1    &$\pm$0.2    &2.5    &$\pm$0.2    &24.9   &$\pm$1.3  &9.5    & \\
40      &"              &18:46:23.0     &-02:15:16      &0.39    &$\pm$0.02   &1.1    &$\pm$0.1    &1.0    &$\pm$0.2    &1.7    &$\pm$0.3  &2.7    &2 \\
41      &18440-0148     &18:46:33.3     &-01:44:52      &0.14    &$\pm$0.06   &0.2    &$\pm$0.1    &       &            &       &          &1.7    & \\
42      &"              &18:46:36.5     &-01:45:22      &0.91    &$\pm$0.06   &2.6    &$\pm$0.4    &5.3    &$\pm$0.3    &19.8   &$\pm$1.1  &2.9    & \\
43      &18445-0222     &18:47:10.0     &-02:18:45      &1.84    &$\pm$0.03   &9.9    &$\pm$0.3    &14.3   &$\pm$0.5    &58.2   &$\pm$2.5  &5.4    & \\
44      &18447-0229     &18:47:20.2     &-02:25:28      &0.36    &$\pm$0.04   &0.8    &$\pm$0.1    &       &            &       &          &2.3    & \\
45      &"              &18:47:21.5     &-02:26:11      &0.86    &$\pm$0.04   &5.2    &$\pm$0.3    &4.2    &$\pm$1.2    &19.7   &$\pm$4.9  &6.0    & \\
46      &"              &18:47:23.5     &-02:26:16      &        &            &       &            &2.4    &$\pm$1.2    &13.8   &$\pm$3.9  &       &2 \\
47      &"              &18:47:26.1     &-02:26:57      &0.39    &$\pm$0.04   &1.1    &$\pm$0.2    &       &            &       &          &2.9    & \\
48      &18449-0158     &18:47:35.5     &-01:55:11      &3.27    &$\pm$0.06   &28.6   &$\pm$0.6    &27.7   &$\pm$1.0    &134.7  &$\pm$5.2  &8.7    & \\
49      &"              &18:47:38.7     &-01:55:10      &        &            &       &            &13.9   &$\pm$1.0    &90.2   &$\pm$3.7  &       & \\
50      &18454-0136     &18:48:02.1     &-01:33:30      &1.62    &$\pm$0.02   &7.3    &$\pm$0.2    &8.0    &$\pm$0.4    &37.9   &$\pm$1.8  &4.5    & \\
51      &18460-0307     &18:48:37.8     &-03:03:48      &0.39    &$\pm$0.03   &0.6    &$\pm$0.1    &       &            &       &          &1.5    & \\
52      &"              &18:48:39.7     &-03:04:07      &0.84    &$\pm$0.03   &5.0    &$\pm$0.2    &5.1    &$\pm$0.4    &26.5   &$\pm$1.5  &6.0    & \\
53      &"              &18:48:40.4     &-03:03:57      &        &            &       &            &2.7    &$\pm$0.4    &5.8    &$\pm$0.7  &       &2 \\
54      &18470-0044     &18:49:37.8     &-00:41:00      &1.97    &$\pm$0.04   &6.3    &$\pm$0.3    &10.7   &$\pm$0.6    &36.2   &$\pm$2.0  &3.2    & \\
55      &18472-0022     &18:49:52.4     &-00:18:59      &1.31    &$\pm$0.03   &7.4    &$\pm$0.3    &6.7    &$\pm$0.3    &45.3   &$\pm$1.7  &5.7    & \\
56      &"              &18:49:53.8     &-00:19:48      &0.41    &$\pm$0.03   &1.4    &$\pm$0.1    &       &            &       &          &3.3    & \\
57      &18488+0000     &18:51:24.4     &+00:04:39      &2.21    &$\pm$0.03   &8.5    &$\pm$0.3    &4.6    &$\pm$0.7    &7.9    &$\pm$1.0  &3.8    & \\
58      &"              &18:51:25.5     &+00:04:11      &        &            &       &            &17.6   &$\pm$0.7    &44.0   &$\pm$2.2  &       & \\
59      &18521+0134     &18:54:40.6     &+01:38:05      &1.31    &$\pm$0.02   &3.3    &$\pm$0.1    &6.8    &$\pm$0.3    &21.6   &$\pm$1.1  &2.5    & \\
60      &"              &18:54:44.4     &+01:37:00      &0.26    &$\pm$0.02   &0.3    &$\pm$0.1    &       &            &       &          &1.2    & \\
61      &18530+0215     &18:55:33.7     &+02:19:09      &2.78    &$\pm$0.07   &11.2   &$\pm$0.5    &13.0   &$\pm$0.9    &68.8   &$\pm$4.0  &4.0    & \\
62      &18540+0220     &18:56:36.6     &+02:24:45      &0.38    &$\pm$0.02   &2.6    &$\pm$0.2    &1.6    &$\pm$0.4    &2.6    &$\pm$0.5  &6.8    &2 \\
63      &"              &18:56:40.1     &+02:25:30      &0.22    &$\pm$0.02   &0.5    &$\pm$0.1    &       &            &       &          &2.4    & \\
64      &18553+0414     &18:57:53.5     &+04:18:16      &1.69    &$\pm$0.02   &4.3    &$\pm$0.2    &14.3   &$\pm$0.7    &23.3   &$\pm$1.6  &2.5    & \\
65      &18566+0408     &18:59:10.2     &+04:12:11      &4.14    &$\pm$0.02   &15.7   &$\pm$0.2    &24.2   &$\pm$0.4    &87.2   &$\pm$2.5  &3.8    & \\
66      &19012+0536     &19:03:45.3     &+05:40:43      &2.70    &$\pm$0.02   &5.4    &$\pm$0.2    &17.4   &$\pm$0.4    &34.4   &$\pm$1.4  &2.0    & \\
67      &19035+0641     &19:06:01.5     &+06:46:35      &3.29    &$\pm$0.02   &10.6   &$\pm$0.2    &23.3   &$\pm$0.4    &72.8   &$\pm$2.3  &3.2    & \\
68      &19074+0752     &19:09:53.4     &+07:57:12      &1.27    &$\pm$0.02   &6.8    &$\pm$0.2    &3.1    &$\pm$0.3    &5.3    &$\pm$0.4  &5.4    & \\
69      &"              &19:09:53.9     &+07:56:55      &        &            &       &            &7.3    &$\pm$0.3    &26.2   &$\pm$1.2  &       & \\
70      &19175+1357     &19:19:48.6     &+14:02:26      &        &            &       &            &3.3    &$\pm$0.4    &7.7    &$\pm$0.6  &       & \\
71      &"              &19:19:48.8     &+14:02:46      &1.03    &$\pm$0.04   &3.7    &$\pm$0.2    &5.0    &$\pm$0.4    &16.1   &$\pm$1.3  &3.5    &2 \\
72      &19217+1651     &19:23:58.6     &+16:57:38      &3.91    &$\pm$0.02   &6.9    &$\pm$0.1    &29.7   &$\pm$0.3    &68.5   &$\pm$1.6  &1.8    & \\
73      &19220+1432     &19:24:19.9     &+14:38:02      &1.53    &$\pm$0.04   &7.5    &$\pm$0.3    &6.3    &$\pm$0.9    &20.1   &$\pm$2.0  &4.9    & \\
74      &19266+1745     &19:28:55.5     &+17:52:00      &2.08    &$\pm$0.03   &6.0    &$\pm$0.2    &11.2   &$\pm$0.3    &36.4   &$\pm$1.3  &2.9    & \\
75      &19282+1814     &19:30:23.1     &+18:20:22      &1.76    &$\pm$0.02   &6.7    &$\pm$0.2    &9.2    &$\pm$0.3    &14.4   &$\pm$0.8  &3.8    & \\
76      &"              &19:30:29.7     &+18:20:37      &0.49    &$\pm$0.02   &2.8    &$\pm$0.1    &2.2    &$\pm$0.3    &7.7    &$\pm$0.7  &5.6    & \\
77      &19403+2258     &19:42:28.8     &+23:05:03      &1.01    &$\pm$0.04   &6.1    &$\pm$0.3    &23.0   &$\pm$0.7    &17.8   &$\pm$1.1  &6.0    & \\
78      &19410+2336     &19:43:10.6     &+23:45:02      &1.34    &$\pm$0.05   &2.9    &$\pm$0.2    &4.8    &$\pm$0.5    &15.7   &$\pm$1.3  &2.2    & \\
79      &"              &19:43:11.2     &+23:44:06      &4.79    &$\pm$0.05   &24.8   &$\pm$0.5    &22.5   &$\pm$0.5    &149.8  &$\pm$3.6  &5.2    & \\
80      &19411+2306     &19:43:17.6     &+23:13:57      &1.35    &$\pm$0.05   &7.2    &$\pm$0.4    &4.8    &$\pm$0.8    &7.6    &$\pm$1.1  &5.3    & \\
81      &19413+2332     &19:43:26.3     &+23:40:26      &0.53    &$\pm$0.03   &1.9    &$\pm$0.1    &2.3    &$\pm$0.3    &6.8    &$\pm$0.8  &3.6    &2 \\
82      &"              &19:43:29.0     &+23:40:19      &0.97    &$\pm$0.03   &5.8    &$\pm$0.3    &4.1    &$\pm$0.3    &27.5   &$\pm$1.6  &6.0    & \\
83      &19471+2641     &19:49:10.1     &+26:49:10      &0.34    &$\pm$0.05   &1.1    &$\pm$0.2    &       &            &       &          &3.2    & \\
84      &"              &19:49:11.8     &+26:49:38      &0.37    &$\pm$0.05   &0.8    &$\pm$0.2    &       &            &       &          &2.3    & \\
85      &20051+3435     &20:07:04.5     &+34:44:45      &1.07    &$\pm$0.02   &6.9    &$\pm$0.2    &5.7    &$\pm$0.7    &23.4   &$\pm$2.0  &6.5    & \\
86      &20081+2720     &20:10:12.6     &+27:29:13      &0.36    &$\pm$0.02   &2.5    &$\pm$0.1    &3.4    &$\pm$0.9    &7.3    &$\pm$2.0  &7.1    & \\
87      &"              &20:10:13.3     &+27:28:21      &0.77    &$\pm$0.02   &3.1    &$\pm$0.1    &3.0    &$\pm$0.9    &12.5   &$\pm$2.7  &4.0    & \\
88      &"              &20:10:16.0     &+27:28:12      &0.70    &$\pm$0.02   &3.2    &$\pm$0.1    &2.8    &$\pm$0.9    &19.3   &$\pm$2.6  &4.5    & \\
89      &"              &20:10:18.7     &+27:27:18      &0.26    &$\pm$0.02   &0.5    &$\pm$0.1    &       &            &       &          &2.0    & \\
90      &20126+4104     &20:14:25.7     &+41:13:30      &5.57    &$\pm$0.04   &21.9   &$\pm$0.5    &29.0   &$\pm$0.9    &91.1   &$\pm$3.9  &3.9    & \\
91      &20205+3948     &20:22:20.0     &+39:58:21      &1.12    &$\pm$0.02   &9.3    &$\pm$0.2    &3.9    &$\pm$0.3    &22.3   &$\pm$1.5  &8.3    & \\
92      &"              &20:22:24.9     &+39:57:55      &0.55    &$\pm$0.02   &4.8    &$\pm$0.2    &4.0    &$\pm$0.3    &10.2   &$\pm$0.8  &8.7    &2 \\
93      &20216+4107     &20:23:23.9     &+41:17:42      &1.44    &$\pm$0.04   &6.1    &$\pm$0.3    &       &            &       &          &4.2    & \\
94      &20293+3952     &20:31:12.9     &+40:03:21      &3.30    &$\pm$0.05   &22.8   &$\pm$0.6    &34.8   &$\pm$5.0    &148.6  &$\pm$16.2 &6.9    &2 \\
95      &20319+3958     &20:33:49.4     &+40:08:32      &1.06    &$\pm$0.02   &3.1    &$\pm$0.1    &       &            &       &          &3.0    & \\
96      &20332+4124     &20:34:58.7     &+41:34:46      &0.68    &$\pm$0.03   &1.5    &$\pm$0.1    &10.1   &$\pm$0.3    &50.6   &$\pm$1.8  &2.2    & \\
97      &"              &20:35:01.1     &+41:34:59      &1.83    &$\pm$0.03   &14.9   &$\pm$0.3    &5.9    &$\pm$0.3    &28.7   &$\pm$1.3  &8.1    & \\
98      &20343+4129     &20:36:03.4     &+41:39:44      &        &            &       &            &4.1    &$\pm$0.4    &18.7   &$\pm$1.3  &       & \\
99      &"              &20:36:06.3     &+41:39:59      &1.99    &$\pm$0.02   &19.0   &$\pm$0.3    &9.7    &$\pm$0.4    &27.5   &$\pm$1.1  &9.5    & \\
100     &"              &20:36:08.1     &+41:39:58      &        &            &       &            &9.2    &$\pm$0.4    &35.0   &$\pm$1.6  &       & \\
101     &22134+5834     &22:15:08.9     &+58:49:08      &1.81    &$\pm$0.03   &8.7    &$\pm$0.3    &9.1    &$\pm$1.1    &30.0   &$\pm$2.8  &4.8    & \\
102     &22551+6221     &22:57:04.3     &+62:37:44      &        &            &       &            &2.4    &$\pm$0.3    &9.4    &$\pm$0.8  &       & \\
103     &"              &22:57:07.4     &+62:37:29      &0.85    &$\pm$0.03   &12.3   &$\pm$0.4    &4.9    &$\pm$0.3    &21.7   &$\pm$1.0  &14.5   & \\
104     &"              &22:57:11.6     &+62:36:46      &0.51    &$\pm$0.03   &1.6    &$\pm$0.1    &       &            &       &          &3.2    & \\
105     &22570+5912     &22:58:55.3     &+59:28:42      &0.53    &$\pm$0.03   &2.0    &$\pm$0.2    &       &            &       &          &3.7    & \\
106     &"              &22:58:59.2     &+59:27:41      &1.04    &$\pm$0.03   &2.8    &$\pm$0.2    &6.0    &$\pm$0.3    &15.5   &$\pm$0.8  &2.7    & \\
107     &"              &22:59:05.0     &+59:28:23      &1.58    &$\pm$0.03   &7.6    &$\pm$0.3    &7.6    &$\pm$0.3    &50.8   &$\pm$1.8  &4.8    & \\
108     &23033+5951     &23:05:24.8     &+60:08:14      &3.38    &$\pm$0.02   &10.4   &$\pm$0.2    &16.2   &$\pm$0.8    &48.6   &$\pm$2.6  &3.1    & \\
109     &23139+5939     &23:16:09.8     &+59:55:31      &3.19    &$\pm$0.02   &8.1    &$\pm$0.2    &15.6   &$\pm$1.1    &38.1   &$\pm$2.6  &2.5    & \\
110     &23151+5912     &23:17:20.4     &+59:28:51      &1.89    &$\pm$0.02   &6.5    &$\pm$0.2    &10.0   &$\pm$0.9    &19.1   &$\pm$1.7  &3.4    & \\
111     &23545+6508     &23:57:02.1     &+65:24:38      &1.06    &$\pm$0.03   &3.7    &$\pm$0.2    &7.6    &$\pm$0.3    &26.8   &$\pm$1.1  &3.4    & \\
112     &"              &23:57:06.4     &+65:24:49      &1.05    &$\pm$0.03   &4.2    &$\pm$0.2    &4.3    &$\pm$0.3    &18.8   &$\pm$1.2  &4.0    & \\

    \end{supertabular}
    \label{tab:flux}
  \end{center}
\end{scriptsize}
\end{landscape}

 \twocolumn}

The target sources IRAS 19266+1745 and IRAS 18553+0414 displayed a gas+dust
mass incompatible with the luminosity of the driving protostar, unless these
sources are projected to the far kinematic distance (\S\ref{sec:distance
  resolved}). Therefore, we reject the near kinematic distance for these
objects and consider them resolved to the far kinematic distance for all
subsequent analysis.

\subsection{Companion Clump Fraction}
\label{sec:ccf}
The majority of candidate HMPOs were detected as companionless, compact and
approximately spherically symmetric submm clumps (eg. IRAS 05553+1631),
although a significant number exhibited submm nebulosity (eg. IRAS
18566+0408), appeared in filaments of emission (eg. IRAS 18437-0216) and/or
existed with multiple additional detections within the field of view (eg. IRAS
23545+6508). While our sample was explicitly constructed to consist of
isolated, companionless candidates away from sources of confusion, we found
only 38 of the 68 target IRAS fields contained a single, companionless clump.
The remaining IRAS fields contained more than one submm clump, usually two
detections, with the mosaic map towards IRAS 18089-1732 containing the most
companions, where 5 separate clumps were resolved. This demonstrates the
difficulty in locating truly isolated candidate HMPOs. We can characterise the
multiplicity of our detections by calculating the companion clump fraction
(CCF), expressed by the formula
\begin{equation}
  CCF = \frac{B+2T+3Q+4P}{S+B+T+Q+P},
  \label{eq:ccf}
\end{equation}
where $S, B, T$, $Q$ and $P$ are the number of single, binary, triple,
quadruple and quintuplet clumps in our sample. If all clumps in our sample
were solitary clumps, the CCF would be 0.0, while if all clumps had one
companion the CCF would be 1.0. 

The CCF for our sample is $0.65\pm0.1$, where the uncertainty comes from
$\sqrt{N}$ counting statistics. In reality, the absolute value of the CCF and
the quoted uncertainty are both lower limits, as they are calculated assuming
we have detected, and are uniformly sensitive, to all companions. This is not
the case, as our limited angular resolution precludes the detection of
companions closer than around a beamsize, plus the finite field of view means
companions of greater than $\sim60''$ separation (assuming a clump central in
the jigglemap) will not be detected. Additionally, when coupled with the large
difference in projected distance (the most distant sources being more than 15
times further away than the closest sources), our companion mass sensitivity
also bears a dependence on distance.
\label{sec:sensitivity concerns}

We examined the effect of different distance projections by sorting our
candidate HMPOs into four bins, containing sources $<2$ kpc, 2 to 4 kpc, 4 to
8 kpc and $>8$ kpc distant, respectively. The CCF of these subsamples remains
remarkably consistent, each section in agreement with the full sample CCF
within the uncertainty limits. This is true regardless of whether
distance-unresolved sources are projected to the near or far kinematic
distance, with the exception of the $<2$ kpc bin projected to the far
kinematic distance, and suggests clumps have a similar number of companions
over a wide range of distance scales.

The expression of multiplicity given in Equation \ref{eq:ccf} is usually used
as a diagnostic of more evolved stars, in particular to quantify the number of
companions a low-mass star is born with (eg. Beck et al.  \cite{beck2003},
Patience et al. \cite{patience2002}), whereas in this study the CCF can be
interpreted as the likelihood of finding additional potentially star-forming
clumps when observing Galactic HMPOs identified by a similar flux-limited
criteria. While it remains difficult to constrain the statistics of such a
disparate sample, the CCF does emphasise that most clumps do not form in
isolation, and that a single IRAS detection is usually resolved into several
submm clumps. New samples of protostars comparable to our sample are hard to
compile, but the strong likelihood of detecting additional clumps in the
locality of our sample suggests that wide-field surveys towards existing
high-mass protostars may also be a productive way of locating new protostellar
candidates.

\subsection{Clump morphology statistics}
Forming an unbiased statistical analysis of the morphological features is
difficult, as distinguishing features are mainly found in the appearance of
low-level extended emission. As such, measuring the FWHM of detections is of
limited use, as it is not sensitive to the faint emission features we wish to
characterise. Instead, we formed a simple statistic that indicates how much
mass lies outside the central beam by measuring $Y$, the ratio of integrated
flux ($S_{int}$) to peak flux ($S_{peak}$) at 850 $\mu$m.  We measure at 850
$\mu$m because of higher signal-to-noise than in the corresponding 450 $\mu$m
maps.  For a point source, $Y$ equals unity. The value of $Y$ for each
detection is listed in column 9 of Table \ref{tab:flux}.

\begin{figure}
\resizebox{\hsize}{!}{
  \includegraphics[angle=-90,width=\hsize]{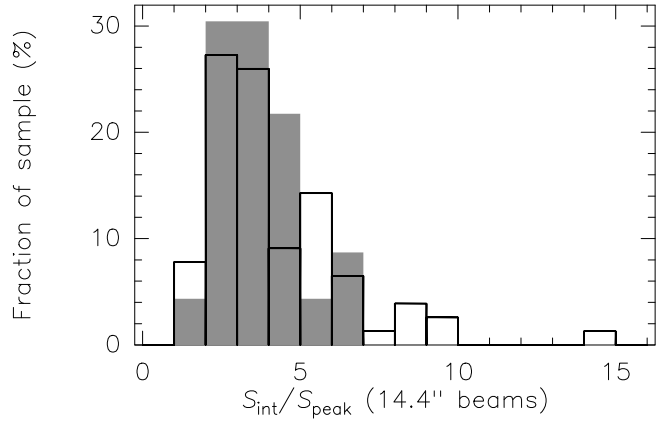}
}
\caption{
  A histogram of the $S_{int}/S_{peak}$ flux ratio for `confirmed'
  isolated sources (sample $A$: shaded bars) and sources potentially
  with companions (sample $B$: outline).}
  \label{fig:fluxratio histogram}
\end{figure}

\label{label:subsample definition}
In Figure \ref{fig:fluxratio histogram} we plot the $Y$ distribution for our
submm detections, dividing our sample into two groups: subsample $A$,
containing detections with a high confidence of being solitary cores (having
just one detection within an IRAS, MSX, SCUBA and IRAM field of view) and
subsample $B$, the remainder. While the solitary detections in sample $A$ do
not display the extended distribution tail seen in subsample $B$, we see both
groups peak at an intensity ratio of $\simeq 3$, which in light of the factor
15 range in distance suggests that the envelope structures may be scale-free.

\begin{figure*}
\resizebox{\hsize}{!}{
  \includegraphics[angle=-90,width=0.33\hsize]{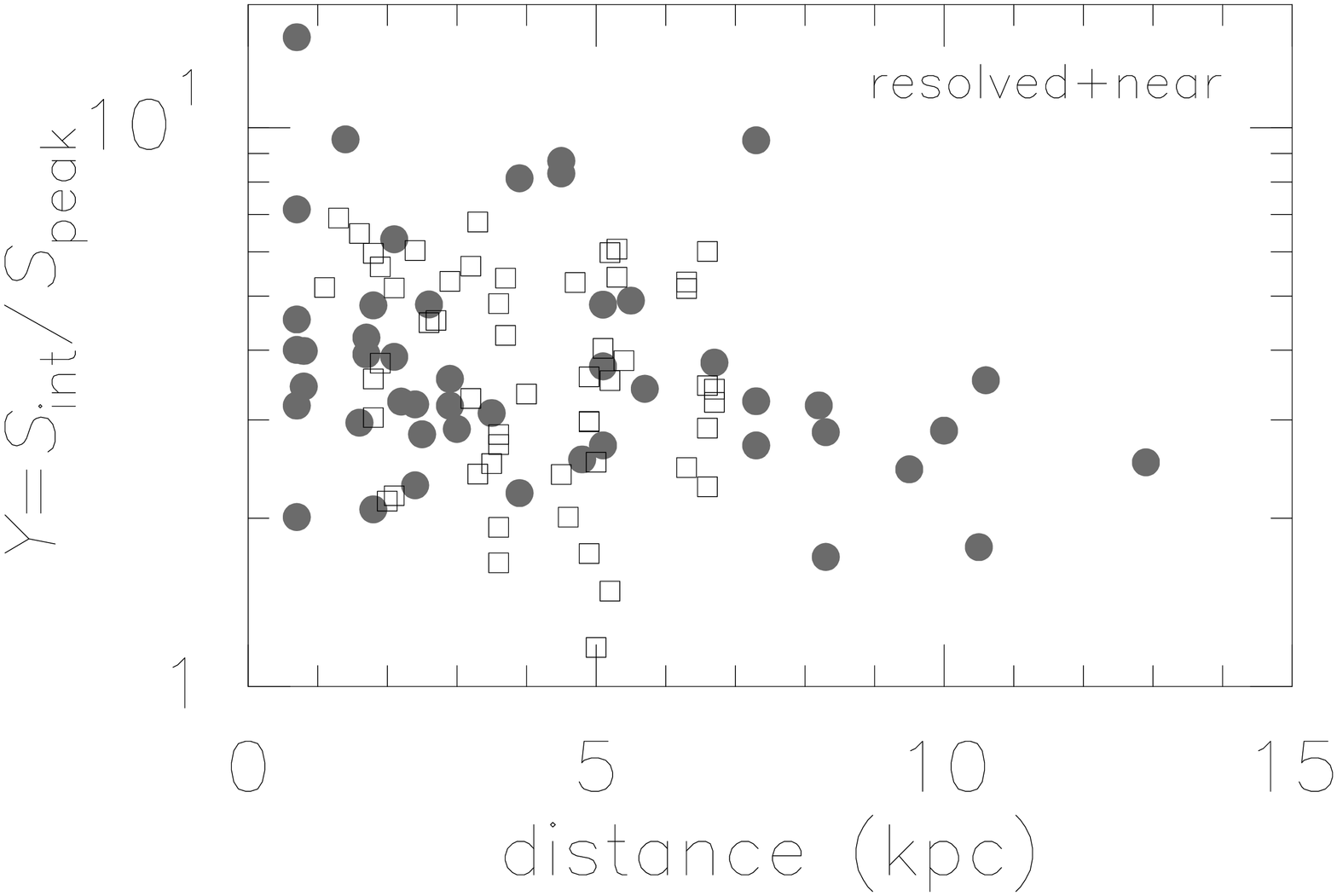}
  \includegraphics[angle=-90,width=0.33\hsize]{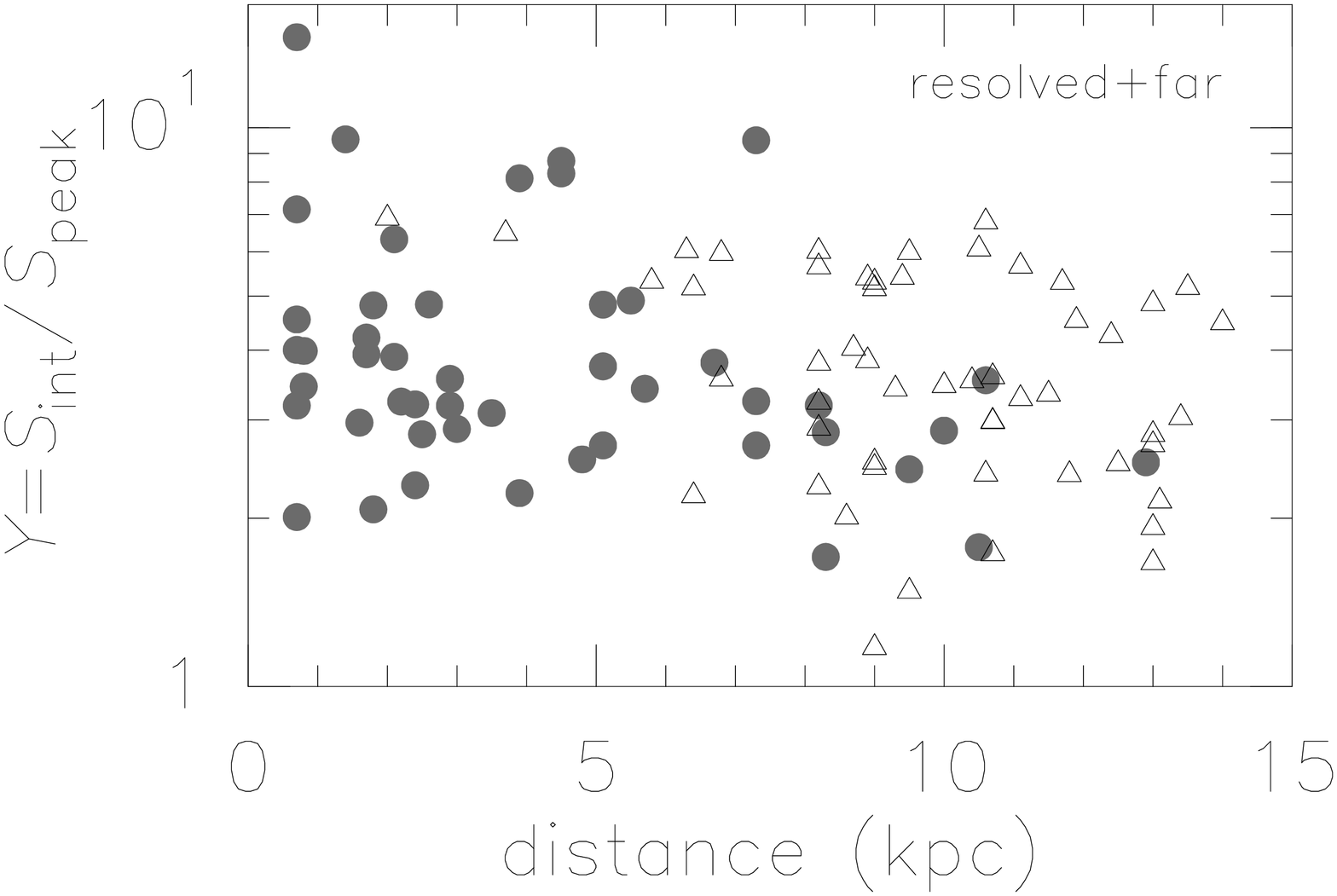}
  \includegraphics[angle=-90,width=0.33\hsize]{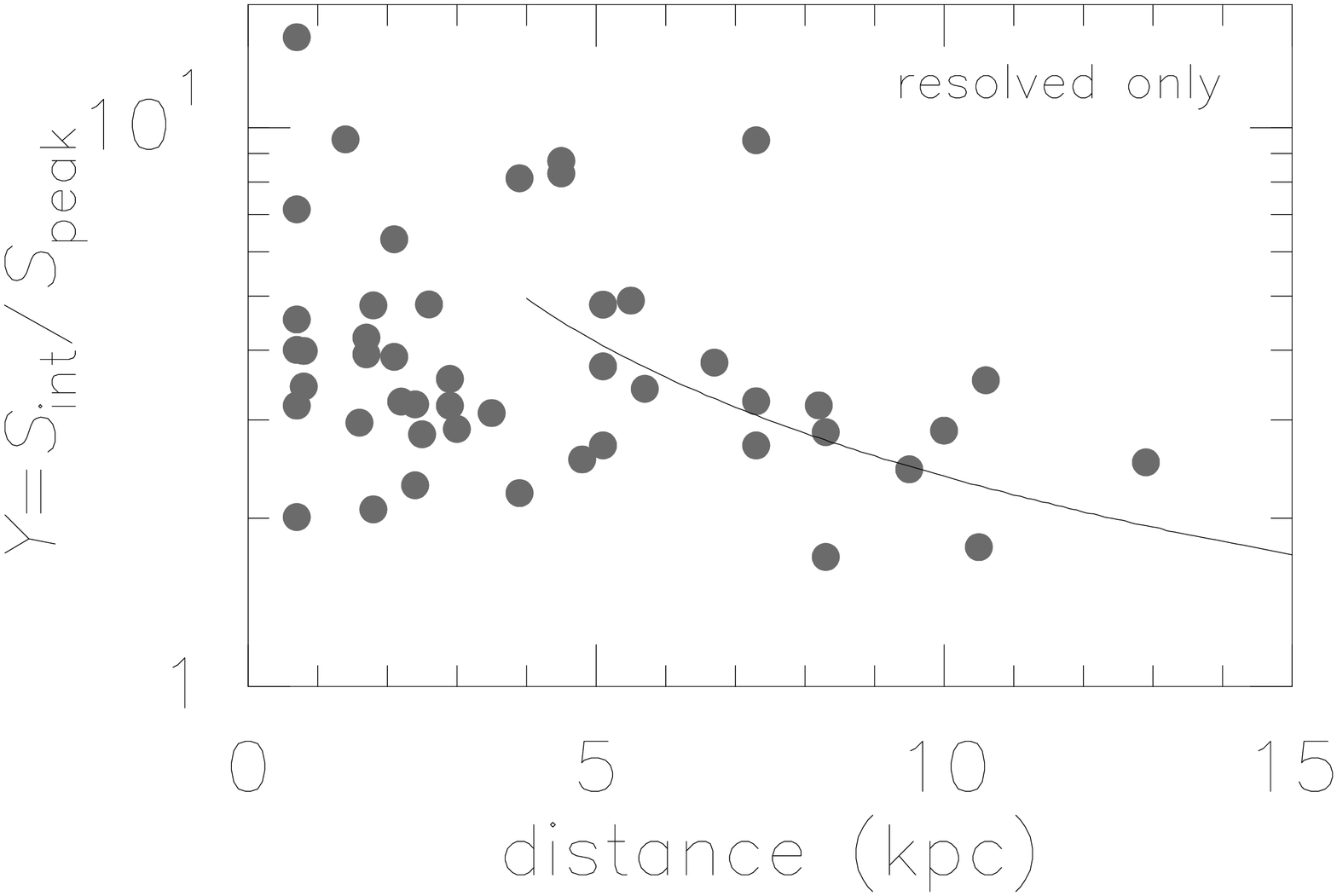}
}
\caption{
  A plot of $Y=S_{int}/S_{peak}$, the ratio of 850 $\mu$m integrated flux to
  peak 850 $\mu$m flux measured in a 14.4$''$ beam, against the kinematic
  distance of each detection. Distance resolved sources are plotted by filled
  circles, while distance unresolved sources are projected to near and far
  kinematic distances and plotted with open rectangles and triangles.  The
  curve in the distance-resolved plot displays a power law of the form
  $S_{int}/S_{peak} \propto d^{-0.8}$.}
\label{fig:fluxratio vs distance}
\end{figure*}

The large scatter in Figure \ref{fig:fluxratio vs distance} means we do not
find any strong correlation with distance at the near distance projection
(which is the most likely projection for our sample), although considering the
distance-resolved sources alone does reveal a trend beyond $d=4$ kpc, where $Y
\propto d^{-0.8}$. This fall-off does not reveal a physical change, but
reflects the diminishing level of integrated emission as the 3-$\sigma$
isophote encloses less of the envelope for more distant sources.

We must qualify a number of uncertainties that could affect the distribution
of $Y$, not least our variable sensitivity to additional embedded sources.
While the large-scale envelope structures appear to be scale-free, on the
small scale there are indications that we are still undersampling the number
of companions separated by less than a beam width. A number of apparently
single detections at 850 $\mu$m are barely resolved as multiple sources at 450
$\mu$m (eg. IRAS 05490+2658); a reminder that further clustering on size
scales less than a 450 $\mu$m beam width may also be present. Overall, it is
inevitable that with limited resolution we misclassify some multiple cores as
solitary detections, a point demonstrated in Figure \ref{fig:integrated vs
  peak}, where all sources (bar one) with 850 $\mu$m flux ratios larger than 7
are resolved as multiple detections at 450 $\mu$m. IRAS 22551+6221 provides
the most visible demonstration of this effect, where the high 850 $\mu$m flux
ratio arises from the inclusion of flux from a bright neighbouring source that
is only fully resolved at 450 $\mu$m.

\begin{figure}
\resizebox{\hsize}{!}{
  \includegraphics[width=\hsize,angle=270]{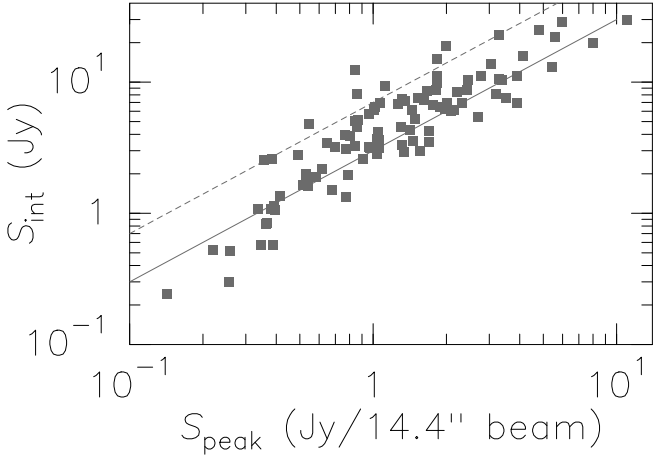}
}
\caption{
  A comparison of the peak flux and integrated flux of each 850 $\mu$m
  detection. The solid line traces the $Y=S_{int}/S_{peak}=3$ distribution
  peak found in Figure \ref{fig:fluxratio histogram}, while the dotted line
  traces the ratio $Y=7$.}
  \label{fig:integrated vs peak}
\end{figure}

We also tend to overestimate the flux of multiple detections, as the
elliptical apertures used for photometry could include emission shared with a
companion source. Although the intersection of apertures around adjacent
components was minimised where possible, it remains a potential cause of
uncertainty. Finally, the ratio for extended sources is likely to be a lower
limit, as emission from a large, extended envelopes is more likely to project
emission onto a noisy bolometer, and flux incident on these noisy bolometers
is masked during jigglemap reduction. As a result, the quoted integrated
emission is a lower limit, and the flux ratio is underestimated.

To conclude, while these concerns affect the quantative results,
\emph{qualitatively} we still observe that a significant fraction of the total
mass lies outside the central `core' at this stage of evolution.

\subsection{Surface Density}
Stars generally form in association with other stars, and the spatial
distribution of these groups of stars can provide information on how the natal
molecular cloud fragmented. One way to probe the distribution of sources is to
use the mean surface density of companion sources (MSDC), a method which has
been successfully used to probe the transition from the formation of binary
stars to star clusters in low mass star forming regions (eg. Gomez et al.
\cite{gomez93}; Larson \cite{larson95}; Nakajima et al.  \cite{nakajima98}).

\begin{figure}
  \resizebox{\hsize}{!}{
    \includegraphics[width=\hsize,angle=-90]{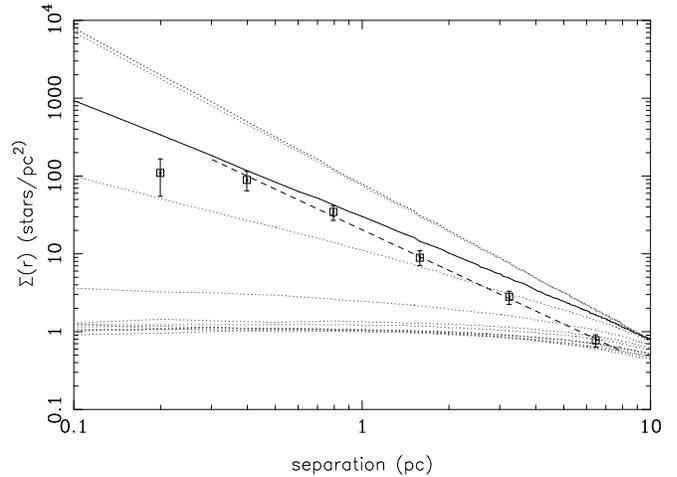}
  }
  \caption{MSDC for single power-law distribution models (light dotted
    lines), with p$($N$<r)\propto r^{\gamma}$ from $\gamma$=-2 (uppermost
    dotted line) to $\gamma$=3.5 (bottommost dotted line) with
    $\delta\gamma=0.5$. The square symbols represent the observed MCSD
    (multiplied by 100), with a thick dashed line plotting the line of best
    fit for the observed MSDC above the break-point. The best fit power-law
    distribution ($\gamma$=-0.75) is plotted by a thick black line.}
  \label{fig:surface density}
\end{figure}

The MSDC of our detections is shown in Figure \ref{fig:surface density}.  The
MSDC was calculated by measuring the linear separation $r$ of each detection
to its companions. The separation of each companion pair was binned into
annuli of separation $r$ to $r+\delta r$. The number of pairs $N$ within each
annulus was then divided by the area of the annulus and the total number of
sources $N_{*}$ to give the MSDC, $\Sigma(r)$, as
\begin{equation}
  \Sigma(r) = N / (2\pi r \delta r N_{*})
\end{equation}
Above 0.4 pc, the MSDC of our detections has a measured gradient $\gamma$ of
-1.7, which is roughly halfway between the power-law indices of binary
pre-main-sequence populations (where $\gamma \sim -0.5$; Nakajima et al.
\cite{nakajima98}) and that of more distant companions (where
$\gamma\sim-2.2$; Nakajima et al. \cite{nakajima98}). The MSDC we observe
appears to turn over below 0.4 pc, but the validity of this turnover is
questionable as we have few measurements in this region. The clump MSDC break
is also quite distinct from the stellar MSDC power-law break found at $\sim
0.04$ pc (Gomez et al. \cite{gomez93}; Larson \cite{larson95}, Nakajima et al.
\cite{nakajima98}), below which the MSDC steepens from the inclusion of close
binaries, as the stellar MSDC breakpoint occurs on much smaller scales than
are detectable by our survey.

To identify the space density distribution consistent with the clump MSDC, we
modelled a number of systems, each containing a collection of $10^{3}$ sources
randomly distributed according to a number of power-law space density
distributions. The projected MSDC of these simulations are also shown in
Figure \ref{fig:surface density}. For separations above 0.4 pc, the MSDC
power-law slope of -1.7 most closely corresponds to a space density
distribution with the number density per unit volume $N_{vol}(r) \propto
r^{-0.75}$.

We must consider the MSDC statistic cautiously, for the MSDC is constructed
from observations of widely different companion seperation sensitivities and
companion flux sensitivities, in a similar vein to the CCF
(\S\ref{sec:sensitivity concerns}). Complications arise from the wide range of
projected distances to our HMPO candidates: specifically, our observations are
sensitive to angular separations from around a beamwidth up to the upper limit
of a 120$''$ field of view, but as most HMPO candidates in our sample are less
than 4 kpc away, there are only a very small range of uniformly sampled linear
separations for our candidates. Towards more distant IRAS fields, we are
uniformly sensitive to a larger range of linear separations, but we then
suffer from fewer measurements and from a reduced sensitivity to close
companions.

We examined the significance of variable sensitivity using a procedure similar
to that used for the CCF (\S\ref{sec:sensitivity concerns}), comparing the
MSDC slope of groups of sources with similar distances, finding the slope of
each MSDC segment agrees with the MSDC of the whole sample within the
uncertainty limits. We suggest the MSDC as calculated provides at least a
basic estimate of the clustering properties of these clumps. Ultimately, the
scarcity of high-mass protostars means there will always be a large range of
distances in samples of HMPOs, and we may never be able to construct a set of
uniformly sampled observations to the extent possible with low-mass
protostars.

\subsection{Clump positions}
\label{sec:position and morphology}
For each candidate HMPO, the telescope was pointed so that the IRAS point
source was central in the field of view, hence the location of the IRAS source
in each map is given implicitly by the map centre. We indicate the position of
neighbouring MSX sources in the field of view with triangular symbols in
Figure \ref{fig:jigglemaps}, finding that many IRAS sources and MSX detections
are roughly coincident with the submm clumps, but in agreement with SBSMW we
also find that some submm detections and IR detections cannot be coincident
within the positional uncertainties of our survey and the IRAS/MSX surveys.
For example, towards IRAS 23545+6508 there are two MSX point sources found
within $\sim$ 30$''$ of the two submm clumps, an offset greater than the
expected absolute uncertainty, plus the MSX point sources also have smaller
relative separation between components than the corresponding SCUBA
detections.  Sources with large SCUBA/MSX offsets do not appear to have
$\beta$ or $S_{850}$ characteristics different to more coincident detections.

Considering each clump as a protocluster may explain displaced IRAS/MSX and
submm detections, as additional stars embedded in the less dense, more
transparent outer reaches of the envelope will not encounter the same degree
of opacity, providing a mechanism for shorter wavelength photons to pass. To
examine this possiblity and resolve whether MSX and SCUBA detections trace the
same body of material will require further high-resolution IR observations.

\section{Dust optical depth}
\label{sec:dust optical depth}
The dust optical depth $\tau$ can be determined from the flux density using
the expression
\begin{equation}
  \tau_{\nu} = -\ln \left[1-\frac{S_{\nu}}{B_{\nu}(T)\Omega}\right],
  \label{eq:tau}
\end{equation}
where $B_{\nu}(T)$ is the Planck function at temperature $T$, and $\Omega$ is
the solid angle subtended by the telescope beam, all measured at frequency
$\nu$. This relationship assumes the emission fills the telescope beam, which
may not be the case, so the derived value of $\tau_{\nu}$ must be considered
the beam-averaged optical depth.

To calculate the dust optical depth for our detections using Equation
\ref{eq:tau} we assumed dust temperatures equal to the SBSMW cold-component
dust temperatures. In the SBSMW study, the spectral energy distribution (SED)
of each IRAS source was successfully modelled as a composite of two
greybodies: one greybody representing a cold dust component, accounting for
the $>60$ $\mu$m flux, while a separate hot dust component contributes the
majority of near-IR flux. As SCUBA is only sensitive to emission from the cold
dust greybody, we set $T$ equal to the temperature of the cold component
($T_{cd}$) as given in Table 1 of SBSMW.
\label{sec:tk sed}

The beam-averaged optical depth and other parameters derived from the flux
density are listed in Table \ref{tab:derived}. While our sources consist of
very dense clumps, the beam-averaged 450 $\mu$m and 850 $\mu$m optical depths
show that they are usually optically thin at submm wavelengths. The optical
depth at 850 $\mu$m spans almost two orders of magnitude, from $10^{-3} \le
\tau_{850} \le 10^{-1}$. At 450 $\mu$m, $\tau_{450}$ is found within the range
$2\times10^{-1} \le \tau_{450} \le 10^{-2}$ with three exceptions: detection
\#8 (IRAS 18089-1732) appears optically thick at 450 $\mu$m with
$\tau_{450}=1.1$ but this is by far the brightest detection of our survey. Two
other detections (\#16: IRAS 18182-1433 and \#19: IRAS 18264-1152) have high
$S_{450}$, leading to higher $\tau_{450}$ than the majority of the detections,
but they remain with $\tau_{450}<1$.

\afterpage{\onecolumn \begin{landscape}
\begin{scriptsize}
  \begin{center}
    \tablefirsthead{
      \toprule%
      \toprule%
      &
      &
      \multicolumn{2}{c}{N$_{gas}$}             &
      \multicolumn{3}{c}{Optical Depth}         &
      \multicolumn{4}{c}{Mass (M$_{\odot}$)}    &
      \multicolumn{2}{c}{$\alpha$}              & 
      \multicolumn{1}{c}{$\beta$}               
      \\
      \cmidrule(r){3-4}\cmidrule(lr){5-7}\cmidrule(lr){8-11}\cmidrule(lr){12-13}
      \multicolumn{1}{c}{WFS}                   &
      \multicolumn{1}{c}{IRAS field}            &
      \multicolumn{1}{c}{850$\mu$m}             &
      \multicolumn{1}{c}{450$\mu$m}             &
      \multicolumn{1}{c}{$\tau_{850}$}          &
      \multicolumn{1}{c}{$\tau_{450}$}          &
      \multicolumn{1}{c}{$\tau=1$}              &
      \multicolumn{2}{c}{850$\mu$m}             &
      \multicolumn{2}{c}{450$\mu$m}             &
      \multicolumn{1}{c}{mean}                  &
      \multicolumn{1}{c}{At $S_{peak}$}         &
      \multicolumn{1}{c}{At $S_{peak}$}         
      \\
      &
      &
      \multicolumn{2}{c}{$\times 10^{22}$ cm$^{-2}$ }     &
      \multicolumn{1}{c}{$\times10^{-3}$}                 &
      \multicolumn{1}{c}{$\times10^{-3}$}                 &
      \multicolumn{1}{c}{$\mu$m}                          &
      \multicolumn{1}{c}{far}                             &
      \multicolumn{1}{c}{near}                            &
      \multicolumn{1}{c}{far}                             &
      \multicolumn{1}{c}{near}                            &
      &
      &
      \\
      \midrule
    }

    \tablehead{%
      \toprule%
      \multicolumn{14}{l}{\small\sl continued from previous page}\\
      \toprule%
      &
      &
      \multicolumn{2}{c}{N$_{gas}$}             &
      \multicolumn{3}{c}{Optical Depth}         &
      \multicolumn{4}{c}{Mass (M$_{\odot}$)}    &
      \multicolumn{2}{c}{$\alpha$}              & 
      \multicolumn{1}{c}{$\beta$}               
      \\
      \cmidrule(r){3-4}\cmidrule(lr){5-7}\cmidrule(lr){8-11}\cmidrule(lr){12-13}
      \multicolumn{1}{c}{WFS}                   &
      \multicolumn{1}{c}{IRAS field}            &
      \multicolumn{1}{c}{850$\mu$m}             &
      \multicolumn{1}{c}{450$\mu$m}             &
      \multicolumn{1}{c}{$\tau_{850}$}          &
      \multicolumn{1}{c}{$\tau_{450}$}          &
      \multicolumn{1}{c}{$\tau=1$}              &
      \multicolumn{2}{c}{850$\mu$m}             &
      \multicolumn{2}{c}{450$\mu$m}             &
      \multicolumn{1}{c}{mean}                  &
      \multicolumn{1}{c}{At $S_{peak}$}         &
      \multicolumn{1}{c}{At $S_{peak}$}         
      \\
      &
      &
      \multicolumn{2}{c}{$\times 10^{22}$ cm$^{-2}$ }     &
      \multicolumn{1}{c}{$\times10^{-3}$}                 &
      \multicolumn{1}{c}{$\times10^{-3}$}                 &
      \multicolumn{1}{c}{$\mu$m}                          &
      \multicolumn{1}{c}{far}                             &
      \multicolumn{1}{c}{near}                            &
      \multicolumn{1}{c}{far}                             &
      \multicolumn{1}{c}{near}                            &
      &
      &
      \\
      \midrule
    }

    \tabletail{%
      \midrule
      \multicolumn{14}{r}{\small\sl continued on next page}\\
      \midrule}
    \tablelasttail{\bottomrule}
    
    \tablecaption[Derived parameters of the submm detections]{ Derived
      parameters of the submm detections resolved by this survey.  The mass of
      each clump is calculated from the 850 $\mu$m integrated flux value,
      assuming the dust grains have thin ice mantles, and using a 100:1
      gas-to-dust ratio. The column density refers to the total gas column
      density (ie. $n$(H+H$_{2}$)).}

    \begin{supertabular}{%
      @{}c@{}
      c
      d{0}@{}d{0}@{}
      d{0}@{}d{0}@{}d{0}@{}
      d{0}@{}d{0}@{}
      d{0}@{}d{0}@{}
      d{-1}@{}d{-1}@{}
      d{-1}@{}
    }

1       &05358+3543     &46     &77     &12     &53     &89     &24     &       &12     &       &2.0     &2.3    &0.5    \\
2       &"              &376    &792    &43     &228    &166    &195    &       &126    &       &2.0     &2.6    &0.8    \\
3       &05490+2658     &91     &43     &8      &27     &72     &64     &       &9      &       &2.2     &2.3    &0.6    \\
4       &"              &       &63     &       &21     &66     &       &       &14     &       &2.2     &2.3    &0.6    \\
5       &"              &47     &69     &6      &22     &66     &33     &       &15     &       &2.2     &2.1    &0.4    \\
6       &05553+1631     &65     &86     &12     &65     &90     &65     &       &26     &       &1.7     &2.5    &0.7    \\
7       &18089-1732     &62     &       &7      &       &68     &1664   &128    &       &       &2.8     &       &       \\
8       &"              &478    &1327   &99     &1146   &250    &12893  &989    &11043  &847    &2.8     &3.0    &1.4    \\
9       &"              &48     &114    &13     &92     &94     &1291   &99     &949    &73     &2.8     &2.9    &1.3    \\
10      &"              &9      &       &3      &       &43     &247    &19     &       &       &2.8     &       &       \\
11      &"              &17     &       &3      &       &46     &460    &35     &       &       &2.8     &       &       \\
12      &18090-1832     &82     &53     &13     &105    &92     &1308   &570    &259    &113    &2.1     &2.1    &0.5    \\
13      &18102-1800     &257    &88     &31     &77     &142    &8049   &278    &853    &29     &-0.6    &0.1    &-1.5   \\
14      &18151-1208     &147    &152    &28     &125    &134    &211    &       &67     &       &1.8     &2.2    &0.5    \\
15      &18159-1550     &50     &46     &5      &33     &58     &1081   &174    &309    &50     &2.2     &2.6    &0.8    \\
16      &18182-1433     &189    &547    &43     &493    &167    &4198   &611    &3747   &545    &2.9     &3.3    &1.6    \\
17      &18223-1243     &126    &210    &16     &103    &102    &3102   &276    &1587   &141    &2.3     &2.7    &1.0    \\
18      &18247-1147     &129    &205    &20     &184    &115    &1784   &926    &873    &453    &2.9     &3.2    &1.6    \\
19      &18264-1152     &376    &680    &84     &631    &230    &9369   &734    &5227   &410    &2.0     &2.6    &1.0    \\
20      &18272-1217     &20     &34     &3      &31     &47     &27     &       &14     &       &2.4     &2.4    &0.6    \\
21      &"              &25     &32     &4      &23     &50     &34     &       &13     &       &2.4     &2.3    &0.5    \\
22      &18290-0924     &176    &142    &16     &66     &101    &3096   &789    &770    &196    &2.4     &2.2    &0.6    \\
23      &"              &       &95     &       &72     &120    &       &       &516    &131    &2.4     &2.4    &0.8    \\
24      &18306-0835     &26     &       &8      &       &73     &476    &100    &       &       &2.6     &2.4    &0.8    \\
25      &"              &135    &370    &24     &202    &126    &2459   &516    &2083   &437    &2.6     &3.0    &1.4    \\
26      &18308-0841     &27     &       &5      &       &58     &499    &105    &       &       &2.2     &       &       \\
27      &"              &148    &262    &22     &146    &120    &2697   &566    &1476   &309    &2.2     &2.5    &0.9    \\
28      &18310-0825     &83     &143    &13     &85     &91     &1439   &360    &760    &190    &2.8     &2.9    &1.2    \\
29      &18337-0743     &56     &187    &9      &53     &78     &1189   &144    &1216   &147    &2.8     &2.8    &1.2    \\
30      &18345-0641     &65     &86     &14     &95     &96     &931    &       &381    &       &2.8     &2.7    &1.1    \\
31      &18348-0616     &56     &       &6      &       &62     &724    &355    &       &       &2.0     &       &       \\
32      &"              &32     &       &7      &       &68     &415    &203    &       &       &2.0     &       &       \\
33      &"              &141    &112    &15     &97     &98     &1821   &892    &446    &218    &2.0     &2.2    &0.6    \\
34      &18372-0541     &87     &90     &16     &129    &100    &2498   &45     &798    &14     &2.7     &       &       \\
35      &18385-0512     &94     &185    &24     &186    &124    &2569   &60     &1565   &36     &2.4     &2.9    &1.1    \\
36      &18426-0204     &84     &94     &9      &50     &76     &2437   &16     &840    &6      &2.5     &2.7    &1.1    \\
37      &18431-0312     &61     &114    &10     &46     &82     &653    &436    &378    &252    &2.5     &2.5    &0.9    \\
38      &18437-0216     &41     &69     &7      &30     &68     &352    &       &180    &       &1.9     &2.3    &0.7    \\
39      &"              &179    &192    &10     &36     &82     &1518   &       &504    &       &1.9     &2.1    &0.5    \\
40      &"              &23     &13     &5      &13     &56     &199    &       &35     &       &1.9     &1.8    &0.3    \\
41      &18440-0148     &1      &       &0      &       &       &15     &       &       &       &2.6     &       &       \\
42      &"              &15     &33     &3      &16     &42     &164    &       &112    &       &2.6     &2.8    &1.0    \\
43      &18445-0222     &137    &257    &14     &122    &95     &1932   &614    &1120   &356    &3.0     &3.4    &1.7    \\
44      &18447-0229     &15     &       &4      &       &48     &161    &104    &       &       &2.5     &2.4    &0.7    \\
45      &"              &94     &120    &8      &48     &75     &1010   &654    &398    &258    &2.5     &2.7    &1.1    \\
46      &"              &       &84     &       &26     &74     &       &       &279    &181    &2.5     &2.4    &0.8    \\
47      &"              &21     &       &4      &       &50     &221    &143    &       &       &2.5     &1.6    &0.0    \\
48      &18449-0158     &394    &595    &24     &252    &126    &4762   &2190   &2218   &1020   &3.0     &3.4    &1.8    \\
49      &"              &       &399    &       &118    &154    &       &       &1485   &683    &3.0     &3.5    &1.9    \\
50      &18454-0136     &161    &293    &19     &119    &112    &3629   &187    &2040   &105    &2.2     &2.7    &1.1    \\
51      &18460-0307     &8      &       &3      &       &44     &121    &36     &       &       &2.3     &2.3    &0.6    \\
52      &"              &74     &125    &7      &44     &67     &1059   &317    &556    &167    &2.3     &2.5    &0.8    \\
53      &"              &       &27     &       &23     &69     &       &       &121    &36     &2.3     &2.8    &1.1    \\
54      &18470-0044     &67     &119    &11     &66     &86     &718    &       &394    &       &2.3     &2.8    &1.1    \\
55      &18472-0022     &92     &178    &9      &49     &76     &1817   &151    &1081   &90     &2.5     &2.9    &1.2    \\
56      &"              &17     &       &3      &       &42     &332    &28     &       &       &2.5     &2.3    &0.6    \\
57      &18488+0000     &127    &39     &18     &42     &108    &1606   &591    &151    &55     &2.6     &2.7    &1.0    \\
58      &"              &       &214    &       &170    &184    &       &       &835    &308    &2.6     &3.2    &1.5    \\
59      &18521+0134     &58     &127    &12     &76     &90     &750    &231    &506    &156    &2.4     &2.7    &1.0    \\
60      &"              &5      &       &2      &       &37     &68     &21     &       &       &2.4     &       &       \\
61      &18530+0215     &133    &256    &18     &92     &108    &1610   &553    &955    &328    &2.6     &2.6    &0.8    \\
62      &18540+0220     &44     &14     &3      &17     &48     &794    &77     &80     &8      &1.6     &1.6    &-0.1   \\
63      &"              &9      &       &2      &       &29     &160    &16     &       &       &1.6     &       &       \\
64      &18553+0414     &64     &114    &14     &135    &95     &1696   &       &932    &       &2.8     &3.3    &1.6    \\
65      &18566+0408     &187    &325    &27     &179    &132    &1342   &       &718    &       &2.4     &2.9    &1.2    \\
66      &19012+0536     &79     &162    &21     &161    &118    &934    &267    &591    &169    &2.3     &3.0    &1.3    \\
67      &19035+0641     &127    &271    &21     &171    &117    &98     &       &65     &       &2.3     &3.1    &1.3    \\
68      &19074+0752     &94     &23     &9      &25     &79     &1190   &206    &91     &16     &2.4     &3.1    &1.4    \\
69      &"              &       &116    &       &60     &110    &       &       &451    &78     &2.4     &2.3    &0.6    \\
70      &19175+1357     &       &47     &       &37     &87     &       &       &261    &       &2.3     &2.2    &0.6    \\
71      &"              &66     &99     &10     &58     &82     &1190   &       &546    &       &2.3     &2.6    &0.9    \\
72      &19217+1651     &118    &386    &36     &363    &153    &2074   &       &2094   &       &2.7     &3.3    &1.6    \\
73      &19220+1432     &110    &95     &12     &56     &89     &530    &       &141    &       &1.9     &2.0    &0.3    \\
74      &19266+1745     &126    &267    &24     &162    &124    &2011   &       &1313   &       &2.5     &2.9    &1.3    \\
75      &19282+1814     &121    &88     &17     &108    &106    &1301   &70     &291    &16     &2.3     &2.5    &0.8    \\
76      &"              &50     &47     &5      &25     &57     &541    &29     &156    &8      &2.3     &2.6    &1.0    \\
77      &19403+2258     &64     &57     &6      &144    &62     &404    &59     &112    &16     &2.0     &2.0    &0.3    \\
78      &19410+2336     &40     &67     &10     &38     &80     &259    &28     &136    &15     &2.2     &2.3    &0.6    \\
79      &"              &333    &642    &35     &193    &151    &2178   &234    &1295   &139    &2.2     &2.6    &0.9    \\
80      &19411+2306     &111    &38     &11     &45     &86     &595    &149    &63     &16     &1.1     &1.1    &-0.6   \\
81      &19413+2332     &29     &34     &4      &21     &54     &215    &15     &79     &6      &2.5     &2.3    &0.6    \\
82      &"              &89     &139    &8      &39     &73     &657    &46     &316    &22     &2.5     &2.6    &1.0    \\
83      &19471+2641     &18     &       &3      &       &45     &17     &       &       &       &3.1     &       &       \\
84      &"              &14     &       &3      &       &47     &13     &       &       &       &3.1     &       &       \\
85      &20051+3435     &87     &92     &7      &42     &69     &189    &35     &62     &12     &2.8     &2.7    &0.9    \\
86      &20081+2720     &50     &49     &4      &43     &50     &4      &       &1      &       &2.7     &3.5    &1.9    \\
87      &"              &60     &83     &8      &37     &74     &5      &       &2      &       &2.7     &2.2    &0.6    \\
88      &"              &62     &128    &7      &35     &70     &5      &       &3      &       &2.7     &2.2    &0.6    \\
89      &"              &10     &       &3      &       &41     &1      &       &       &       &2.7     &       &       \\
90      &20126+4104     &208    &262    &29     &165    &137    &96     &       &37     &       &2.2     &2.6    &0.8    \\
91      &20205+3948     &119    &90     &8      &29     &72     &383    &       &90     &       &2.4     &2.5    &0.7    \\
92      &"              &61     &41     &4      &30     &50     &196    &       &41     &       &2.4     &2.4    &0.6    \\
93      &20216+4107     &82     &       &10     &       &83     &38     &       &       &       &        &       &       \\
94      &20293+3952     &243    &489    &19     &233    &111    &155    &66     &96     &41     &3.5     &3.7    &1.9    \\
95      &20319+3958     &25     &       &4      &       &55     &10     &       &       &       &        &       &       \\
96      &20332+4124     &16     &166    &4      &63     &51     &39     &       &125    &       &2.5     &2.9    &1.1    \\
97      &"              &159    &94     &11     &36     &83     &385    &       &71     &       &2.5     &2.7    &1.0    \\
98      &20343+4129     &       &86     &       &35     &85     &       &       &8      &       &2.4     &2.8    &1.1    \\
99      &"              &269    &125    &15     &84     &99     &84     &       &12     &       &2.4     &2.9    &1.2    \\
100     &"              &       &160    &       &80     &127    &       &       &15     &       &2.4     &2.7    &1.0    \\
101     &22134+5834     &85     &88     &9      &50     &79     &91     &       &29     &       &2.4     &2.6    &0.8    \\
102     &22551+6221     &       &41     &       &20     &64     &       &       &1      &       &2.4     &2.7    &1.0    \\
103     &"              &170    &96     &6      &40     &65     &13     &       &2      &       &2.4     &2.4    &0.7    \\
104     &"              &23     &       &4      &       &50     &2      &       &       &       &2.4     &2.5    &0.7    \\
105     &22570+5912     &22     &       &3      &       &45     &92     &       &       &       &2.4     &       &       \\
106     &"              &31     &53     &6      &38     &65     &130    &       &68     &       &2.4     &2.8    &1.0    \\
107     &"              &85     &175    &9      &49     &79     &352    &       &224    &       &2.4     &2.6    &0.8    \\
108     &23033+5951     &121    &176    &21     &113    &118    &237    &       &106    &       &2.2     &2.6    &0.9    \\
109     &23139+5939     &126    &192    &27     &155    &132    &462    &       &218    &       &2.3     &2.6    &1.0    \\
110     &23151+5912     &55     &49     &9      &48     &76     &286    &       &78     &       &2.3     &2.5    &0.7    \\
111     &23545+6508     &44     &102    &7      &55     &68     &5      &       &3      &       &2.3     &2.7    &0.9    \\
112     &"              &51     &72     &7      &30     &68     &5      &       &2      &       &2.3     &2.1    &0.4    \\

    \end{supertabular}
    \label{tab:derived}
  \end{center}
\end{scriptsize}
\end{landscape}

 \twocolumn}

We have modelled the submm emission seen towards our sample using a
one-dimensional radiative-transfer code (details and results can be found in
the companion to this paper: Williams et al. \cite{williams2004}). These
models assume a fixed dust grain chemical composition, with a
silicon-to-graphite ratio half that of the interstellar medium (Mathis et al.
\cite{mrn}) and a standard Draine \& Lee (\cite{draine}) dust grain size
distribution. We used these model dust grains to predict the optical depth of
the cores as a function of wavelength, which when scaled to match the observed
850 $\mu$m and 450 $\mu$m optical depths gives an estimate of the wavelength
at which the submm detections become optically thick (listed in column 7 of
Table \ref{tab:derived}). We find that for the average clump, $\tau>1$ for
wavelengths shorter than $\sim$90 $\mu$m.

\subsection{Spectral index of the dust emission}
\label{sec:alpha}
As noted in \S\ref{sec:tk sed}, the dust emission at submm wavelengths can be
well represented by a greybody, with intensity varying smoothly as a function
of frequency.  Having measured the intensity of dust emission at two submm
frequencies, we can characterise the SED using $\alpha$, the spectral index of
the dust emission. This is defined
\begin{equation}
  \alpha = \frac{\ln(S_{1}/S_{2})}{\ln(\nu_{1}/\nu_{2})},
  \label{eq:alpha}
\end{equation}
where $S_{1}$ and $S_{2}$ are flux densities at frequencies $\nu_{1}$ and
$\nu_{2}$ respectively. By dividing calibrated images measured at frequencies
$\nu_{1}$ and $\nu_{2}$, we can determine the spatial distribution of $\alpha$
and examine its relationship with the intensity of emission and density of gas
and dust.

However, first we must take into account the different JCMT response and beam
patterns at 850 $\mu$m and 450 $\mu$m, for as seen in Figure \ref{fig:psf},
more flux lies outside the main beam at 450 $\mu$m compared to the response at
850 $\mu$m. We accounted for these differences by following the procedures
defined in Hogerheijde \& Sandell (\cite{hogerheijde}), normalizing the images
to a common response before finally determining $\alpha$ as the ratio of
images. In detail, we described the JCMT beam at each wavelength as a
superposition of three Gaussians, the parameters of which were found by a fit
to the azimuthal average response to Uranus. The amplitude and FWHM of these
components are listed in Table \ref{tab:gaussian components}.  We then
deconvolved the 850$\mu$m and 450$\mu$m SCUBA images with the corresponding
beam model, smoothing the deconvolved images with a single Gaussian to achieve
a final, uniform, spatial resolution of 15.0$''$ before forming $\alpha$ as
given in Equation \ref{eq:alpha}.

\begin{table}
\begin{center}
  \begin{tabular}{d{-1}d{-1}d{-1}d{-1}}
    \toprule
    \toprule
    \multicolumn{2}{c}{850 $\mu$m} &\multicolumn{2}{c}{450 $\mu$m} \\
    \cmidrule(r){1-2}\cmidrule(l){3-4}
    \multicolumn{1}{c}{Relative} &\multicolumn{1}{c}{FWHM} &\multicolumn{1}{c}{Relative} &\multicolumn{1}{c}{FWHM} \\
    \multicolumn{1}{c}{Amplitude} &\multicolumn{1}{c}{(``)} &\multicolumn{1}{c}{Amplitude} &\multicolumn{1}{c}{(``)} \\
    \midrule
    0.93 &14.4 &0.81 &8.0  \\
    0.05 &44.3 &0.17 &24.7 \\
    0.02 &62.9 &0.02 &71.3 \\
    \bottomrule
  \end{tabular}
\end{center}
\caption{Parameters used for a three Gaussian component description of the JCMT
beam.}
\label{tab:gaussian components}
\end{table}

The spatial distribution of $\alpha$ can be seen to the right of the submm
emission maps in Figure \ref{fig:jigglemaps}; $\alpha$ is blanked for emission
lying outside the first 850 $\mu$m and 450 $\mu$m contours, so only the
intersecting area of the maps with good signal to noise is both displayed and
analysed. In Table \ref{tab:derived} we list the spectral index both in terms
of the index at the position of peak 850 $\mu$m flux ($\alpha_{peak}$; column
13) and the mean value of the $\alpha$ map ($\alpha_{mean}$; column 12).
Unless we state otherwise, the spectral index is discussed in terms of
$\alpha_{peak}$, the index at the location of peak submm flux. Also, we do not
include detection \#13 (IRAS 18102-1800) in the analysis, as it appears an
outlier with significantly lower $\alpha$ than any other detection; we suspect
this is caused by the suspiciously weak 450 $\mu$m emission seen towards this
source, most likely due to a transient telescope or calibration error as the
450 $\mu$m flux appears inconsistent with the 100 $\mu$m, 850 $\mu$m and 1 mm
flux constraints.

The spectral index measured at the position of peak 850 $\mu$m emission varies
from $\alpha_{peak}=1.1$ (IRAS 19411+2306) to $\alpha_{peak}=3.7$ (IRAS
20293+3952), though overall this index is fairly uniform with a sample mean of
$\overline{\alpha_{peak}} = 2.6 \pm 0.4$. Averaging all spectral index data
around a detection slightly reduces the statistical variability, so that
$\alpha_{mean}$ ranges from 1.1 to 3.5, and the sample mean falls to
$\overline{\alpha_{mean}} = 2.4 \pm 0.3$. As implied by these statistics, the
majority of detections display an $\alpha$ distribution that peaks towards the
location of maximum 850 $\mu$m emission, although some sources display an
anticorrelation with intensity.

\begin{figure*}
\resizebox{\hsize}{!}{
  \includegraphics[angle=0,width=0.49\hsize]{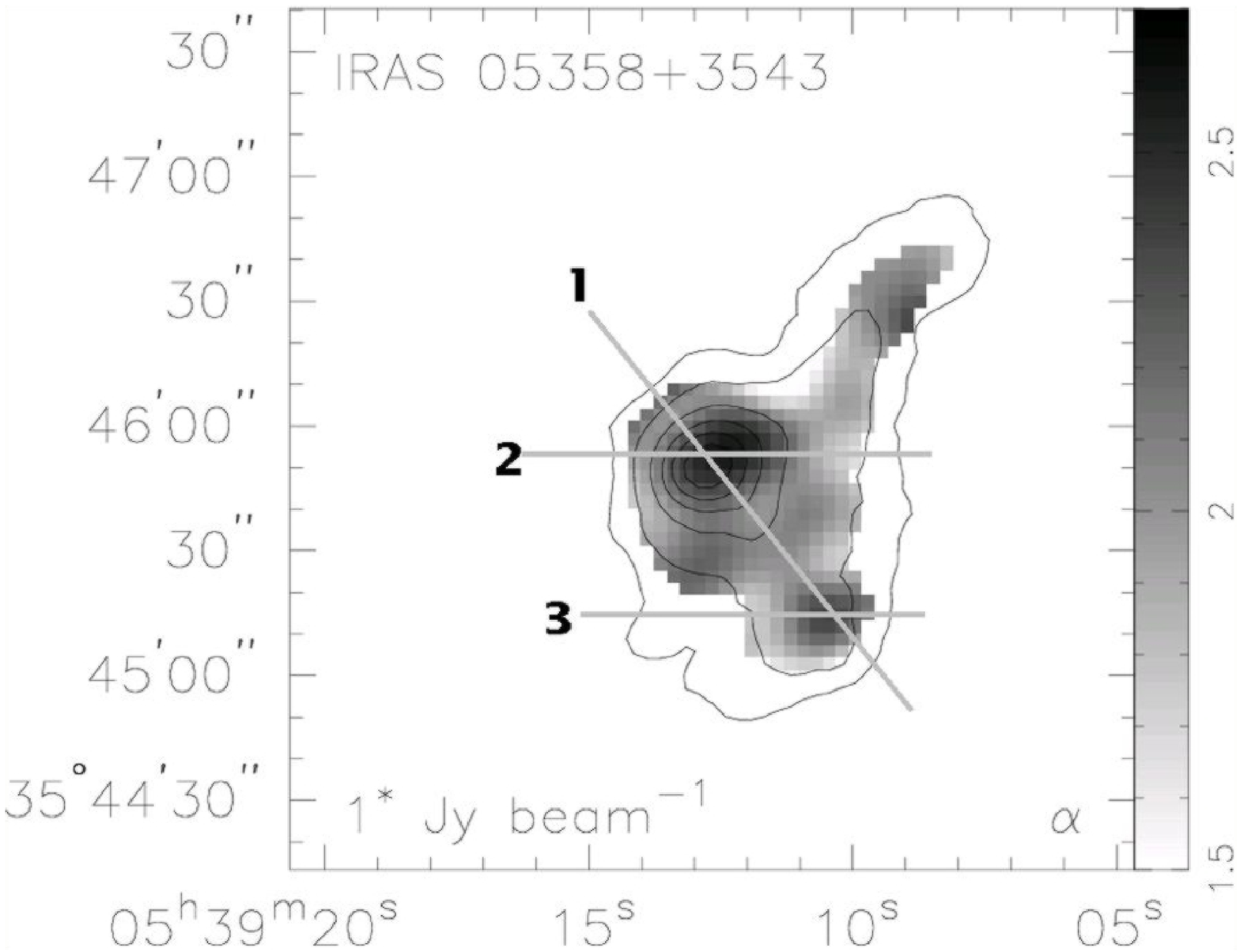}
  \includegraphics[angle=0,width=0.49\hsize]{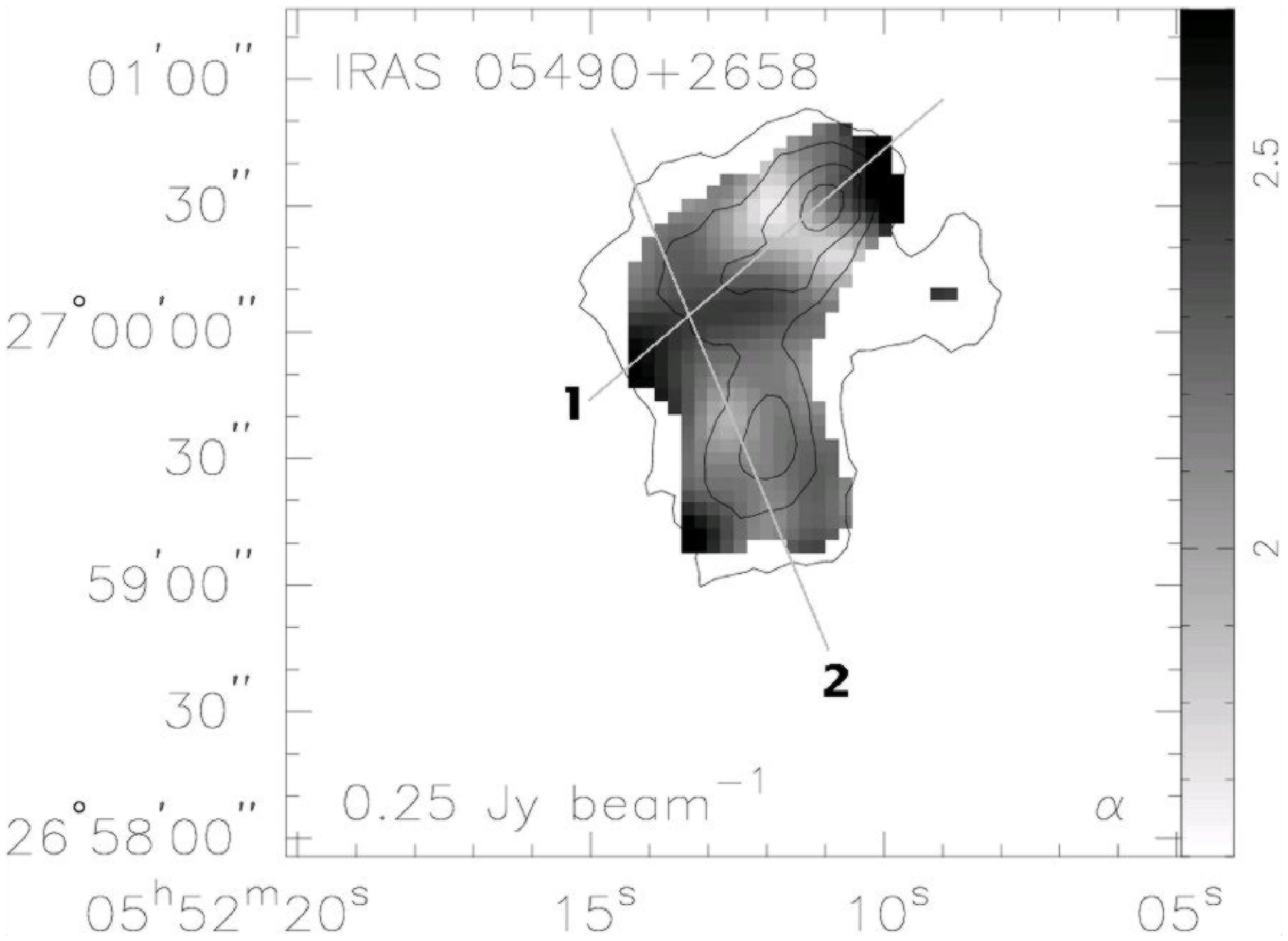}
} 
\resizebox{\hsize}{!}{
  \includegraphics[angle=-90,width=0.49\hsize]{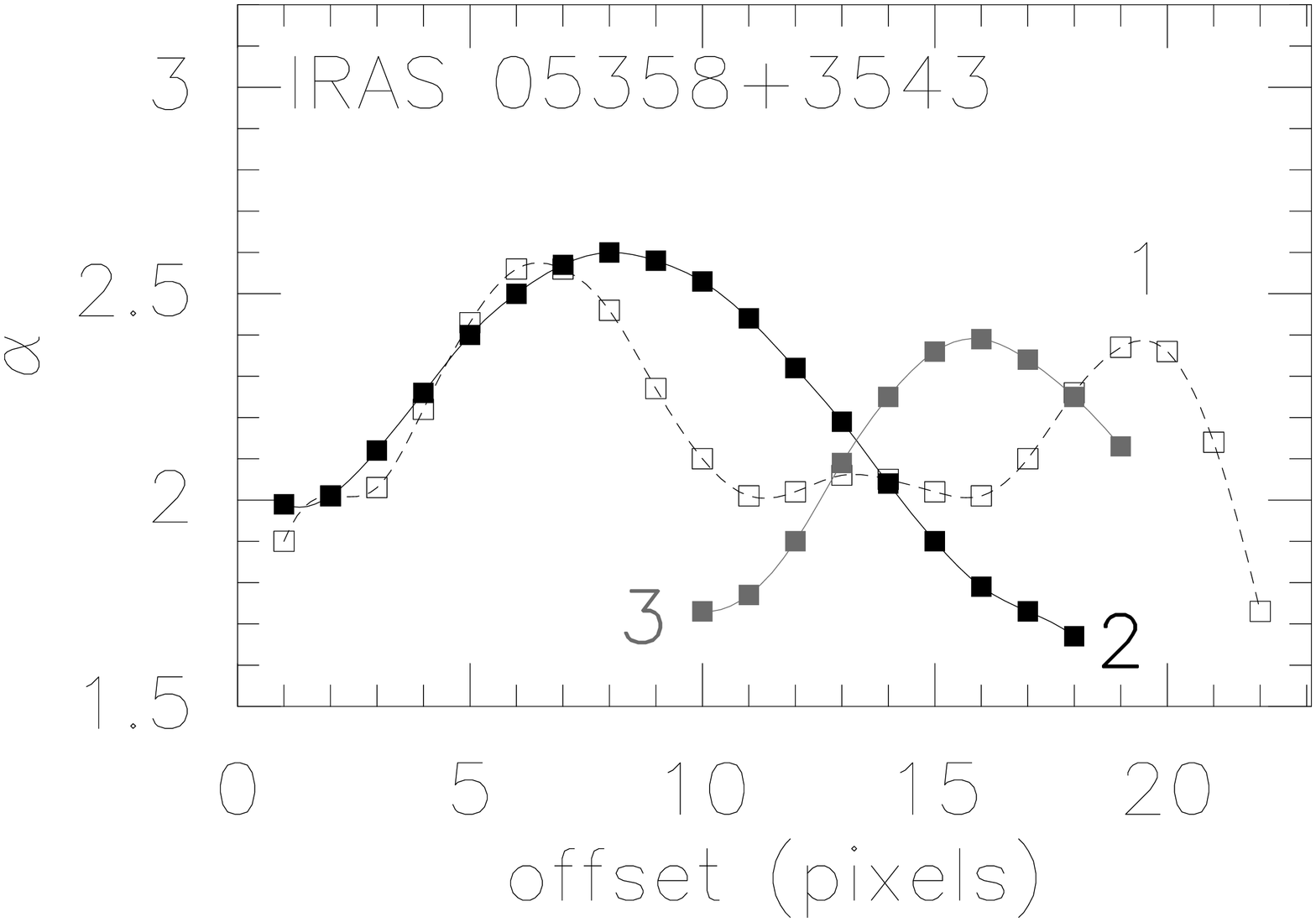}
  \includegraphics[angle=-90,width=0.49\hsize]{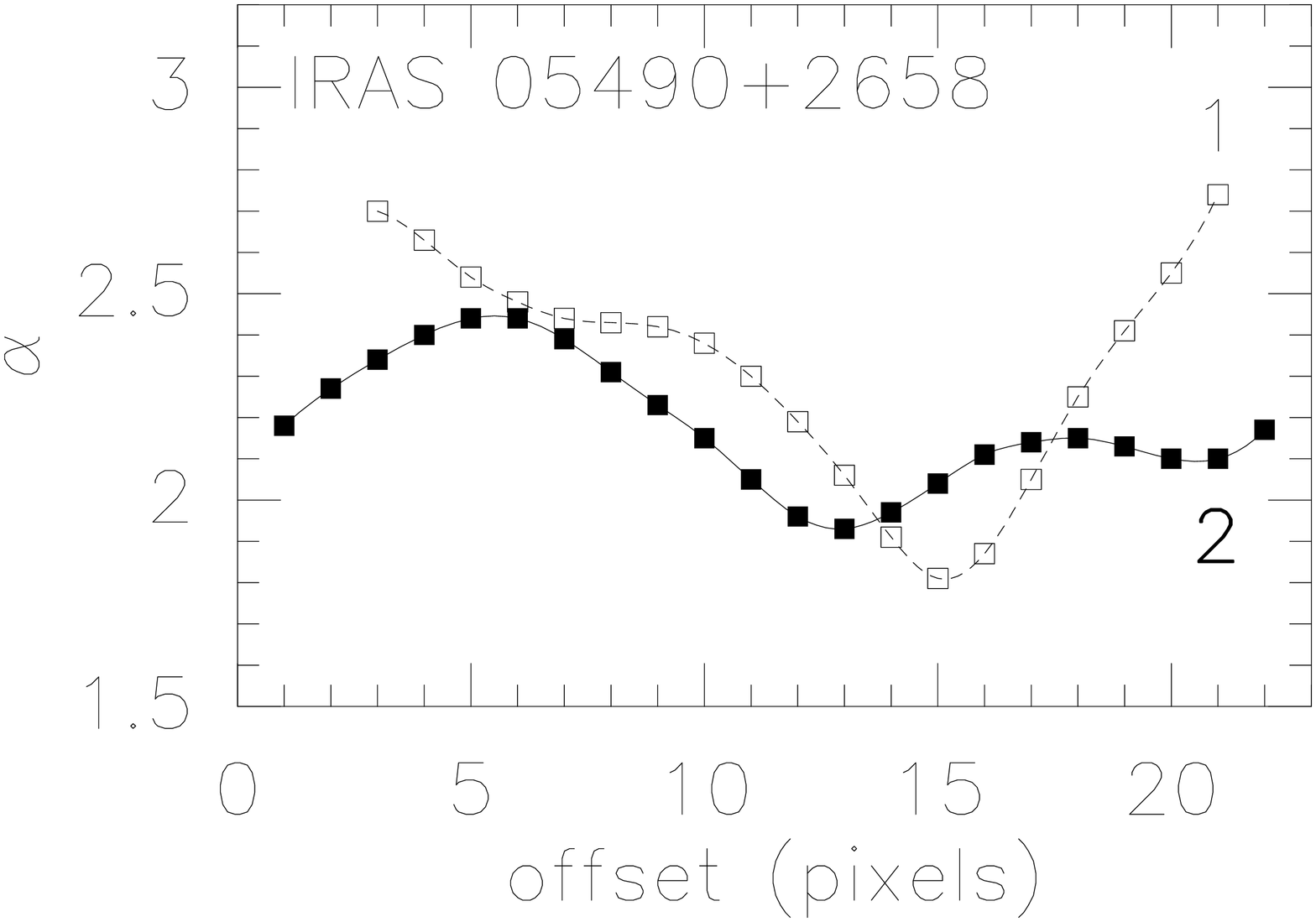}
} 
\caption{A plot of the typical cross-sections seen for peaked $\alpha$
  morphologies (IRAS 05358+3543; left-hand plots) and negative-dip
  morphologies (IRAS 05490+2658; right-hand plot). The upper row displays the
  orientation of the cross-sections, while the bottom row displays the
  $\alpha$ index measured along the labelled cut, using a pixel scale of $3''$
  per pixel.}
\label{fig:alpha slices}
\end{figure*}

Cross-sections of the $\alpha$ distributions towards IRAS 05358+3543 and IRAS
05490+2658 are presented in Figure \ref{fig:alpha slices}. Although these
sources display very different $\alpha$ morphologies, we note that the
positive and negative features seen towards the location of peak submm
emission are roughly comparable in depth and width. As these sources are at
roughly the same kinematic distance, the cause of these features could
potentially occur on a similar spatial scale.

\subsubsection{Peaked $\alpha$ distributions}

The spectral index of the dust emission depends on a combination of the
beam-averaged values of dust temperature, opacity, and spectral index of the
dust opacity ($\beta$, defined in the sense $\tau \propto \nu^{\beta}$). As a
result, there are three mechanisms which may explain the spatial distributions
we observed:
\begin{enumerate}
\item Temperature variations through the dust envelope.
\item Emission originating from optically thick regions.
\item Changes in the composition of the dust grains themselves.
\end{enumerate}

The only currently available estimate of dust temperature was derived by SBSMW
using greybody fits to the SED at long wavelengths. The majority of dust cores
have a temperature similar to the mean of the sample ($\overline{T_{dust}}$=44
K), although a number of cores are associated with higher temperatures: IRAS
18440-0148 with $T_{dust}$=97 K, IRAS 20319+3958 with $T_{dust}$=73 K, and
IRAS 23151+5912 with $T_{dust}$=68 K are the most prominent higher temperature
cores. However, with only a single dust temperature estimate for each source,
the magnitude of any dust temperature gradient across the protostellar
envelopes remains unknown, which leaves the contribution of any
temperature-dependent mechanism to variations in $\alpha$ unclear.

Single-dish NH$_{3}$ observations have also been conducted towards our sample,
tracing gas within the cooler, extended envelope (SBSMW). From these
observations, SBSMW found a mean temperature of $T_{NH_{3}}$=19 K, around 25 K
lower than the dust temperature in an average core. However, a core containing
warm dust and an extended envelope characterised by cool gas does not prove
the existence of a temperature gradient, for it is very difficult to make the
gas temperature close to that of warm dust, even with the high densities
($N_{H}\simeq 10^{6}$ cm$^{-3}$) seen towards typical protostellar candidates
(Goldsmith et al.  \cite{goldsmith}).  However, CH$_{3}$OH and CH$_{3}$CN
molecular tracers, pointing to high temperature, high density regions and
indicating the presence of a hot core, have been detected towards a number of
our candidate HMPOs (SBSMW). These detections imply that there are regions
within the clumps of much higher temperature than the beam-averaged dust
temperature alone would suggest, so we expect a strong temperature gradient
must be present towards at least some of our sources.

We examined the significance of a temperature gradient by forming simulated
$\alpha$ maps using the 450 $\mu$m and 850 $\mu$m continuum images created by
our best-fit radiative transfer models (Williams et al.  \cite{williams2004}).
These models assume a single luminous protostar embedded in a dense, dusty
envelope, and form excellent fits to the observed emission while maintaining
constant dust grain characteristics (ie. opacity and variable grain
composition are not a factor in the simulated $\alpha$ map). In general, our
best fit models suggest the presence of dust envelopes with temperatures
around $300-500$ K at the inner boundary, falling to around $10-15$ K at the
outer boundary. With a temperature gradient as the only factor, the simulated
$\alpha$ distribution peaks towards the hottest, densest, most central
regions. These centrally peaked $\alpha$ morphologies are similar to those
seen towards the majority of our sample (eg. IRAS 18247-1147; IRAS
18306-0835), suggesting the observed $\alpha$ features are dominated by
temperature gradients across the envelope.  This result emphasises that we
must know the spatial temperature distribution of the clumps if we are to
refine our investigation and accurately quantify the contribution of other
factors towards these sources.

\subsubsection{$\alpha$-dip distributions}

On the other hand, peaked $\alpha$ distributions are not the only morphology
observed: the IRAS sources 05490+2658, 18290-0924, 18530+0215, 19413+2332 and
20051+3435 form notable exceptions where $\alpha$ falls towards the location
of maximum 850 $\mu$m emission. In terms of temperature gradients, these
$\alpha$ morphologies run counter to the $\alpha$ distribution expected for an
internally heated core: if no other factors are involved, they imply that the
inner core must be cooler than the surrounding envelope. But is the formation
of a hot envelope and cool inner core a realistic possibility? To form a
typical negative dip morphology with an $\alpha$ valley depth of $\Delta
\alpha=-0.3$, while maintaining an average dust temperature of 44 K, would
require an inner core temperature of around 26 K with a surrounding envelope
of $\sim60$ K. It would be hard to explain such low inner temperatures in the
presence of large, luminous protostars when `hot-cores' associated with
typical pre-UCHII protostars have temperatures of $>100$ K (Kurtz et al.
\cite{kurtz:ppiv}).  Alternatively, external heating could warm the outer
layers of the clump relative to the inner core, but the interstellar radiation
field alone is not capable of heating such dense dust to such high
temperatures, and while nearby luminous stars could conceivably heat the
exterior to higher temperatures our radiative transfer modelling shows the
submm emission profiles are well matched by \emph{low} temperatures
($\sim10-15$ K) at the external envelope boundary (Williams et al.
\cite{williams2004}). Furthermore, low core temperatures may preclude the very
formation of a massive protostar, as the Jeans mass becomes much lower within
cooler cores, suggesting that a series of lower mass protostars would form
instead. This does not exclude the possiblity that a massive star could form
through the coalescence of discrete low-mass protostars, but considering the
weight of evidence it is hard to envisage how `cool cores' may cause the
observed variations.

An alternative explanation is that these $\alpha$-dip cores are optically
thick. This possibility is unlikely, as \S\ref{sec:dust optical depth} shows
that all cores (with the exception of Source \#8, the main component towards
IRAS 18089-1732) are optically thin even at 450 $\mu$m. Even so, we recognise
that this statement is based on the beam-averaged values, and there may be
much denser, optically thick regions present on scales smaller than our
observations can probe. For example, a circumstellar disk would lead to a
large density concentration in the very inner envelope, but would remain
unresolved by our observations.  Then again, the presence of circumstellar
disks has been confirmed towards IRAS 20126+4104 (Cesaroni et al.
\cite{cesaroni1997}) and IRAS 05553+1631 (Shepherd \& Kurtz
\cite{shepherd1999}), and the $\alpha$ distribution towards these sources
remains strongly peaked. Overall, we conclude that optically thick regions do
not significantly affect the $\alpha$ distribution at the spatial resolution
of our measurements.

Finally, variations in the properties of the dust-grains themselves could help
explain the trends. The optical properties of dust grains can be quantified by
$\beta$, the spectral index of the dust opacity. This is often a quantity of
interest as it may give information on the composition and evolutionary
history of dust grains within the envelope. There are many models that predict
$\beta$ for different grain characteristics, and the majority of grain
compositions result in a spectral index of $\beta\sim1.5-2$ (eg. Gezari, Joyce
\& Simon \cite{gerazi}; Draine \& Lee \cite{draine}; Kr\"{u}gel \&
Siebenmorgen \cite{krugel}), although it may range from $\beta\sim1$ (Mathis
\& Whiffen \cite{mathis}) up to $\beta=3$ (Aannestad \cite{aannestad}). To
observe an $\alpha$-dip morphology, dust grains within the central core must
be of lower $\beta$ than grains in the surrounding envelope. Low $\beta$ and
$\beta$ distributions that fall towards regions of high density are usually
attributed to grain growth in these dense, innermost regions (eg.  Mannings \&
Emerson \cite{mannings}; Beckwith \& Sargent \cite{beckwith1991}; Goldsmith et
al.  \cite{goldsmith}), and our observed $\alpha$-dip morphologies are
generally consistent with this grain growth interpretation, as $\alpha$ (and
thus $\beta$) fall preferentially towards the centre of the dense cores we
have observed.

However, the models of Ossenkopf \& Henning (\cite{ossenkopf}) predict that
$\beta$ will only change if the dust grains do not have ice mantles, which
would require inner cores with dust temperatures $\geq$ 100 K. The detection
of CH$_{3}$OH and CH$_{3}$CN towards the $\alpha$-dip detections IRAS
19413+2332 and IRAS 18530+0215 signifies the presence of a hot core of
sufficient temperature to melt ice mantles, thus permitting grain growth.
However, the detection of CH$_{3}$OH and CH$_{3}$CN towards a large number of
candidate protostars with centrally peaked $\alpha$ morphologies raises an
interesting question: these molecular tracers imply a high central temperature
- certainly high enough to melt ice mantles and permit changes in $\beta$, so
why are there no signs of grain growth?  Why do the majority of these sources
have positively peaked $\alpha$ morphologies?  The strongly peaked $\alpha$
distributions we observe suggests that the temperature gradient effect
outweighs any contribution from grain evolution.  On the other hand, as it
takes time for ice mantles to melt, perhaps these icy grains remain towards
cores only recently heated, and perhaps these cores are younger than those
associated with $\alpha$-dip distributions.

Also, CH$_{3}$OH and CH$_{3}$CN have not been detected towards IRAS 05490+2658
and IRAS 18290-0924, suggesting a hot core has not formed, yet these objects
are still found with $\alpha$-dip morphologies. Clearly, neither temperature
gradients or variable dust grain compoaition taken alone cannot fully explain
the observed $\alpha$ morphologies. It is not clear that hot cores are
strongly correlated with grain growth nor with $\alpha$-dip morphologies, and
accurate high-resolution measurements of the temperature of the clumps are
vital if we are to determine the magnitude of grain growth towards our sample.

\subsection{$\beta$ and signs of grain growth}
\label{sec:beta}
As we have an estimate of the temperature of dust grains within the cores, we
can calculate the spectral index of the dust opacity using the equation
\begin{equation}
  \beta = (\alpha + \Delta\alpha) - 2,
\end{equation} 
where $\Delta\alpha$ is a Rayleigh-Jeans correction factor, necessary because
the dust body temperatures are closer to the equivalent temperature $T_{\nu}$
of our observations than a Rayleigh-Jeans approximation would permit (where
temperature $T_{\nu}$ at frequency $\nu$ is $T_{\nu}=h\nu/k$, giving $T_{850}
\approx 17$ K and $T_{450} \approx 32$ K). Assuming the equivalent
temperatures $T_{\nu_{1}}$ and $T_{\nu_{2}}$ for the observed frequencies
$\nu_{1}$ and $\nu_{2}$, the required correction can be expressed
\begin{equation}
  \Delta\alpha = \frac{
    \ln \left[ \frac{\exp(T_{\nu_{2}} / T_{dust})-1}
      {\exp(T_{\nu_{1}}/T_{dust})-1}\right]
  }{\ln (T_{\nu_{2}}/T_{\nu_{1}})}
  - 1.
  \label{eq:correction}
\end{equation}
For example, with a median cold-component dust temperature of 43 K, the
average correction is around +0.3. By taking our $\alpha$ maps and adding
$(\Delta \alpha - 2)$ to each image, the $\alpha$ maps are transformed to
display the spatial distribution of $\beta$. However, as we only have an
estimate of the dust temperature in the inner core (setting $T_{dust} =
T_{cd}$, the cold-component dust temperature from SBSMW), likewise the
estimate of $\beta$ obtained from these maps is only valid towards the
innermost regions.

\begin{figure}
\includegraphics[angle=-90,width=\hsize]{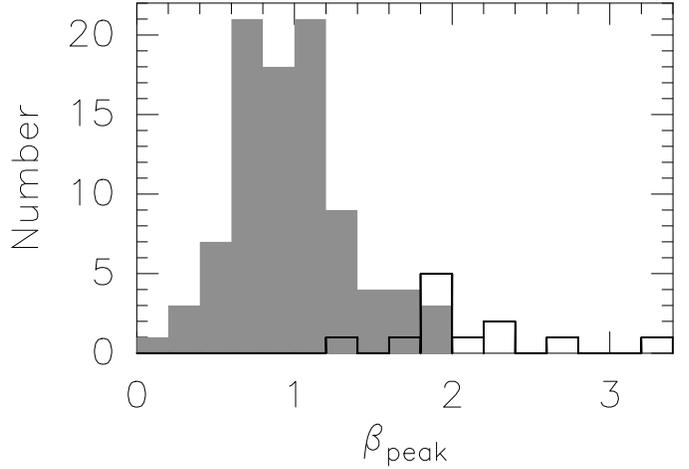}
\caption{
  A histogram of the $\beta_{peak}$ distributions for our sample of
  HMPOs (filled histogram) and for young high-mass stars associated
  UCHII regions (clear bars). The latter distribution is derived from
  the cold-component $\beta$ values listed in Table 3.5 in Hunter
  (\cite{hunter thesis}).}
\label{fig:beta histogram} 
\end{figure}

\begin{figure*}
\resizebox{\hsize}{!}{
  \includegraphics[angle=-90,width=0.33\hsize]{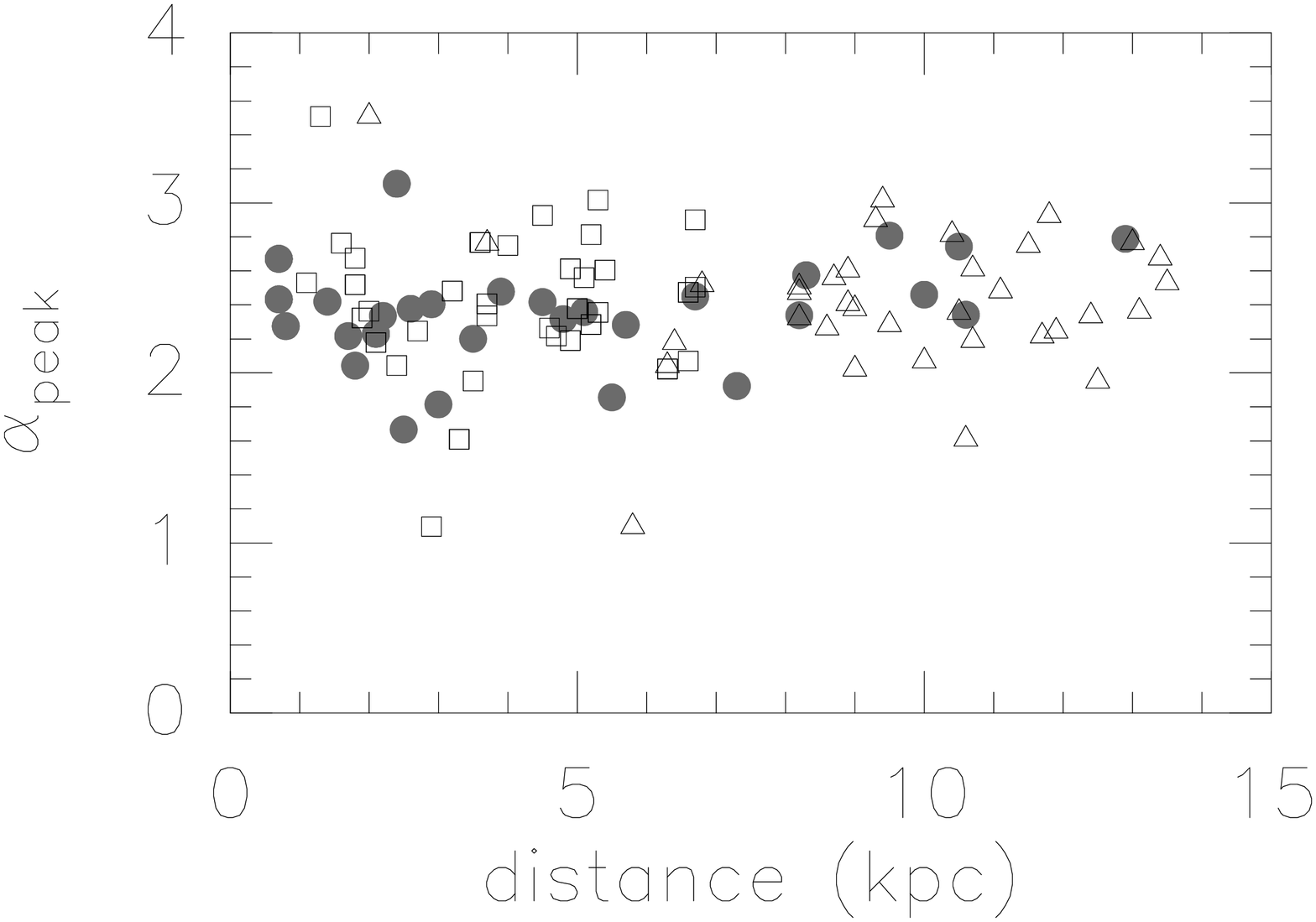}
  \includegraphics[angle=-90,width=0.33\hsize]{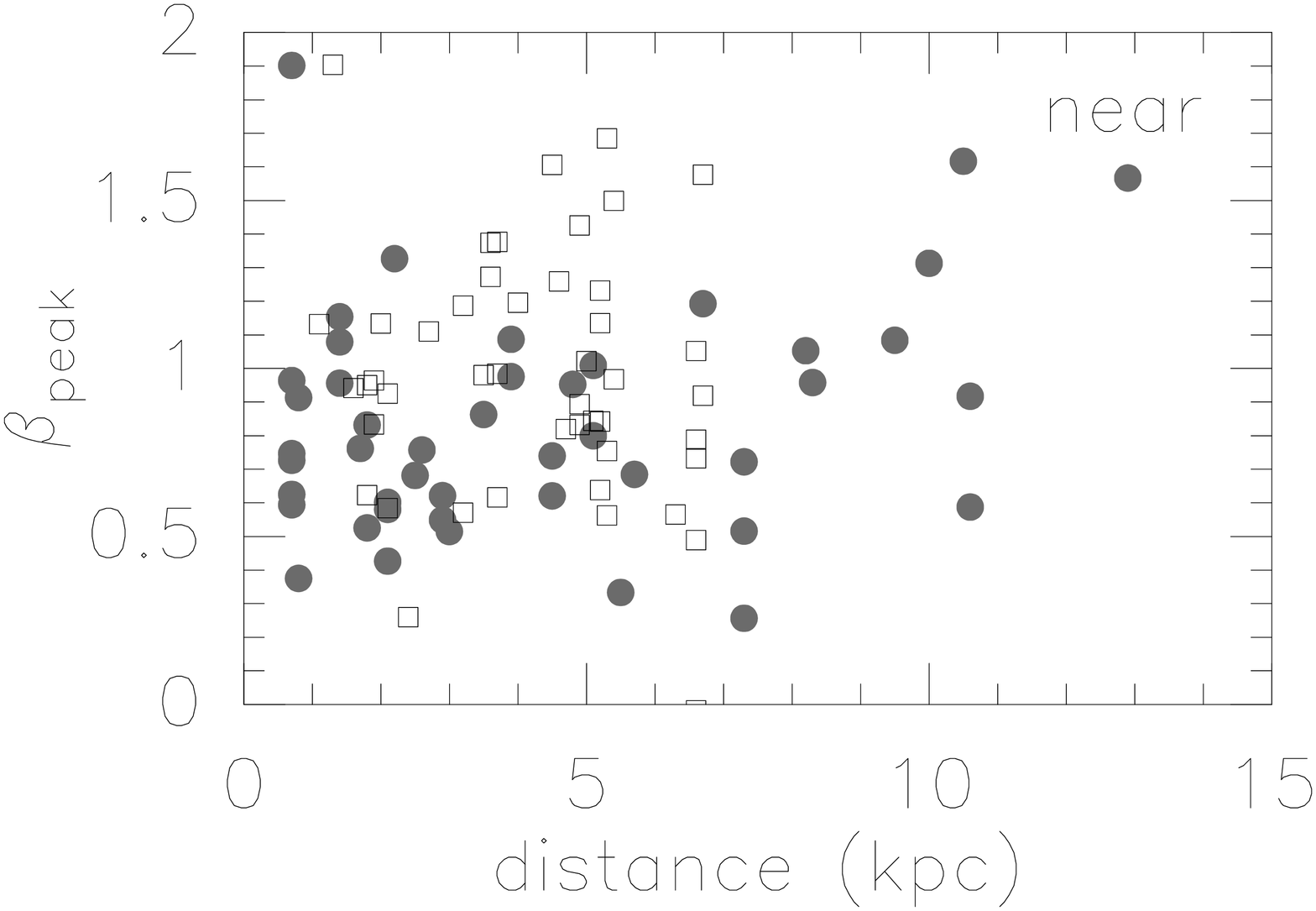}
  \includegraphics[angle=-90,width=0.33\hsize]{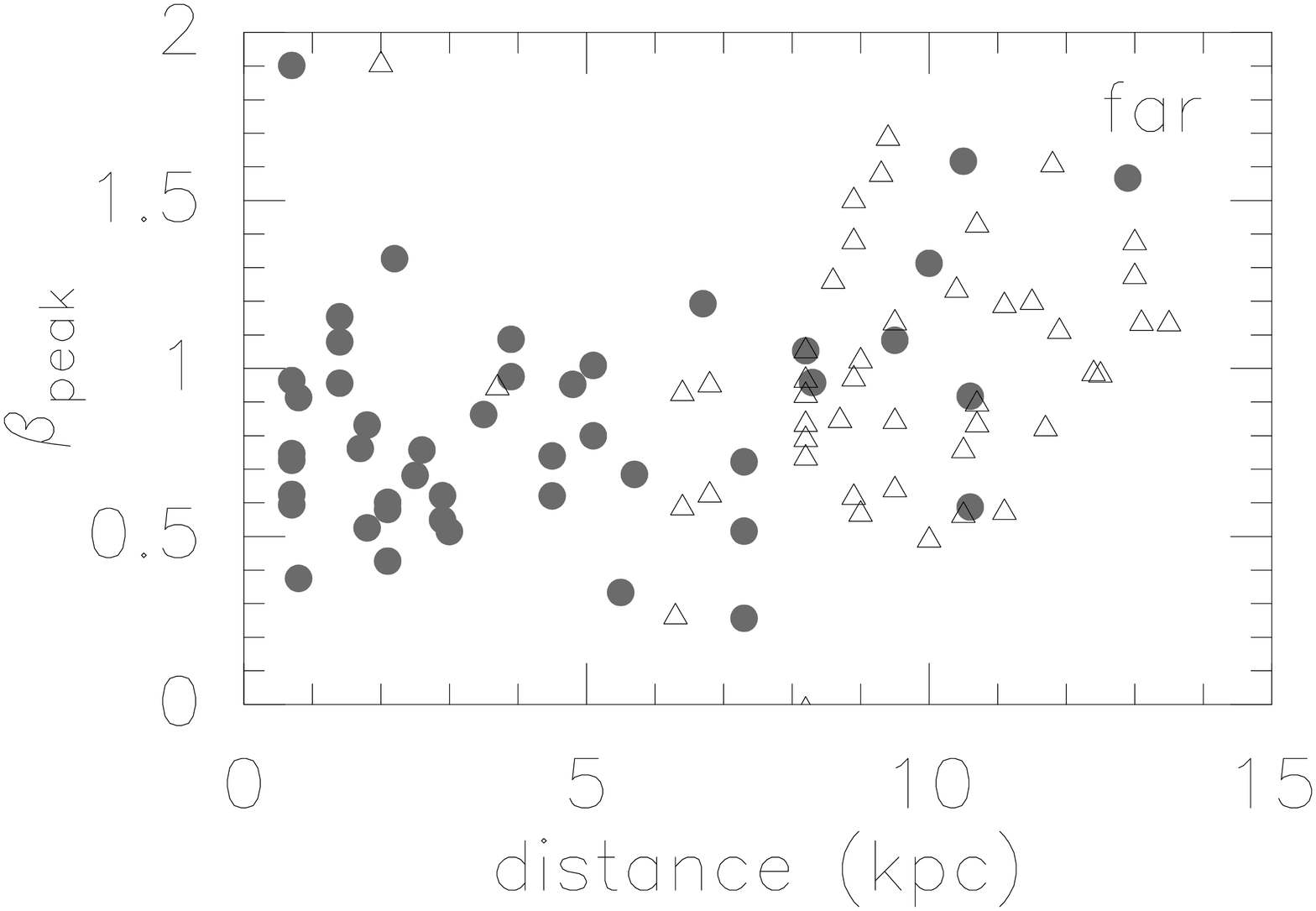}
}
\caption{
  A plot of $\alpha_{peak}$ (left-hand plot) and $\beta_{peak}$
  (centre and right-hand plot) against distance. Distance resolved
  sources are plotted by filled circles, while the square and
  triangular symbols represent distance unresolved sources projected
  to the near and far kinematic distance respectively.}
\label{fig:beta against distance}
\end{figure*}

The value of $\beta$ at the location of peak 850 $\mu$m submm emission is
listed individually for each detection in Table \ref{tab:derived}.  The
average grain opacity index for our sample is $\overline{\beta} = 0.9 \pm
0.4$. This index is smaller than for that seen towards high-mass stars
associated with UCHII regions (Hunter \cite{hunter thesis}), and the $\beta$
distribution is substantially shifted to lower indices compared to the more
evolved objects (Figure \ref{fig:beta histogram}). Low $\beta$ is often
associated with young, less evolved sources, which would point to further
evidence that our sample of young high-mass stars are at an earlier stage of
evolution than their UCHII counterparts. An inadequate Rayleigh-Jeans
correction could raise our estimate of $\beta$, but even if we have globally
overestimated core temperatures by 20 K, $\overline{\beta}$ would only to rise
to $\sim 1.2$, still lower than $\beta$ towards more evolved UCHII sources. No
relationship is found between $\alpha$, $\beta$ and distance (Figure
\ref{fig:beta against distance}), suggesting resolution is not an issue.

\begin{figure*}
\resizebox{\hsize}{!}{
  \includegraphics[angle=-90,width=0.49\hsize]{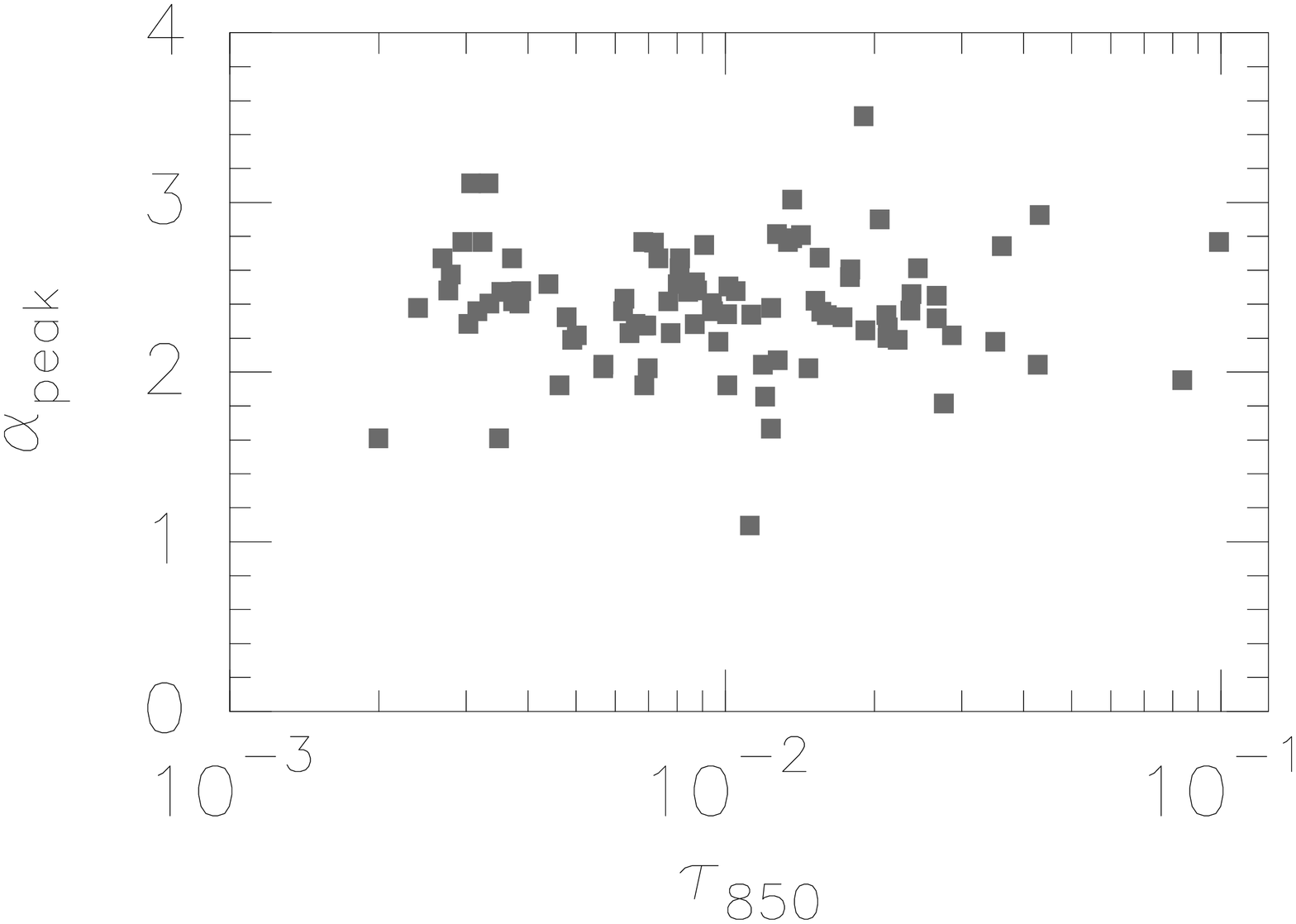}
  \includegraphics[angle=-90,width=0.49\hsize]{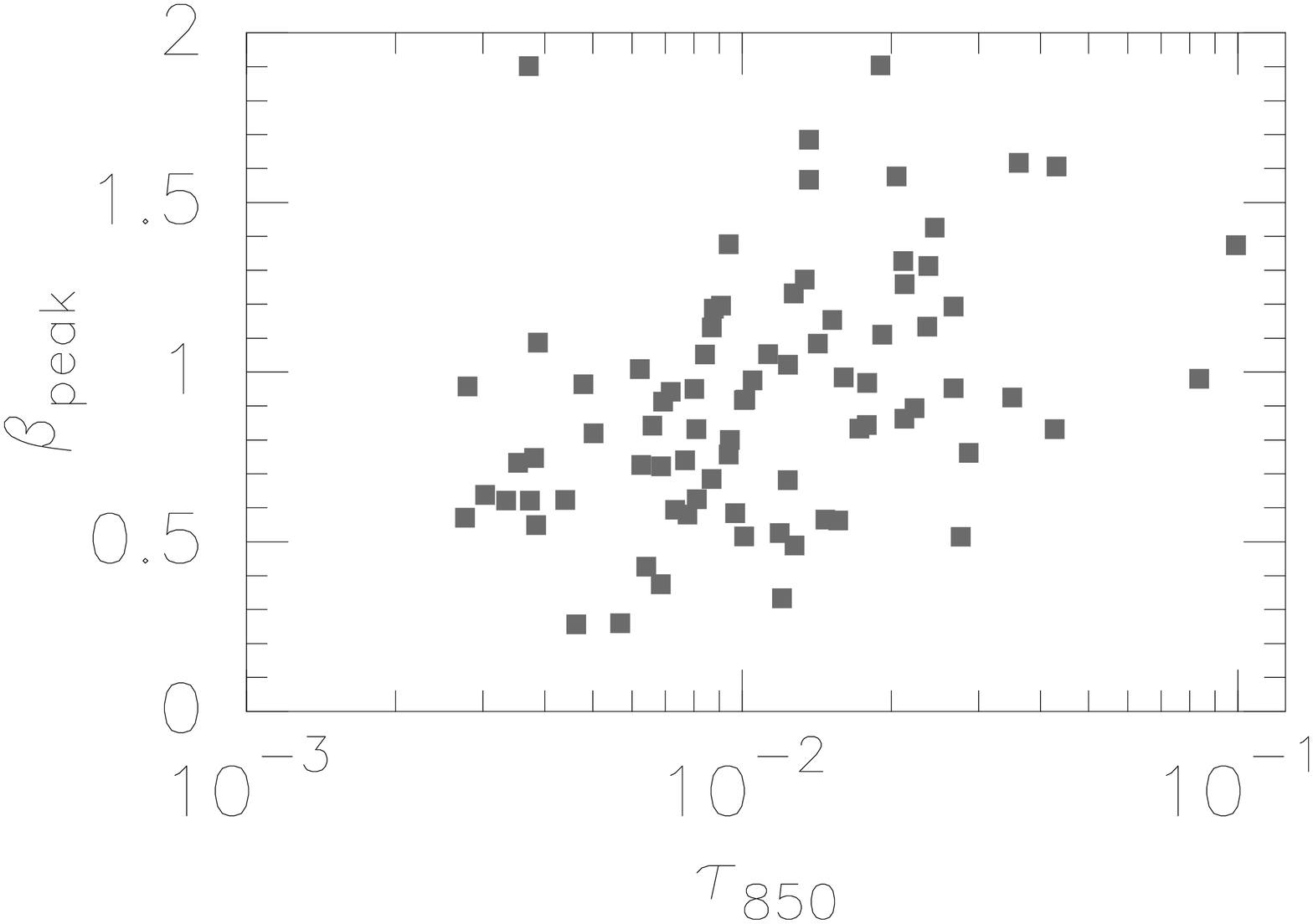}
}
\caption{A plot of the relationship of $\alpha_{peak}$ (left-hand plot)
  and $\beta_{peak}$ (right-hand plot) to $\tau_{850}$, the optical depth at
  850 $\mu$m.}
\label{fig:alpha vs 850 tau}
\end{figure*}

Figure \ref{fig:alpha vs 850 tau} shows that despite the absence of any
correlation between $\alpha$ and $\tau_{850}$ or $\alpha$, $\beta$,
$\tau_{850}$ or $T_{dust}$ and source distance, there is significant
correlation between $\beta$ and $\tau_{850}$, albeit with a large scatter.
Our estimate of $\beta$ is dependent on the SBSMW dust temperature, which was
calculated assuming $\beta$=2. Had SBSMW used lower values of $\beta$, they
would have derived higher dust temperatures, which in turn both reduces the
Rayleigh-Jeans correction and decreases the implied dust optical depth.  As a
result, low-$\beta$ points in Figure \ref{fig:alpha vs 850 tau} would move
down (due to the smaller Rayleigh-Jeans correction) and to the left (due to
decreased optical depth), a shift greatest for points towards the bottom-left
quadrant as these objects would be associated with the largest dust
temperature increase, hence strengthening the correlation between $\beta$ and
$\tau_{850}$.

Low values of $\beta$ are typically attributed to growth and evolution of the
dust grains within dense, dusty regions, so the observed trend of low values
of $\beta$ at low optical depths is initially surprising.  We would expect the
densest, most massive clumps to undergo the most significant grain growth,
resulting in lowest $\beta$ for these cores, whereas actually we seem to
observe the opposite trend. In addition, Ossenkopf \& Henning
(\cite{ossenkopf}) found that dust grains within a protostellar core remain
below the Rayleigh scattering size limit after 10$^{5}$ years (the typical age
expected for our sample; Behrend \& Maeder \cite{behrend2001}), with the dust
grain opacity changing only by a factor of $\sim 2$ at 850 $\mu$m. However,
the results in Figure \ref{fig:alpha vs 850 tau} suggest a change of close to
an order of magnitude in opacity: for values of $\beta$ around 0.5,
$\tau_{850}$ has a value of around $4 \times 10^{-3}$, increasing to
$\tau_{850} \sim 3 \times 10^{-2}$ for dust grains with $\beta = 1.5$. Only in
the extremely dense and cold regions within circumstellar disks is it believed
that a significant number of grains can grow beyond the Rayleigh limit,
allowing the large shift in opacity (e.g. Schmitt et al.  \cite{schmitt1997}).

\begin{figure}
\resizebox{\hsize}{!}{
\includegraphics[width=\hsize,angle=270]{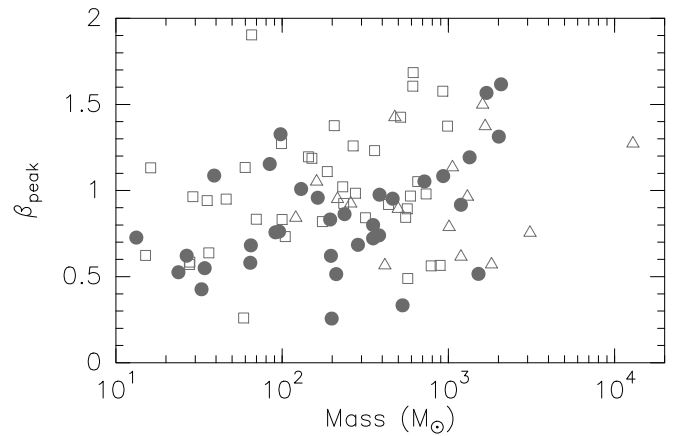}
}
\caption{A comparison of detection mass and $\beta$ at the location
  of peak 850 $\mu$m emission ($\beta_{peak}$). Sources with known distance
  are plotted by filled circles, while sources whose distance remains
  unresolved are projected to the near kinematic distance (square symbols) and
  the far kinematic distance (triangular symbols).}
  \label{fig:mass vs beta}
\end{figure}

These inconsistencies can be understood by considering the inhomogeneous
nature of our sample, and how the clumps we have observed vary in mass
(\S\ref{sec:mass}) and most probably evolutionary status (SBSMW). Sources
within the high $\tau_{850}$, high $\beta$ quadrant of Figure \ref{fig:alpha
  vs 850 tau} show considerable `excess' mass compared to the mass of an
equivalent luminosity main-sequence star (cf. Figure \ref{fig:mass vs beta}),
suggesting these clumps could easily be forming protogroups or protoclusters
(\S\ref{sec:cluster considerations}).  In any case, the majority of dust
grains within these high-mass clumps will not be intimately associated with
the high-mass protostar, and will most likely remain outside the $T > 100$ K
boundary necessary to melt ice mantles and allow large variations in $\beta$.

In contrast, the much lower mass of the low $\tau_{850}$, low $\beta$
detections means these clumps may be forming solitary high-mass stars
(solitary for high-mass stars being a relative term, which we interpret as
existing with only a small number of lower-mass stellar companions), with
fractionally much more dust lying inside the massive protostar's sphere of
influence. This distinction becomes important when we consider that the
$\tau_{850}$ and $\beta$ we observe are actually the optical depth and dust
opacity index \emph{averaged} along the line of sight. Within the largest
clumps, low $\beta$ grains in the vicinity of the high-mass protostar will be
rendered less detectable, swamped by the higher $\beta$ dust grains lying
within the envelope of the proto-group/cluster, whereas fractionally there
will be many more low-$\beta$ grains along the line of sight towards low mass,
low $\tau_{850}$ cores, making these evolved grains appear more prominently
towards less massive cores.

Secondly, grain growth will occur in parallel with evolution of the high-mass
protostar. The mass of material in the core will decrease as material either
collapses to form stars or has been removed from the core by the action of
stellar jets and winds. These mechanisms will act to reduce the opacity
towards more evolved cores by removing dust and gas. For example, the outflows
of high-mass stars may disperse up to eight times the mass of material that
falls onto the star (Churchwell \cite{churchwell97}), allowing the dust
opacity to change by a larger degree than that possible via grain growth
alone.

\section{Mass}
\label{sec:mass}
The majority of photons we detect with SCUBA originate from optically thin
dust emission, so flux contours also trace the mass and column density within
each field of view. The combined mass of the gas and dust can be calculated
from the expression
\begin{equation}
  M = g S_{\nu}d^{2}/\kappa_{\nu}B_{\nu}(T_{dust}),
\end{equation}
where $S_{\nu}$ is the flux density, $d$ is the distance to the source,
$\kappa_{\nu}$ is the absorption coefficient per unit mass of dust, $g$ the
gas-to-dust ratio and $B_{\nu}(T_{dust})$ represents the Planck function for a
blackbody of temperature $T_{dust}$, all measured at frequency $\nu$. We adopt
SBSMW kinematic distances and cold-component dust temperatures, but as
companion clumps do not have individual temperature measurements we have to
assume all multiple detections in a field of view have the same dust
temperature. We derive the absorption coefficient at 850 $\mu$m
($\kappa_{850}$) from the evolutionary opacity models of Ossenkopf \& Henning
(\cite{ossenkopf}), assuming an initial hydrogen number density of
$n_{H}=10^{6}$ cm$^{-3}$, thin ice mantles and a formation timescale of
$10^{5}$ years to get $\kappa_{850}=1.54 \times 10^{-2}$ cm$^{2}$ g$^{-1}$ and
$\kappa_{450}=5.23 \times 10^{-2}$ cm$^{2}$ g$^{-1}$. For parity with other
continuum observations, we assume a gas-to-dust ratio of 100:1, and we list
the derived clump masses in Table \ref{tab:derived}. However, note that the
mass may be uncertain by a factor of 2.5 or more, due to large variation in
the gas-to-dust ratio (eg. Hildebrand \cite{hildebrand}; McCutcheon et al.
\cite{mccutcheon}).

\begin{figure*}
\resizebox{\hsize}{!}{
  \includegraphics[angle=-90,width=0.33\hsize]{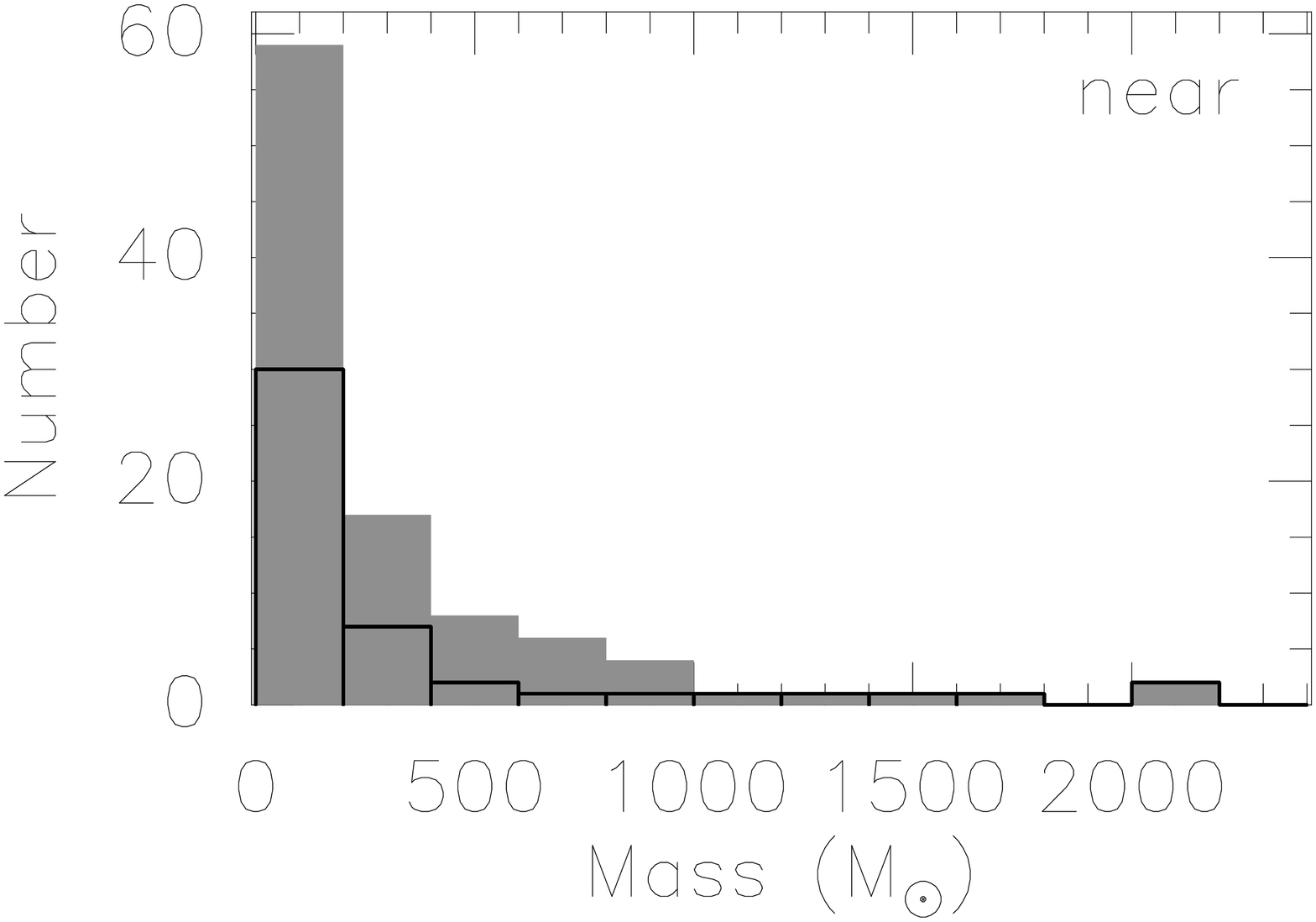}
  \includegraphics[angle=-90,width=0.33\hsize]{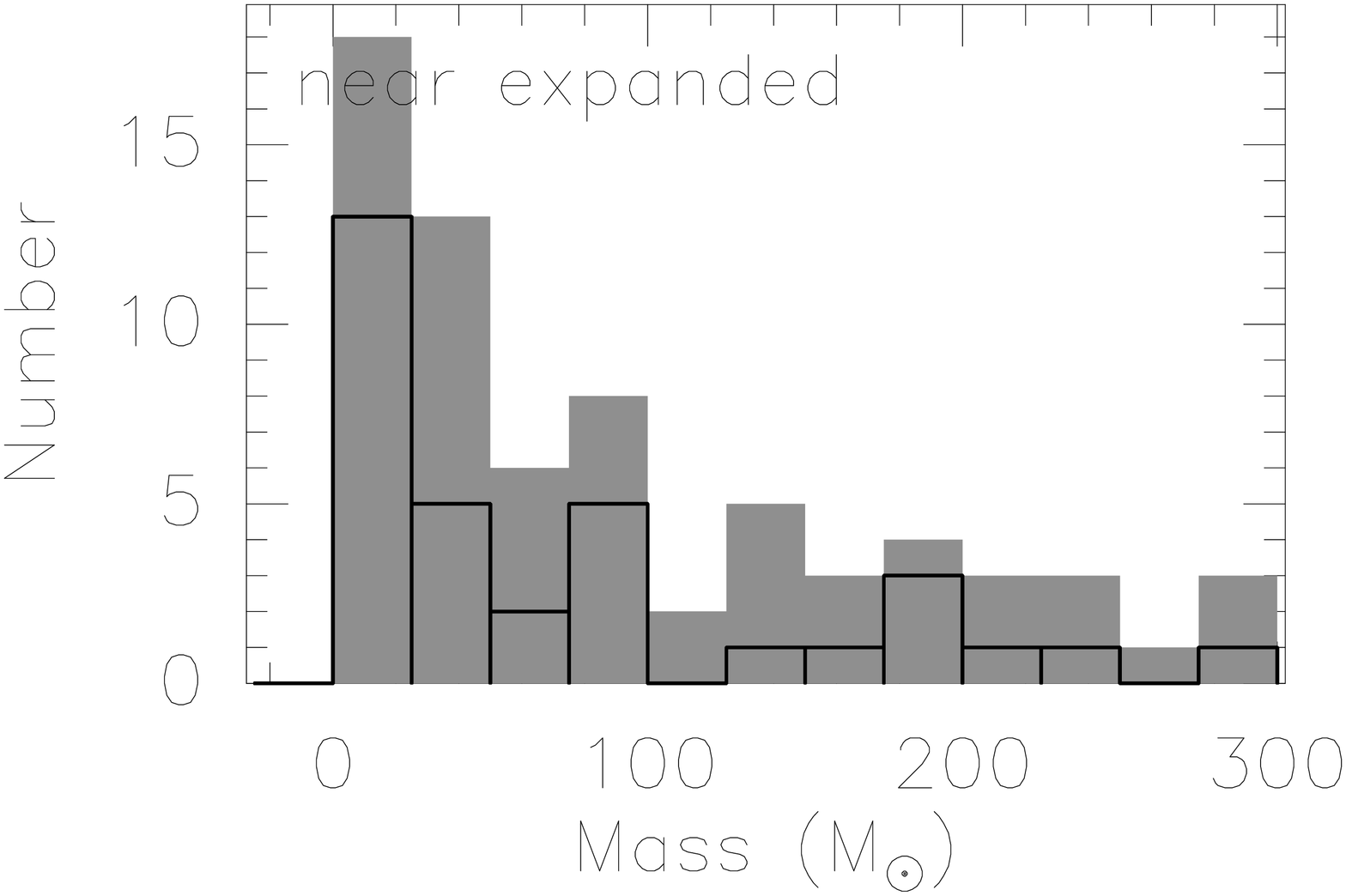}
  \includegraphics[angle=-90,width=0.33\hsize]{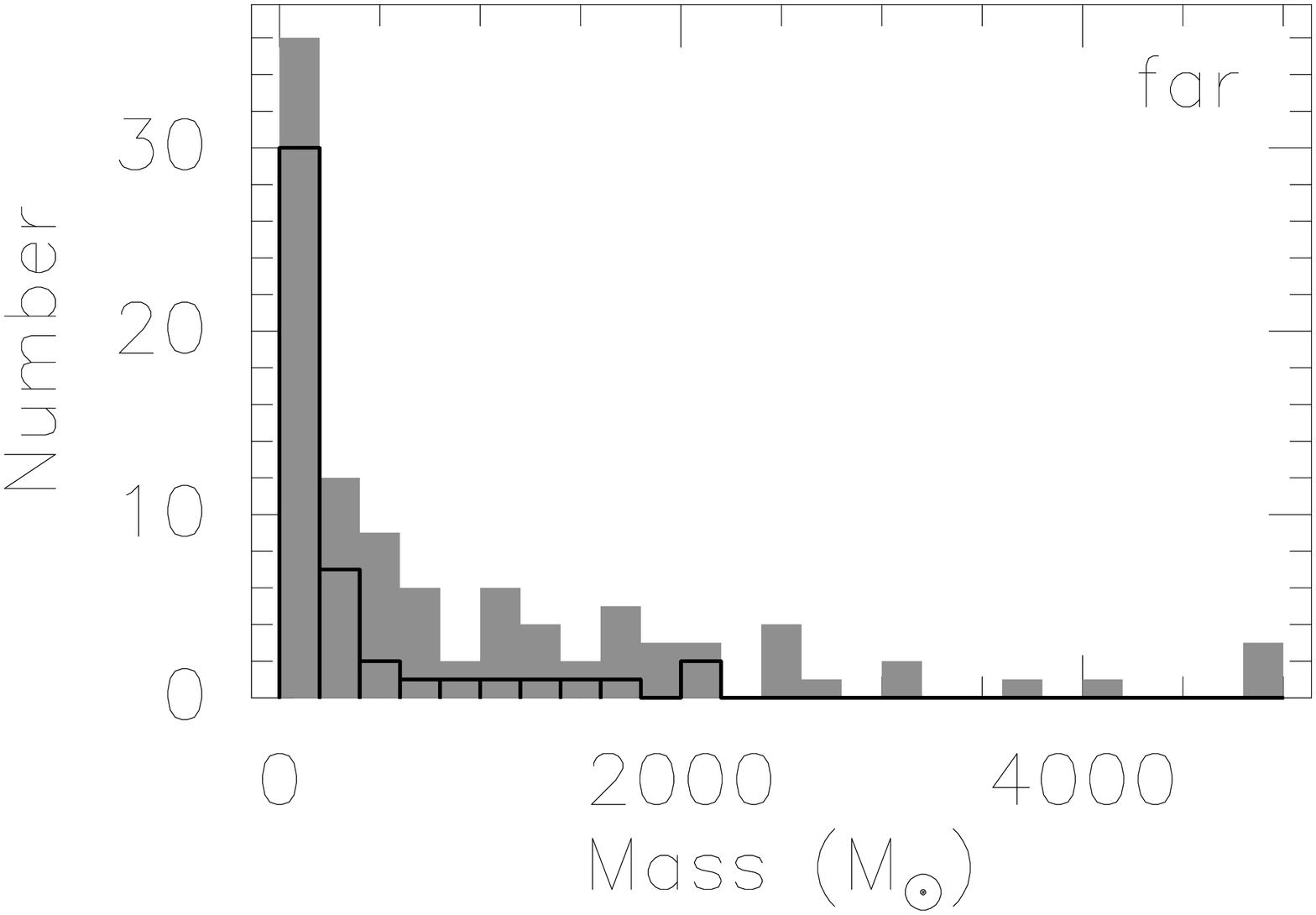}
}
\caption{A histogram of the object mass, assuming the near
  kinematic distance (left-hand and centre plot) and far kinematic distance
  (right-hand plot) for distance-unresolved IRAS fields. The contribution of
  distance resolved detections to each histogram is plotted by an outlined
  histogram. The mass of each detection is calculated from 850 $\mu$m
  continuum emission using a 100:1 gas-to-dust ratio.  The centre panel
  display an expanded view of the near kinematic distance projection for
  clumps with $M<300M_{\odot}$.}
\label{fig:mass histogram}
\end{figure*}

Figure \ref{fig:mass histogram} shows the histogram of clump mass, where we
see the majority of clumps have a mass of less than 500 M$_{\odot}$,
regardless of whether distance-unresolved cores are projected to the near
kinematic distance or the far kinematic distance. The distance-resolved
detections in our sample have a mean clump mass of around 350 M$_{\odot}$,
although the median mass is less at around 100 M$_{\odot}$.  Assuming the near
kinematic distance for distance-unresolved sources results in a mean clump
mass of $\sim$ 330 M$_{\odot}$, with a median roughly half this value at 143
M$_{\odot}$, while projecting to the far kinematic distance results in a mean
clump mass of 1120 M$_{\odot}$ and a median mass of 460 M$_{\odot}$.

The mass of the distance-resolved and near distance-projected clumps are
comparable to other continuum-derived mass estimates of similar massive
protostars. For example, Mueller et al (\cite{mueller}) find an average clump
mass of 209 M$_{\odot}$, while Molinari et al. (\cite{molinari}) measure
$\overline{M}=235$ M$_{\odot}$. These values support claims that natal clumps
bearing massive stars may contain up to 100 times the mass of the most massive
adult star that emerges (Churchwell \cite{churchwell97}). This is a very
different scenario to that seen in the primarily low-mass star-forming region
of Rho Ophiucus (Motte et al \cite{motte98}), where a high fraction of the
initial clump mass is seen to transfer onto the resulting low-mass protostars.

Using the mass and column density relationships defined in Hildebrand
(\cite{hildebrand}), we can also translate the mass of each detection to a
beam-averaged gas column density. The values we derive are presented in Table
\ref{tab:derived}, where we find an average H+H$_{2}$ column density of
$9\times10^{23}$ cm$^{-2}$. With an average clump diameter of 0.6 pc and
projected distance of 4 kpc, this translates to a mean hydrogen number density
of $3\times10^{5}$ cm$^{-3}$ through the clump.

\subsection{Cumulative mass spectrum}
\label{sec:mass spectrum}
\begin{figure}
\includegraphics[angle=-90,width=\hsize]{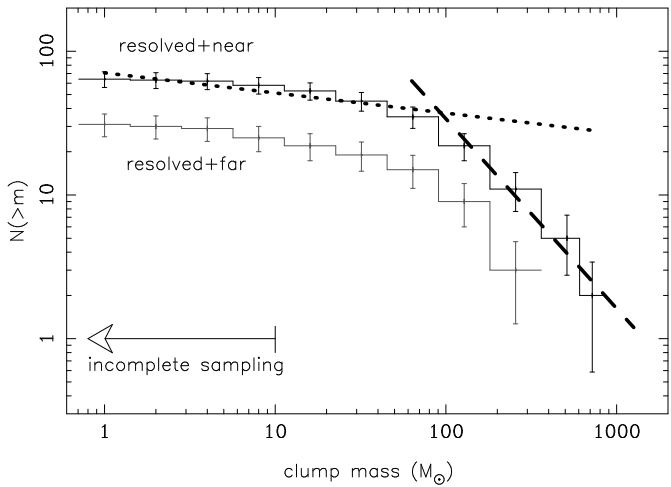}
\caption{Cumulative mass distribution of all detections found at a 
  kinematic distance of $<5$ kpc, incorporating 65 clumps when assuming the
  near kinematic distance (upper curve) and 32 clumps for the far kinematic
  distance (lower curve). The mass of each object is calculated from the 850
  $\mu$m emission using a 100:1 gas-to-dust ratio. The error bars correspond
  to $\sqrt[]{N}$ counting statistics. The thick dotted and thick dashed lines
  plot the lines of best fit for the first seven bins and the last four bins
  respectively.}
\label{fig:mass spectrum}
\end{figure}

Figure \ref{fig:mass spectrum} presents the cumulative mass spectrum of our
850 $\mu$m detections, distance-limited to IRAS fields less than 5 kpc distant
to increase the region of complete sampling. We estimate a completeness limit
of 10 M$_{\odot}$, calculated by determing the mass of a 3-$\sigma$ detection
at the upper distance limit of 5 kpc, assuming the dust temperature of the
detection equals that of the sample average dust temperature, with
$\overline{T_{dust}}$=44 K. The best fit power-law to the mass spectrum is
fairly flat below $\sim 80$ M$_{\odot}$, with $N(>m) \propto m^{-0.14}$. A
break in the spectrum is seen around 100 M$_{\odot}$, above which the mass
spectrum steepens to $N(>m) \propto m^{-1.32}$. The mass distribution
breakpoint is found comfortably above the completeness limit, suggesting this
is not an observational artefact.

We compared the index of these power-law fits to other observations, finding
the distribution below 100 M$_{\odot}$ to be significantly flatter than that
seen in other studies. This points to either an absence or accelerated
evolution of the lower-mass clumps, both of which would reduce the submm
emission observed in the lower mass range. Most likely, this reflects an
absence of low-mass clumps as we have only observed high-mass candidates,
preferentially sampling only the high-mass tail of the initial mass
distribution, and additionally many close low-mass clumps undoubtedly lie
unresolved.  Correcting for these factors could raise the power-law index to a
more typical IMF-like power-law slope, whereas at the moment the most similar
power-law index for this region is that of clumps within molecular clouds,
where the index averages around -0.60 (Kramer et al.  \cite{kramer98}; Kramer
et al.  \cite{kramer}).  This is still a factor of four greater than our
study, and without knowing how far we can minimize this difference we must
question their true degree of similarity.

Above the 100 M$_{\odot}$ breakpoint, the slope of our sample becomes very
similar to that of the field star IMF ($N(>m) \propto m^{-1.3}$ for
$m/$M$_{\odot} \ge 0.5$; Kroupa \cite{kroupa2001}). If the IMF-like
distribution is valid, the apparently similar power-law index of main-sequence
and massive pre-stellar clumps would suggest the star formation efficiency
within these protoclusters is relatively mass invariant; it would appear that
just the breakpoint shifts to lower mass as the core fragments and additional
stars form.

\label{sec:cluster considerations}

Assuming the multiple power-law IMF of Kroupa \cite{kroupa2001}, we may
estimate the number of sources within an average 350 M$_{\odot}$ clump. Even
if the star-formation efficiency $\epsilon$ for a 10M$_{\odot}$ star is as low
as 5\%, this still leaves sufficient mass to form an additional $\sim 100$
lower-mass objects created with efficiency $\epsilon=30\%$, distributed via
the IMF number ratios given in Table \ref{tab:imf}.  With a typical clump
radius of $\sim0.5'$, this would result in a typical stellar volume density of
$\sim150$ stars pc$^{-3}$.

\begin{table}
\begin{small}
  \begin{center}
    \begin{tabular}{lcd{-1}d{-1}}
      \toprule
      \toprule
      &Mass Range  &\multicolumn{2}{c}{Contribution (\%)} \\
      \cmidrule{3-4}
      \multicolumn{1}{c}{Type} &$M_{\odot}$ &\multicolumn{1}{c}{Population} &\multicolumn{1}{c}{Mass} \\
      \midrule
      Brown dwarfs &0.01-0.08   &37   &4.3 \\
      M dwarfs     &0.08-0.5    &48   &28  \\
      K dwarfs     &0.5-1.0     &8.9  &17  \\
      Intermediate &1.0-8.0     &5.7  &34  \\
      O stars      &$>8$        &0.37 &17  \\
      \bottomrule
    \end{tabular}
  \end{center}
\end{small}
\caption{Stellar IMF, taken from Kroupa (\cite{kroupa2001}). The
  third and fourth columns tabulate the contribution of each object type
  to the population and mass total.}
\label{tab:imf}
\end{table}

With a median mass of $\sim$143 M$_{\odot}$, the typical median clump could
easily form at least one $>8$ M$_{\odot}$ star assuming an average 50\%
star-formation efficiency and an IMF-like mass distribution, assuming a
high-mass star accounts for 17\% of the total cluster mass (Table
\ref{tab:imf}).  However, the limited mass reservoir of the lowest-mass
companionless clumps means the star-formation efficiency $\epsilon$ within
these objects must be fairly high, as the total mass of gas and dust is close
to that of a high-luminosity protostar. The limited residual mass reservoir
would also mean that not many lower-mass protostars can co-exist within the
clumps. As a result, the stellar mass spectrum within these clumps must be
skewed compared to the IMF of field stars or that within higher mass clumps.
Although an isolated high-mass protostar has yet to be found, we suggest these
clumps form ideal candidates of high-mass protostars with a minimal number of
stellar companions.

\section{Discussion}
\subsection{Comparison to 1.2mm observations}
\label{sec:consistent calibration}
\label{sec:discussion}
The 850 $\mu$m and 450 $\mu$m emission we observe with the JCMT is found to be
similar to the 1.2 mm IRAM continuum observations of Beuther et al.
(\cite{beuther2002a}). The detections are comparable in size and morphology,
although the extended IRAM field of view reveals more sub-clumps per target
field, strengthening the observation that high-mass star-forming clumps do not
exist in isolation.

\begin{figure}
\resizebox{\hsize}{!}{
  \includegraphics[angle=-90]{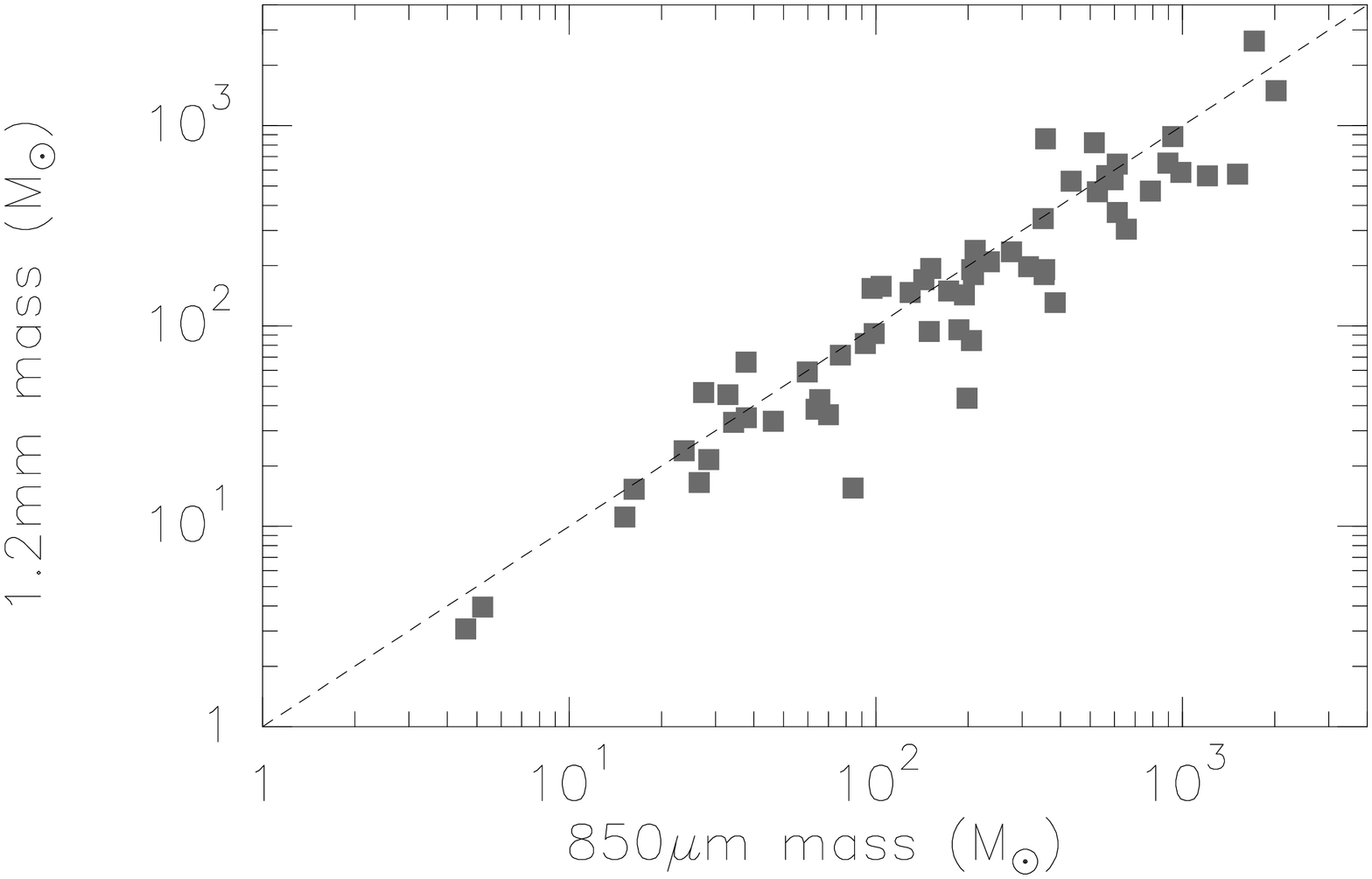}
}
\resizebox{\hsize}{!}{
  \includegraphics[angle=-90]{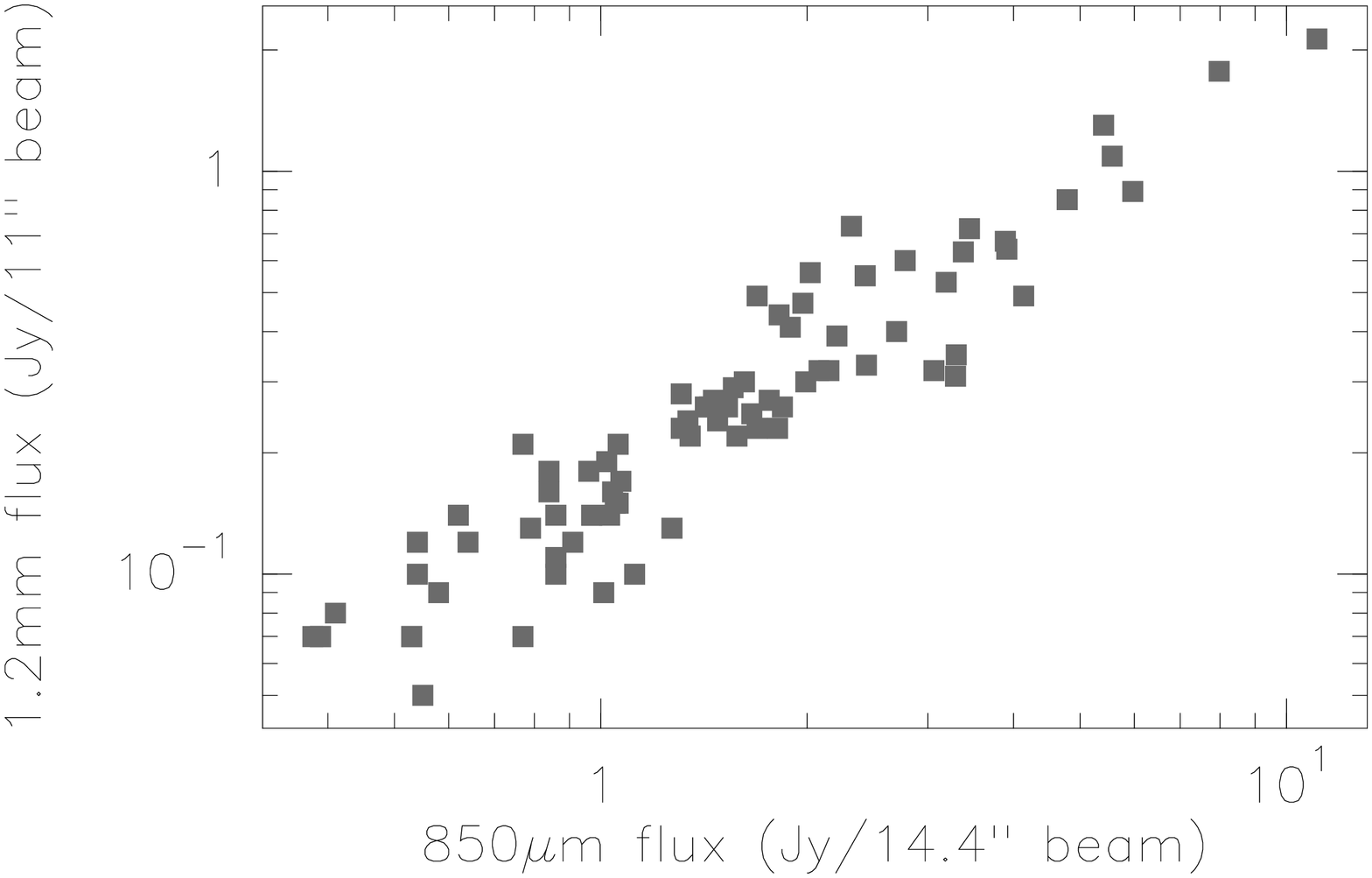}
} 

\caption{A plot displaying the correlation between mass (upper plot)
  and peak flux (lower plot) of coincident IRAM 1.2mm detections and
  JCMT 850 $\mu$m detections. The dashed line in the lower plot traces
  the 1:1 mass ratio.}
\label{fig:flux comparison}
\end{figure}

The mass of coincident 850 $\mu$m detections and 1.2 mm detections correlate
well when the same gas and dust characteristics are assumed, as can be seen in
the upper plot of Figure \ref{fig:flux comparison}. Points lying away from the
main trend generally represent sources with companions lying in extended
emission, suggesting the difference originates in the different techniques
used to measure the integrated emission around extended and multiple component
sources. Admittedly, the choice of where once source ends and another begins
can be subjective, and we do not consider this a cause for concern.

There is good agreement in the peak flux of 850 $\mu$m and 1.2 mm detections,
which implies there is no large optical depth gradient between these
wavelengths, and additionally that the studies are calibrated consistently
with respect to one another, measuring approximately 7 Jy/14.4$''$ beam at
850 $\mu$m per 1 Jy/11$''$ beam measured at 1.2 mm (lower plot of Figure
\ref{fig:flux comparison}).  This is another indication that the 850 $\mu$m
and 1.2mm observations are detecting the same material and thus trace the same
amount of mass. By confirming the mass of the clumps, we may place more
confidence in the mass-luminosity relationship derived by Sridharan et al.
(\cite{sridharan}).

\subsection{Distance ambiguities}
\label{sec:distance resolved}
We remain conscious that the distance ambiguities remaining towards a number
of sources could alter the conclusions we are able to draw.  Attempts to
resolve the uncertainties using the C$^{34}$S(2-1) and CS(2-1) linewidths of
Beuther et al. (\cite{beuther2002a}), comparing the continuum-derived mass
estimate with the virial mass at both kinematic distances, proved fruitless.
However, the simulations of Walsh et al. (\cite{walsh2001}) require the
majority of our distance unresolved sources to be placed at the near kinematic
distance if we are to obtain $M/L$ ratios similiar to those expected for the
majority of $10^{2}-10^{3} M_{\odot}$ clusters.

The greybody analysis of SBSMW also gives an estimate of the bolometric
luminosity of each IRAS source. Assuming $M_{*} \propto L_{bol}^{0.25}$ (where
$M_{*}$ is the mass of the protostar and $L_{bol}$ is the SBSMW luminosity),
similar to the mass-luminosity relationship for massive stars on the main
sequence, it is simple to calculate the mass of the illuminating source. In
comparing $M_{*}$ to the mass of the clump at both the near and far kinematic
distance projections, we found the luminosity of detections \#64 (IRAS
18553+0414) and \#74 (IRAS 19266+1745) to be incompatible with their
near-distance clump mass, suggesting that these sources are actually located
at the far kinematic distance.

\section{Conclusions}
\label{sec:conclusion}

We observed a sample of candidate high-mass protostars with the JCMT.
Dust continuum emission was detected towards all sources, and from
analysis of the 850 $\mu$m and 450 $\mu$m maps we reach the following
conclusions:

\begin{itemize}
  
\item The average clump has a mass of $\sim330$ M$_{\odot}$, a total hydrogen
  column density of $\sim9\times10^{23}$ cm$^{-2}$, and an angular diameter of
  30$''$, which projected to the mean projected distance of 4 kpc equals a
  linear diameter of 0.6 pc. Assuming spherical symmetry, these values
  translate to an average hydrogen number density of $3\times10^{5}$ cm$^{-3}$
  in a typical clump.
  
\item The clumps we detect vary in mass from around 1 M$_{\odot}$ to over 1000
  M$_{\odot}$. The high luminosity and low mass of the smallest clumps
  suggests they will only form a minimal number of stellar companions, and
  that we are close to probing the final mass of the most luminous protostar.
  These small, isolated clumps may be analogues of the low-mass NH$_{3}$ cores
  (Myers \& Benson \cite{myers}), from which we have learned a great deal
  about low mass star formation. As such, these small high luminosity cores
  may be the most promising sites for further high-resolution observations.
  The extremely large mass of the largest clumps suggest they may be forming
  proto-groups or proto-clusters.
  
\item Above 100 M$_{\odot}$, the mass spectrum of the submm detections
  displays a distribution very similar to that of the field star IMF,
  suggesting the clump-mass to stellar-mass transfer efficiency is relatively
  mass invariant within these protoclusters.
    
\item A large fraction of the submm clumps we detect are roughly coincident
  with IRAS and MSX detections in the region. This suggests the short
  wavelength photons and submm photons have the same origin, despite the
  typical envelopes being optically thick at wavelengths below 90 $\mu$m. The
  coincident detections could be conceivably be reconciled by non-uniformity
  of the envelope, eg.  where outflow cavities provide a low-opacity escape
  route for short wavelength photons.
 
\item We also find some submm detections are significantly offset from their
  associated MSX and IRAS counterparts, in agreement with Sridharan et al.
  (\cite{sridharan}). This raises the question of whether these objects are
  related, or whether they are actually unique objects with different submm
  and IR properties. If the clumps we detect are potential proto-groups or
  proto-clusters, it is possible that young stars in the less dense, more
  transparent extremeties of the envelope may be more visible at shorter
  wavelengths. This possibility could explain both coincident and
  non-coincident long/short wavelength detections.
  
  The degree of coincidence may point to evidence of evolution, from the
  oldest, most evolved sources with MSX detections and no coincident submm
  flux, to MSX detections with some degree of submm flux, to the youngest,
  most embedded sources with potentially no MSX detection and high IR optical
  depth. Further investigation of these possibilities will require high
  resolution observations at submm and far infra-red wavelengths.
  
\item We find a companion clump fraction of $\sim 0.6$, emphasising that
  clumps bearing high-mass stars do not form in isolation, and that they may
  lie in close proximity to other potentially star-forming clumps.
  
\item We measure the mean surface density of companions, and find the clump
  spatial density distribution peaks at a separation of $r \sim 0.4$ pc. The
  projected distribution ($N_{proj}(r)\propto r^{-1.7}$) corresponds to a
  volume density distribution such that $N_{vol}(r)\propto r^{-0.75}$.
  
\item The mean spectral index of the dust emission, $\alpha$, at the position
  of peak submm emission is 2.6 $\pm$ 0.4, and we observe both $\alpha$
  morphologies both correlated and anti-correlated with submm intensity.
  
  Peaked $\alpha$ morphologies can be reproduced with simple internal
  heating of a dusty envelope, with the observed $\alpha$ gradient
  resulting from the corresponding temperature gradient through the
  envelope.
  
  An $\alpha$-dip morphology could occur if the inner region of the
  clumps are cool with respect to their surroundings or if substantial
  grain growth occurs in these dense central regions. Although we
  suggest grain growth is the most likely factor, identifying the
  dominant mechanism will require high-resolution observations of
  temperature tracers towards the regions.

\item The mean spectral index of the optical depth, $\beta$, at the
  position of peak submm emission is 0.9 $\pm$ 0.4. This is lower than
  $\beta$ towards in most dusty regions, but could be consistent with
  grain growth occuring in the dense clumps we observe.

\end{itemize}

We would like to thank Claire Chandler for her insight and invaluable comments
that helped improve this paper.

The James Clerk Maxwell Telescope is operated on a joint basis between the
United Kingdom Particle Physics and Astronomy Research Council (PPARC), the
Netherlands Organization for the Advancement of Pure Research (ZWO), the
Canadian National Research Council (NRC), and the University of Hawaii (UH).
This research has made use of NASA's Astrophysics Data System Bibliographic
Services and the SIMBAD database, operated at CDS, Strasbourg, France.



\begin{thebibliography}{}
\bibitem[1975]{aannestad} Aannestad, P.~A.,
  1975, ApJ 200, 30

\bibitem[2003]{beck2003} Beck, T.~L., Simon, M., \& Close, L.~M.,
  2003, ApJ 583, 358

\bibitem[1991]{beckwith1991} Beckwith, S.~V.~W., \& Sargent, A.~I.,
  1991, ApJ 381, 250

\bibitem[2001]{behrend2001} Behrend, R., \& Maeder, A.,
  2001, A\&A 373, 190
  
\bibitem[2002a]{beuther2002a} Beuther, H., Schilke, P., Menten, K.~M., Motte,
  F., Sridharan, T.~K., \& Wyrowski, F., 2002a, ApJ 566, 945
  
\bibitem[2002b]{beuther2002b} Beuther, H., Schilke, P., Sridharan, T.~K.,
  Menten, K.~M., Walmsley, C.~M., \& Wyrowski, F., 2002b, A\&A 383, 892

\bibitem[2001]{brand2001} Brand, J., Cesaroni, R., Palla, F. \& Molinari, S.,
  2001, A\&A 370, 230

\bibitem[1996]{bronfman} Bronfman, L., Nyman, L.~A., \& May, J.
  1996, A\&AS, 115, 81
  
\bibitem[1997]{cesaroni1997} Cesaroni, R., Felli, M., Testi, L., Walmsley, C.~M., \& Olmi, L.,
  1997, A\&A 325, 725

%
\bibitem[2001]{sun226} Chipperfield, A.~J., \& Draper, P.~W.,
  2001, Starlink User Note 226
  CCLRC / Rutherford Appleton Laboratory / Particle Physics \& Astronomy 
  Research Council

%
\bibitem[1997]{churchwell97} Churchwell, E.,
  1997, ApJ 479, L59

%

\bibitem[1984]{draine} Draine, B.~T., \& Lee, H.~M.,
  1984, ApJ 285, 89

%
%
%
\bibitem[2004]{fuller2003} Fuller, G.~A., Williams, S.~J., \& Sridharan, T.~K.,
  2004, in. prep.

\bibitem[1993]{galli} Galli, D., \& Shu, F.~H.,
  1993, ApJ 417, 220

%
\bibitem[1973]{gerazi} Gerazi, D., Joyce, R., \& Simon, M.,
  1973, ApJ 179, L67

\bibitem[1997]{goldsmith} Goldsmith, P.~F., Bergin, E.~A., \& Lis, D.~C.,
  1997, ApJ 491, 615

\bibitem[1993]{gomez93} Gomez, M., Hartmann, L., Kenyon, S., Hewett, R.,
  1993, AJ 105, 1927

\bibitem[1979]{habing1979} Habing, H.~J, \& Israel, F.~P., 
  1979, ARA\&A 17, 345

%
\bibitem[1983]{hildebrand} Hildebrand, R.~H.,
  1983, QJRAS 24, 267

\bibitem[2000]{hogerheijde} Hogerheijde, M.~R., \& Sandell, G.,
  2000, ApJ 534, 880

\bibitem[1999]{holland} Holland, W.~S., Robson, E.~I., Gear, W.~K., et al.,
  1999, MNRAS 303, 659

\bibitem[1997]{hunter thesis} Hunter, T.,
  1997, PhD Thesis, Caltech

%
\bibitem[1998]{jenness} Jenness, T., \& Lightfoot, J.~F., 
  in Astronomical Data Analysis Software and Systems VII,
  ASP Conf. Ser. 145 
  ed.\ Albrecht R., Hook R.~N., Bushouse H.~A. 

\bibitem[1996]{kramer} Kramer, C., Stutzki, J., \& Winnewisser, G.,
  1996, A\&A 307, 915
  
\bibitem[1998]{kramer98} Kramer, C., Stutzki, J., R\"ohrig, R., \&
  Corneliussen, 1998, A\&A 329, 249

\bibitem[1994]{krugel} Kr\"{u}gel, E., \& Siebenmorgen, R.,
  1994, A\&A 288, 929

\bibitem[2001]{kroupa2001} Kroupa, P.,
  2001, MNRAS 322, 231

%
\bibitem[2000]{kurtz:ppiv} Kurtz, S., Cesaroni, R., Churchwell, E., Hofner,
  P., \& Walmsley, C.~M., 2000, in Protostars and Planets IV, ed. V. Mannings,
  A.~P.  Boss, \& S.~S.  Russell (Tucson: Univ. Arizona Press), 299

%
\bibitem[1995]{larson95} Larson, R.~B.,
  1995, MNRAS 272, 213

\bibitem[1994]{mannings} Mannings, V., \& Emerson, J.~P.,
  1994, MNRAS 267, 361

\bibitem[1977]{mrn} Mathis, J.~S., Rumpl, W., \& Nordsieck, K.~H.,
  1977, ApJ 217, 425

\bibitem[1989]{mathis} Mathis, J.~S., \& Whiffen, G.,
  1989, ApJ 341, 808
  
\bibitem[1995]{mccutcheon} McCutcheon, W.~H., Sato, T., Purton, C.~R.,
  Matthews, H.~E., \& Dewdney, P.~E., 1995, AJ 110, 1762

\bibitem[2000]{molinari2000} Molinari, S., Brand, J., Cesaroni, R., \& Palla, F.,
  2000, A\&A 355, 617

\bibitem[2002]{molinari} Molinari, S., Testi, L., Rodr\'{i}guez, L.~F., \& Zhang, Q.,
  2002, AJ 570, 758

\bibitem[1998]{motte98} Motte, F., Andr\'e, P., \& Neri, R.,
  1998, AA 336, 150

\bibitem[2001]{motte01} Motte, F., \& Andr\'e, P.,
  2001, A\&A 365, 440

\bibitem[2002]{mueller} Mueller, K.~E., Shirley, Y.~L., Evans N.~J. II, \& Jacobson, H.~R.,
  in Hot Star Workshop III: The Earliest Phases of Massive Star Birth,
  ASP Conf. Ser.
  ed. Crowther P.~A.

\bibitem[1983]{myers} Myers, P.~C., \& Benson, P.~J,
  1983, ApJ 266, 309

%
\bibitem[1998]{nakajima98} Nakajima, Y., Tachihara, K., Hanawa, T. \& Nakano, M.,
  1998, ApJ

%
\bibitem[1994]{ossenkopf} Ossenkopf, V., \& Henning, Th.,
  1994, A\&A, 291, 943
  
\bibitem[1993]{palla1993} Palla, F., Cesaroni, R., Brand, J., Caselli, P.,
  Comoretto, G., \& Felli, M., 1993, A\&A 280, 599

\bibitem[2002]{patience2002} Patience, J., Ghez, A.~M., Reid, I.~N., \& Matthews, K.,
  2002, AJ 123, 1570

%
\bibitem[1997]{ramesh} Ramesh, B., \& Sridharan, T.~K.,
  1997, MNRAS 284, 1001

\bibitem[1955]{salpeter} Salpeter, E.~E.,
  1955, ApJ 121, 161

\bibitem[2001]{sandell} Sandell, G., Jessop, N., \& Jenness, T.,
  2001, The SCUBA map reduction cookbook, 
  CCLRC / Rutherford Appleton Laboratory / Particle Physics \& Astronomy 
  Research Council
  
\bibitem[1997]{schmitt1997} Schmitt, W., Henning, T., \& Mucha, R., 
  A\&A 325, 569
   
\bibitem[1999]{shepherd1999} Shepherd, D.~S., \& Kurtz, S.~E., 1999, ApJ 523,
  690

%
%
\bibitem[2002]{sridharan} Sridharan, T.~K., Beuther, H., Schilke, P., Menten, K.~M., \& Wyrowski, F.,
  2002, ApJ 566, 933


%
%
\bibitem[2004]{williams2004} Williams, S.~J., Fuller, G.~A., \& Sridharan, T.~K,
  2004, In prep.

%

\bibitem[1989]{wood} Wood, D.~O.~S., \& Churchwell, E.,
  1989, ApJ 340, 265
  
\bibitem[2001]{walsh2001} Walsh, A.~J., Bertoldi, F., Burton, M.~G., \&
  Nikola, T., 2001, MNRAS 326, 36

%
%

\bibitem[2001]{zhang2001} Zhang, Q., Hunter, T.~R., Brand, J., Sridharan, T.~K., Molinari, S., Kramer, M.~A., \& Cesaroni, R.,
  2001, ApJ 552, L167
\end{thebibliography}
\end{document}